\documentclass[12pt,3p,review,doublespacing]{elsarticle/elsarticle}
\journal{Journal of the Taiwan Institute of Chemical Engineers}
\emergencystretch=\maxdimen
\usepackage{natbib}
\setcitestyle{round, square, number}
\usepackage{hyperref}
 \hypersetup{
 	plainpages = false, 
	bookmarksopen = true,
	bookmarksnumbered = true,
	breaklinks = true,
	linktocpage,
	colorlinks = true,
	linkcolor = purple,
	urlcolor  = black,
	citecolor = purple,
	}
\usepackage{multicol}
\usepackage{multirow}
\usepackage{imakeidx}
\usepackage{nomencl}
\makenomenclature
\usepackage[xindy]{glossaries}
\makeglossaries
\usepackage{graphics}
\usepackage{graphicx}
\usepackage{epsf,psfrag}
\usepackage{epstopdf}
\usepackage[para]{threeparttable}
\usepackage{subfig}
\usepackage{epsfig}
\usepackage{pdflscape}
\usepackage{rotating}
\usepackage{lscape}
\usepackage{float}
\usepackage{stfloats}
\usepackage{textgreek}
\usepackage{longtable}
\usepackage{lscape}
\usepackage{capt-of}
\usepackage{comment}
\usepackage{tablefootnote}
\usepackage{footnote}
\usepackage{caption}
\usepackage[autopunct=true]{csquotes}
\usepackage{rotating}
\usepackage{txfonts}
\usepackage[normalem]{ulem}
\usepackage{setspace}
\usepackage{color,xcolor,soul}
\usepackage{amsfonts,amsmath,amssymb}
\usepackage{latexsym,array}
\usepackage{textcomp}

\newcommand{\RomanNumeralCaps}[1]
\linenumbers
\makesavenoteenv{tabular}
\makesavenoteenv{table}
\emergencystretch=\maxdimen
\usepackage[left,displaymath, mathlines]{lineno}
 
\providecommand\bnabla{\boldsymbol{\nabla}}

\DeclareMathAlphabet{\mathpzc}{OT1}{pzc}{m}{it}
\def\fig{Figure~}
\def\figs{Figures~}
\def\eqn{Eq.~}
\def\eqns{Eqs.~}
\def\tab{Table~}

\def\micro{\textmu}
\providecommand\bnabla{\boldsymbol{\nabla}}

\providecommand\p{{\partial}}

\newcommand\cac{Ca_{\text{c}}}
\newcommand\qr{Q_{\text{r}}}

\def\tsc#1{\csdef{#1}{\textsc{\lowercase{#1}}\xspace}}
\tsc{WGM}
\tsc{QE}
\tsc{EP}
\tsc{PMS}
\tsc{BEC}
\tsc{DE}
\newcommand{\rev}[1]{\textcolor{black}{#1}}     

\begin{document}
%
%
%
\begin{frontmatter} 
%
%
%
%
\title{{Effects of capillary number and flow rates on the hydrodynamics of droplet generation in T-junction microfluidic systems}}
\author[labela]{Akepogu Venkateshwarlu}
\author[labela]{Ram Prakash Bharti\corref{coradd}}\ead{rpbharti@iitr.ac.in}
\address[labela]{Complex Fluid Dynamics and Microfluidics (CFDM) Lab, Department of Chemical Engineering, Indian Institute of Technology Roorkee, Roorkee - 247667, Uttarakhand, INDIA}
%
%
\cortext[coradd]{\textit{Corresponding author. }}
%
\begin{abstract}
The control and manipulation of the hydrodynamics of droplets primarily relate to flow governing and geometrical parameters. This study has explored the influences of capillary number ($10^{-4}\le \cac\le 1$) and flow rate ratio ($0.1 \le \qr\le 10$) on the hydrodynamics of droplet generation in two-phase flow through T-junction cross-flow microfluidic device. The finite element method is used to solve the Eulerian framework of a mathematical model based on mass continuity, Navier-Stokes, and \rev{conservative} level set equations at fixed flow ($Re_{c}=0.1$). Results are presented in terms of the instantaneous phase flow field, droplet size, droplet detachment time, and generation frequency as a function of governing parameters ($\cac$ and $\qr$). The flow regimes namely squeezing, first transition, dripping, second transition, parallel, and jet flow are marked. In contrast to reported value of threshold $\cac \approx 10^{-2}$, squeezing regime exists for all $\cac$ and $2\le \qr \le 10$. The flow regimes are also mapped into droplets and non-droplet zones by using threshold $\cac$ which scales quadratically with $\qr$. The droplet length varies linearly with $\qr$ in the squeezing regime. Both droplet size and frequency show a power-law relation $\cac$ and $\qr$ in the droplet zone. Finally, predictive correlations are presented to guide the engineering and design of droplet microfluidics devices.
\end{abstract}
\end{frontmatter}
%
%
\section{Introduction}
\label{sec:intro}
\noindent The study of emulsion generation from two immiscible fluids has drawn much attention in recent years due to its versatile applications in many areas of science and engineering. For instance, emulsions are widely prevalent in many   industrial processes and engineering fields like mixing, pumping, pharmaceuticals,  cosmetics, food, agriculture, coatings and paints \citep{Asua2002,Mulqueen2003,vanderSchaaf2017}, biomedical reagents, inks, oil recovery and mining \citep{Barnes1994,Stone2004,Cristini2004,Whitesides2006,Shi2014,Mansard2016,Zhu2017,Gerecsei2020}.
The droplet generation originates from the fluid\rev{-fluid} interface instability, which is attributed to the imbalance of the interfacial force between the two fluids and the inertial and viscous forces due to flow. From the Rayleigh--Plateau instability, it is understood that there is a pressure difference created between the two phases at the interface, and a combination of the additional inertial forces \rev{(i.e., momentum flux from high to low pressure)} eventually leads to the detachment or breakup of the droplets \citep{Anna2016}. \rev{The interfacial tension acts as a gain on the interface shape divergence ($\nabla\cdot n$) and thus, increasing interfacial tension leads to more pronounced pressure gradient (as per Young-Laplace equation) and faster droplet formation.} 
Microfluidic devices are the most useful platform to generate emulsions having uniform size droplets through various methods on a micro-scale level. 
\\
The commonly used geometries to produce different sizes of the droplets are cross-flow, co-flow, and flow-focusing microfluidic devices \citep{Anna2003,Eggers2008,Glawdel2012a,Glawdel2012b,Glawdel2012c,Zhu2017,Doufene2019}. %
In flow-focusing systems, due to their geometrical constraint, it is difficult to control and maintain the frequency of generation of droplets. Cross-flow (T- or cross-shaped) devices are commonly used in experimental and numerical investigations to study droplet dynamics such as formation, fission, and fusion \citep{Li2012,Liu2009,Kang2019,Liu2019} due to their distinct advantages of simple and efficiently controlled modulation of droplet size and frequency. 
The geometry is extensively used because of its simplicity and capability to produce mono-dispersed droplets \citep{Zhu2017}. 
\\
In the T-shaped microfluidic geometry, the \rev{side} channel (containing a dispersed phase, DP) intersects perpendicular to the main channel  (containing the continuous phase, CP) and thereby forming up a T-shape. In this arrangement, the continuous and dispersed phases, entering through inlets of the main channel and T branch, meet at the junction point and flow downstream of the main channel along with the continuous phase (CP). Depending upon the interplay between various forces acting throughout the flow, either the two-layer flow or the emulsions consisting of droplets or bubbles of the dispersed phase (DP) may be formed. 
Generally, the flow through such microfluidic systems is governed by the two dimensionless numbers: capillary number ($Ca=u \mu/\sigma $) and Reynolds number ($Re=\rho u w/ \mu $) representing the relative importance of viscous over interfacial forces,  and inertial to viscous forces, respectively. The flow dynamics is further governed by the physical properties (density $\rho_{\text{r}} = \rho_{\text{d}}/\rho_{\text{c}}$, viscosity $\mu_{\text{r}} = \mu_{\text{d}}/\mu_{\text{c}}$) and flow rate ($\qr = Q_{\text{d}}/Q_{\text{c}}$) of dispersed and continuous phases. Here, subscripts $c$, $d$  and $r$ denote for the continuous and dispersed  phases, and ratio, respectively.
Depending on the value of $Ca$, generally, three main regimes, i.e., dripping, jetting, and squeezing, are distinguished and reported in the literature. 
Smaller droplets are formed at high $Ca$ and low $Re$ due to stronger shearing (large viscous force) of a dispersed phase by the continuous phase. However, larger droplets are obtained at low $Ca$ due to restricting of the downstream \rev{flow}. 
\\
Many experimental and numerical studies have been conducted to understand the underlying dynamics of the droplet formation in the T-type microchannels. For instance, \citet{Thorsen2001} have performed experiments and reported that the geometry of channels has a significant impact on the droplet generation pattern. \citet{Nisisako2002}  have introduced an experimental method to generate water droplets in oil in the T-junction microchannel and reported that the droplet size is inversely proportional to the flow rate of the continuous phase. \citet{VanderGraaf2006} studied the formation of droplet necking both experimentally and numerically. 
Several studies have reported the influence of viscosity ratio ($\mu_{\text{r}}$) on the droplet size.  \citet{Nekouei2017} have reported that, for a fixed $Ca$, the droplet size increased with increasing  $\mu_{\text{r}} > 1$ and it  does not vary much for $\mu_{\text{r}}<1$. 
{At low $Re$, viscous and interfacial forces play a crucial role in the formation of the droplet due to insignificant inertia.  At very low capillary numbers ($\cac\ll10^{-2}$), the dispersed phase occupies a relatively large flow area in the main channel and resists the continuous phase stream. This leads to a higher pressure buildup in the upstream region, followed by the necking of the dispersed phase, eventually resulting in the formation of the droplet. The droplet breaks up at the junction point towards the downstream length, and it is not affected by the wall shear stress. The length of the droplet is \rev{reported to be} a function of the channel \rev{size (i.e., $w_{\text{r}}$)} and the flow rate ratio ($\qr$) of the two liquid phases.}  
It has been found that the droplet length ($L$) varies linearly with the flow rate ratio ($\qr$) \citep{Garstecki2006} for  a given microfluidic geometry. The droplets with a length ($L$) greater than the width ($w_{\text{c}}$) of the main channel were observed to be stable and uniform in size \citep{Anna2016,Bashir2014,Demenech2008,Garstecki2006}. Such flow governing conditions are called a `squeezing regime' and there is a transition of droplet behavior at $\cac\approx10^{-2}$. 
For $\cac>10^{-2}$,  the interfacial force prevents the dispersed phase from entering into the main channel. In contrast, the inertial and viscous forces assist the same, thereby forming a droplet further downstream by the shear stresses. The droplet size decreased with increasing $\cac$. Such flow governing conditions are called the `dripping regime' 
\citep{Garstecki2006,Christopher2008,Bashir2011}.  
At higher $\cac$, the flow becomes a jet-type, two-layered flow, and chaotic due to the dominance of inertial effects. \cite{Tarchichi2013} reported another flow regime, named as `ballon type', at low velocities of the dispersed phase. 
\\
{In general, it is challenging to generate mono-dispersed droplets due to \rev{intense competition among} inertial, viscous and interfacial forces. Therefore, it is crucial to understand the droplet dynamics for a wide range of parameters.}
It is further acknowledged that the experimental investigations \rev{to understand of the droplet generation and underlying physics at a higher viscosity ratio ($\mu_{\text{r}}>1)$} for such a broad range of governing parameters become \rev{quite complex and} difficult  \citep{Abate2012}.  Simultaneous control of the governing parameters in the experiments is another challenge \citep{Nekouei2017}.  Numerical simulations, therefore, are an alternate way to overcome the limitations encountered with the experiments to study precise \rev{simultaneous} control of the flow parameters upon the droplet generation and dynamics mechanisms.  \rev{For instance, a} simulation study \citep{Cristini2004} has shown great potential to describe the transitions accurately and capable of extending for various other geometries over a wide range of parameters.
\\ \noindent
\tab\ref{tab:1} has summarized the relevant experimental, numerical and theoretical studies of the two-phase flow in a T-junction microfluidic devices. {However, it is found that there is no single correlation known to predict whether or not there is a droplet formation for the specific values of $Ca_{\text{c}}$ and $Q_{\text{r}}$.}
\begin{sidewaystable}
		\caption[Summary of experimental and numerical studies of the droplet formation in a T-junction microchannel.]{Summary of experimental and numerical studies of the droplet formation in a T-junction microchannel. [\textbf{Num}: numerical; \textbf{Exp}: experimental; $^a$Lattice Boltzmann method (LBM), $^b$Phase-field method (PFM), $^c$Level set method (LSM), $^d$Volume of fluid (VOF) method] }
\label{tab:1}
	\scalebox{0.78}
	{
		\centering
		\begin{tabular}{lp{0.55in}p{0.7in}cccccccp{0.75in}p{1.2in}}
			\hline
			Reference               &Type     & Fluids    &$\cac$               &$Re_{\text{c}}$             &$w_{\text{r}}$        &$\qr$     &      &$\mu_{\text{r}}$    &$\theta^{\circ}$        & $\sigma (mN/m)$  &Remarks  \\ 
			\hline    
			\citet{Garstecki2006}    & Exp     & Silicone oil, water &$<10^{-2}$    &$-$                    &$0.25-0.1$     & $0.01-10$ &  & $0.01-0.1$           &$-$            &$36.5$     &f($Q_{r}$, geometry)    \\
			\citet{VanderGraaf2006} & Num$^a$, Exp  & Oil, water & $10^{-2}-8\times10^{-2}$  &$-$   		   & $1$          & $0.05-1$  &   &$3.44$             &$115 -180$ &$1-15$    &  f($\theta$, $\sigma$, $Q_{r}$)  \\
			\citet{Christopher2008}  & Exp     & Silicone oil, water  &$5\times10^{-3}-0.3$   &$-$         &$0.5-2.5$     &$0.05-2.5$ & &$0.003-0.2$ &-         &$46$     & f($\mu_{r}$, geometry)  \\
			\citet{Demenech2008}  & Num$^b$  &-   & $10^{-3}-7\times10^{-2}$   & $-$  		& $1$                 	& $0.01-2$    &    &$0.125-1$             &$150$    &$12.5$        & f($\mu_{r}$, pressure)   \\
			\citet{VanSteijn2010}    & Exp    & Fluorinated oil, water &$<10^{-2}$  & $-$                   &$0.33-3$       & $0.1-10$       &    & $0.01-0.1$  &                  $170$ & $17.9$ &  f(geometry)    \\
			\citet{Bashir2011}     & Num$^c$, Exp & Oil, water & $10^{-3}-10^{-2}$  & $0.4-2.4$  	& $0.6$        & $0.05-0.2$   &     &$0.1-0.8$          &$120-180$ &$1-10$  & f($\theta$, $Q_{r}$)      \\
			\citet{Abate2012}       & Exp  & Oil, water    &$0.03-0.21$   &$1$                  & $-$             &$-$              &    & $0.833$    &$-$             &$4$      &f(pressure)   \\
			\citet{Li2012}         & Num$^d$ &$-$   & $\leq0.1$  		        & $-$      		       &  $1$         &$0.009-0.83$  &    &$0.154$               &$60-180$   &- & f($\mu_{r}$, pressure)       \\
			\citet{Shi2014}        & Num$^a$   &$-$   & $1.5\times10^{-3}-10^{-1}$  & $-$              & $0.25-2$      & $0.1-0.9$    &    &$0.25-1$                 &$60-180$  &-  & f($\theta$, geometry)      \\
			\citet{Wehking2014}     & Exp     & Silicone oil, many fluids &$10^{-3}-0.4$  &$-$          &$0.5$           & $0.084-49$   &    & $0.01-1$    & $-$                 &$5.42-43.12$     &  f($\sigma$)   \\ 
			\citet{Nekouei2017}     & Num$^d$     &$-$   & $10^{-3}-0.5$      & $<0.1$                 & $0.33-3$     & $0.05-10$    &    & $0.01-15$               & $180$ & -            & f($\mu_r$, $Q_{r}$)       \\ 
			\citet{Zhang2018}       & Exp    & Mineral oil, water   &-              &-                   &$1$       & $0.00833-2$  & & $0.19-0.2$  & $146.8$  &$4$     & f($\mu_r$, $Q_{r}$)   \\
			\citet{Wong2019}          & Num$^c$      & Olive oil, water  &$-$        & $-$                   &$0.4$          & $0.04-0.0625$   &    & $0.147-1$             & $180$  & - & f($\mu_r$, $Q_{r}$)     \\
			\citet{Zeng2020}        & Exp    & Silicone oil, water  & $\leq0.1$     & $-$                   &$0.5-1.5$       &$-$         &      & $0.0125 -0.05$ &$-$             &-       &f($\mu_r$)  \\ 
			\hline
			Present   & Num$^c$         &$-$    & $10^{-4}-1$   		 &$0.1$  		    & $1$                 	& $0.1-10$   &     &$0.007143-0.7143$               &$135$    &$1.96\times10^{4}-1.96\times10^{-3}$  &f($Ca, Q_{r}$)\quad  or \quad f($\sigma, \mu_r$)   \\
			\hline
		\end{tabular}}
	\end{sidewaystable}
	%
\\ \noindent
Numerous computational fluid dynamics (CFD) approaches such as the volume of fluid (VOF) method, lattice Boltzmann method (LBM), phase field method (PFM), and level set method (LSM) are used in the literature to explore the hydrodynamics of droplets in emulsion/multiphase flows \citep{Olsson2005,VanderGraaf2006,Demenech2008,Bashir2011,Li2012,AkhlaghiAmiri2013,Shi2014,Nekouei2017,Yu2019,Wong2019}. All the CFD approaches have primarily focused on tracking the topological changes of the interface in motion. Among others, LSM has shown greater accuracy and capability to capture the fluid interface  \citep{Osher1988, Bashir2011} wherein the interface is represented by a level set function ($\phi$). To overcome the problem of mass loss \rev{(i.e., non-conservativeness) with LSM}, tuning parameters such as re-initialization or stabilization parameter ($\gamma$, m/s) and the interface thickness controlling parameter ($\epsilon_{\text{ls}}$, m) are included to maintain the stability and speed of the re-initialization step, respectively.  LSM is also known to provide accurate surface tension effect calculations  \citep{AkhlaghiAmiri2013,Olsson2005}.
\\ \noindent
In summary, the above discussed literature primarily examined different regimes of the \rev{two-phase flow and} droplet formation in a T-\rev{junction} microfluidic systems by varying viscosity ($\mu_{\text{r}}$), flow rate ($\qr$), and the channel dimension  ($w_{\text{r}}=w_{\text{d}}/w_{\text{c}}$). However, the in-depth effect of interfacial tension ($\sigma$) on the droplet generation broadly remains unexplored for a wide range of parameters \citep{Gupta2010,Glawdel2013,Wehking2014,Li2019}. 
Further, the classification of droplet generation zones has not been thoroughly elucidated for a wide range of parameters which helps to choose a microfluidic device for a specific application. For instance, the predictive correlations are known as a function of either $\cac$  or $\qr$. 
Practically, a functional dependence of both $\cac$ and $\qr$ needs to be considered to predict the droplet size and frequency to design a suitable microfluidic device. A thorough study of measuring droplet size and its breakup time can assist in determining the efficiency of a microfluidic device.
\\
In the current framework, thus,  a novel attempt is made to include the effect of interfacial tension ($\sigma$) between the two phases and carried out a comprehensive study for the generation and dynamics of the droplet in cross-flow microfluidic geometry for a practically wide range of flow rate ($\qr$), viscosity ($\mu_{\text{r}}$), and capillary number ($\cac$).  \rev{The conservative level set method (CLSM) and finite element method (FEM)} based CFD approaches have been used to understand in-depth dynamics of each droplet in terms of the droplet length, frequency, formation time, and different flow regimes.
\section{Physical and Mathematical Modeling}
\begin{figure}[h]
	\centering\includegraphics[width=1\linewidth]{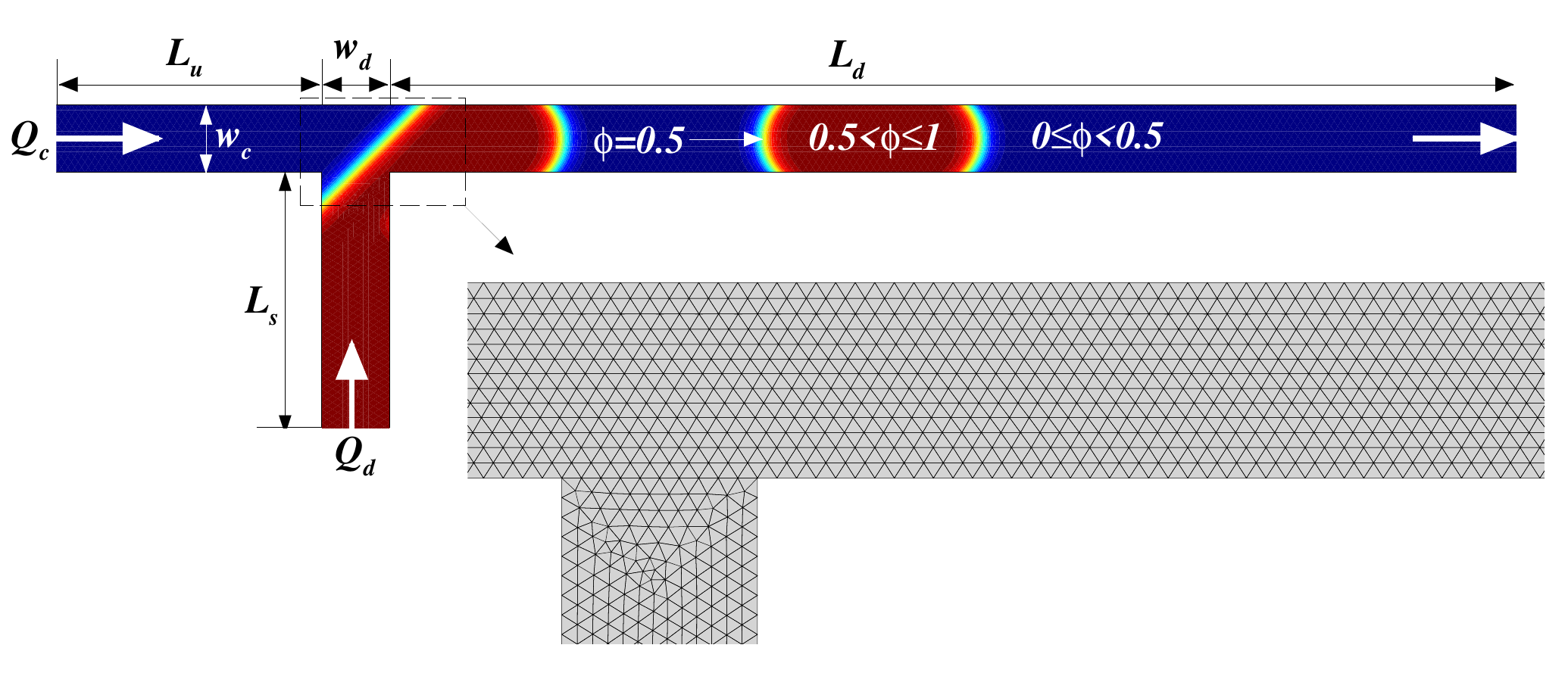}
	\caption{Schematic representation of two-dimensional T-junction microfluidic device  with the triangular mesh.}
	\label{fig:1}
\end{figure}

\noindent Consider the two-phase laminar flow of immiscible fluids through a T-junction \rev{cross-flow} microfluidic device, as shown in \fig\ref{fig:1}.  
The device consists of a main rectangular channel (width $w_{\text{c}}$ \micro m and length $L_{\text{m}}$ \micro m) and a side  branched channel (width $w_{\text{d}}$ \micro m and length $L_{\text{s}}$ \micro m).  The side channel intersects perpendicular to the main channel  and thereby forming up a T-junction. The branched channel is located at $L_{\text{u}}$ (upstream length) and at $L_{\text{d}}$ (downstream length) distances respectively from the inlet and outlet of the main channel. The length of the main channel is thus  $L_{\text{m}}=(L_{\text{u}}+w_{\text{d}}+L_{\text{d}})$.
\\\noindent 
The continuous phase (CP) and dispersed phase (DP) fluids are allowed to flow through the inlets of the main and side channels, respectively. The inlet volumetric flow rates (in \micro L/s) of CP and DP fluids are $Q_{\text{c}}$  and $Q_{\text{d}}$, respectively. The two fluids meet at the junction point and flow downstream of the main channel along with the continuous phase (CP).    The outlet of the main channel is open to ambient (i.e., outlet pressure $p=0$).   The ratios of widths of the two channels and flow rates of the two fluids are defined as  $w_{\text{r}}=(w_{\text{d}}/w_{\text{c}})$ and $\qr=(Q_{\text{d}}/Q_{\text{c}})$. Both immiscible fluids are taken to be isothermal, incompressible and Newtonian, i.e., density, viscosity, and interfacial tension are uniform throughout the flow process.  
\subsection{Governing equations and boundary conditions}
\noindent The flow physics under the above-noted description and approximations can mathematically be governed by mass continuity and momentum equations as follows.
\begin{eqnarray}
\nabla . \boldsymbol{u} &=& 0\label{eqn:l} \\
\rho(\phi)\left[\dfrac{\p \boldsymbol{u}}{\p t}+\boldsymbol{u} \cdot \bnabla\boldsymbol{u}\right]&=& -\bnabla p+\bnabla \cdot \boldsymbol{\tau}+\boldsymbol{F_{\sigma}} \label{eqn:2}
\end{eqnarray}
where $\boldsymbol{u}$, and $p$ are the velocity vector, and pressure fields, respectively. 
The level set function ($\phi$)  is a smooth step function which equals to $\phi =0$ in one fluid domain (say, CP) and $\phi =1$ in other fluid domain (say, DP). The value of  $\phi$ is thus ranging as  $0 \le \phi \le 1$.  Further, $\phi =0.5$ defines the fluid-fluid interface of two-phase flow.
\\
The deviatoric stress tensor ($\boldsymbol{\tau}$) is related with the rate of deformation tensor ($\boldsymbol{D}$) as follows.
\begin{equation}
\boldsymbol{\tau}=2\mu (\phi)\boldsymbol{D} \qquad\text{where}\quad
\boldsymbol{D}=\frac{1}{2}\left[(\bnabla \boldsymbol{u})+(\bnabla \boldsymbol{u})^{T}\right]
\label{eq:tauD} 
\end{equation}
The density ($\rho$) and dynamic viscosity ($\mu$) of the two-phase are expressed as follows. 
\begin{equation}
\rho(\phi)=\rho_{\text{c}}+(\rho_{\text{d}}-\rho_{\text{c}})\phi,\qquad \text{and}\qquad
\mu(\phi)=\mu_{\text{c}}+(\mu_{\text{d}}-\mu_{\text{c}})\phi, 
\end{equation}
where the subscripts `$\text{c}$' and `$\text{d}$' refer to the continuous and dispersed phases, respectively.  
\\
The interfacial force ($\boldsymbol{F}_{\sigma}$) between the two fluids is governed by the following relation. 
\begin{equation}
\mathbf{F}_{\sigma}=\sigma \kappa \delta(\phi) \boldsymbol{n}
\qquad\text{where}\quad 
\boldsymbol{n} = \frac{\mathbf{\bnabla \phi}}{|\mathbf{\bnabla \phi}|}
\label{eq:Fsigma}
\end{equation}
where $\sigma$, $\boldsymbol{n}$, $\kappa$, and $\delta(\phi)$ are the interfacial tension (in mN/m), unit normal, the curvature of the interface, $\kappa={R}^{-1}=- (\mathbf{\bnabla} \cdot \boldsymbol{n})$, and the Dirac Delta function, respectively.
\\
The topological behaviour of the interface in two-phase flow can be tracked by the following additional equation of the \rev{conservative} level set method (\rev{C}LSM). 
\begin{equation}
\dfrac{\p \phi}{\p t}+\boldsymbol{u} \cdot {\bnabla} \phi=\gamma {\bnabla} \cdot \left[\epsilon_{\text{ls}}{\bnabla} \phi- \phi (1-\phi)\boldsymbol{n}\right]
\label{eqn:lsm}
\end{equation}
The left side of \eqn(\ref{eqn:lsm}) accounts for the motion of the interface, and the right side introduces for the necessary numerical stability. 
Here, $\boldsymbol{n}$ is unit normal, $\epsilon_{ls}$ is the  controlling parameter for  thickness of region wherein $\phi$ goes smoothly as $0 \le \phi \le 1$. It is typically of the same order as that of size of mesh elements. The  reinitialization or stabilization parameter ($\gamma$) ascertains the stabilization of level set function ($\phi$).  
\\
The above field governing equations are subjected to the following physically realistic boundary conditions:  (a) The flow rates  ($Q_{\text{c}}$ and $Q_{\text{d}}$) of both CP and DP are imposed at the inlets, 	(b) The outlet of the main channel is open to ambient, i.e., $p=0$. Further, fully developed, i.e., Neumann condition, is imposed for the velocity and phase fields, and 	(c) The channel walls being solid and impermeable are subjected to the no-slip condition. 
\\
The numerical solution of the governing equations with boundary conditions results in instantaneous velocity, pressure, and phase concentration as a function of flow governing parameters.  These fields are analyzed to present the droplet generation and dynamics and further used to deduce the length and diameter of the droplets.   
The essential definitions and parameters used in the present work are given as follows.
\\
The effective droplet diameter ($d_{\text{eff}}$, m)  is calculated from the surface area of the dispersed phase \citep{Liu2011a,Jamalabadi2017,Wong2017,Wong2019} as follows. 
\begin{equation}
d_{\text{eff}}=2 \left[\dfrac{1}{\pi}\int_{\Omega}(\phi>0.5)d \Omega\right]^{1/2}
\label{eq:deff}
\end{equation}
The length of the droplet ($L$) is determined by plotting the phase variable \rev{($\phi$)} along the length of main channel and analyzed the  length occupied by $\phi>0.5$.
\\
The droplet detachment time ($t_{\text{dd}}, \text{s}$) is defined as the time interval between the two subsequent droplet formation.  The droplet detachment frequency (${f_{\text{dd}}, \text{s}^{-1}}$), i.e., the number of droplets generated per unit time, is calculated as the inverse of the droplet detachment time. 
\begin{equation}
t_{\text{dd}} = (t_{i+1} - t_{i}),		\qquad\text{and}\qquad	{f_{\text{dd}}} =\frac{1}{t_{\text{dd}}} 
\label{eq:tfdd}
\end{equation}
where $t_i$ is time taken by $i^{\text{th}}$ droplet generation (or detachment) and $i$ is the droplet number. 
\\
The dimensionless parameters such as capillary number ($Ca$), Reynolds number ($Re$), flow rate ratio ($\qr$), viscosity ratio ($\mu_{\text{r}}$), density ratio ($\rho_{\text{r}}$)  and channel width ratio ($w_{\text{r}}$) defined as follows. 
\begin{eqnarray}
\cac &=& \frac{u_{\text{c}} \mu_{\text{c}}}{\sigma}, \qquad 
Ca_{\text{d}} = \frac{u_{\text{d}} \mu_{\text{d}}}{\sigma}, \qquad 
Re_{\text{c}} = \frac{\rho_{\text{c}} u_{\text{c}} w_{\text{c}}}{\mu_{\text{c}}}, \qquad
Re_{\text{d}} = \frac{\rho_{\text{d}} u_{\text{d}} w_{\text{d}}}{\mu_{\text{d}}}\label{eq:dimp1}\\
Ca_{\text{r}} &=& \frac{Ca_{\text{d}}}{Ca_{\text{c}}},\qquad
Re_{\text{r}} = \frac{Re_{\text{d}}}{Re_{\text{c}}},\qquad
\qr = \frac{Q_{\text{d}}}{Q_{\text{c}}},\qquad
\mu_{\text{r}} = \frac{\mu_{\text{d}}}{\mu_{\text{c}}},\qquad
\rho_{\text{r}} = \frac{\rho_{\text{d}}}{\rho_{\text{c}}},\qquad
w_{\text{r}}=\frac{w_{\text{d}}}{w_{\text{c}}}
\label{eq:dimp2}
\end{eqnarray}
where, subscripts $c$, $d$  and $r$ denote for the continuous and dispersed  phases, and ratio, respectively.
\section{Solution Approach and Numerical Parameters}
\label{sec:sanp}
\noindent
In this work, the preceding mathematical model equations based on the \rev{conservative} level set method (\rev{C}LSM) for two-phase laminar flow are solved using the finite element method (FEM) based  COMSOL multiphysics CFD solver. 
The two-dimensional non-uniform linear triangular mesh has been adopted to discretize the computational domain of T-junction microfluidic device, as shown in \fig\ref{fig:1}. 
\\\noindent \rev{
The two-dimensional (2D), laminar, two-phase flow, level set method modules of COMSOL multiphysics are used to represent the present mathematical model.  
The time-dependent partial differential equations (PDEs) are converted into an implicit system of ordinary differential equations (ODEs) through a finite element spatial discretization. 
The temporal derivatives are approximated by using the time-implicit backward differentiation formula (BDF) with variable order of accuracy from one (i.e., backward Euler method) to five. BDF method, known for their stability, is a Differential-Algebraic Equations (DAE) solver. The steps taken by solver are set to be `free' with `automatic' maximum time-step constraint. COMSOL selects BDF solver by default due to its stable nature for solving complex problems \citep{Bashir2011,Sartipzadeh2020} with variable time step ($\Delta t$). It generally uses higher-order for accurate solution and selects the lower-order to obtain the stable and robust convergence.} 
\\\noindent \rev{Segregated solutions of phase ($\phi$) and flow ($\mathbf{u}$, and $p$) fields  are obtained by using the fastest direct solver PARDISO (PARallel DIrect SOlver) for large sparse system of linear equations \citep{Schenk2011,Bollhofer2020}, and Newton's non-linear solvers}. The sufficiently small time step ($\Delta t = 10^{-4}$ s)  is used in all the simulations. The relative tolerance of $5\times10^{-3}$ is satisfied with the iterative solver to obtain the converged numerical solution.
\\\noindent 
The geometrical parameters for the physical system are taken as follows. Both channel widths are taken to be equal as $w_{\text{c}}=w_{\text{d}}=100$ \micro m . The length independence test resulted the upstream and downstream lengths of main channel and the length of side branch channel as  $L_{\text{u}}=9w_{\text{c}}$ and $L_{\text{d}}=30 w_{\text{c}}$ and $L_{\text{s}}=9w_{\text{c}}$ to be sufficient to eliminate the end effects.
The values of level set parameters $\gamma$=1 m/s and $\epsilon_{\text{ls}}=h_{\text{max}}/2=5$ \micro m are used in the present study, where $h_{\text{max}}$ is the maximum size of mesh element. 
\\
In the present study, the density of both the fluids is assumed to be equal (i.e., $\rho_{\text{r}}=1$) and the contact angle ($\theta$) with respect to the dispersed phase on the walls is taken as $135^{\circ}$. Further, the viscosity and the flow rate of the dispersed phase are kept constant as $\mu_{\text{d}}=10^{-3}$ Pa.s and $Q_{\text{d}}=0.14$ \micro L/s \citep{Garstecki2006, Soh2016}. 
The numerical simulations have been performed for the ranges of conditions mentioned in Table \protect\ref{tab:1} as follows. Reynolds number ($Re_{\text{c}}=0.1$), capillary number ($10^{-4}\le \cac\le 1$), flow rate ($0.1\le \qr\le 10$), and viscosity ratio ($7.143\times 10^{-3}\le \mu_{\text{r}} \le 7.143\times 10^{-1}$),  and  interfacial tension ($1.96\times 10^{4}\le\sigma \le 1.96\times 10^{-3} \text{mN/m}$).  
\subsection{Mesh independence study}
\noindent 
The accuracy and efficacy of the numerical solutions depend on the characteristics of the mesh used to discretize the governing equations and boundary conditions.   While the solutions are expected to be most accurate when the mesh (or element) size approaches zero,  the computational efforts (time and memory requirements) enhance enormously and are strongly dependent on the degree of freedom (DOF = number of nodes $\times$ number of dependent variables). 
Therefore, the optimum mesh is selected through the mesh independence study by analyzing the trade-off between the accuracy of the solution and the computational efforts. 
\begin{table}[!t]
\begin{center}
\caption{Data from the mesh independence study.}\label{tab:2}
	\scalebox{0.95}
	{
\begin{tabular}{p{0.4in}p{0.7in}p{0.65in}p{0.65in}p{0.6in}p{0.5in}p{2in}}
S.No. &$\Delta_\text{{max}}$ (\micro m) &   N$_{\text{e}}$   & DoF & $d_{\text{eff}}$ (\micro m)   &$t_{\text{dd}}$ (ms) & Evolution of the interface \\ \hline
\multicolumn{7}{l}{(a) non-uniform triangular mesh}\\
			TM1 & 13 &  8366  &31891  &168.46    &22.0   & \raisebox{-0.4\totalheight}{\includegraphics[scale = 0.25]{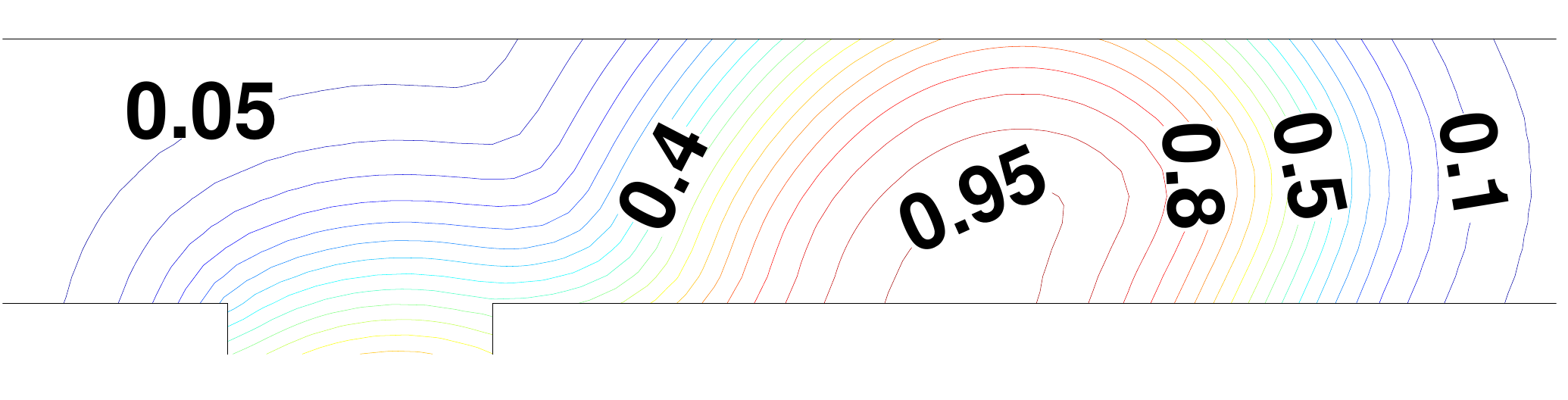}}\\
			TM2 & 12.5 &  8666  &33037  &164.24    &21.2  & \raisebox{-0.4\totalheight}{\includegraphics[scale = 0.25]{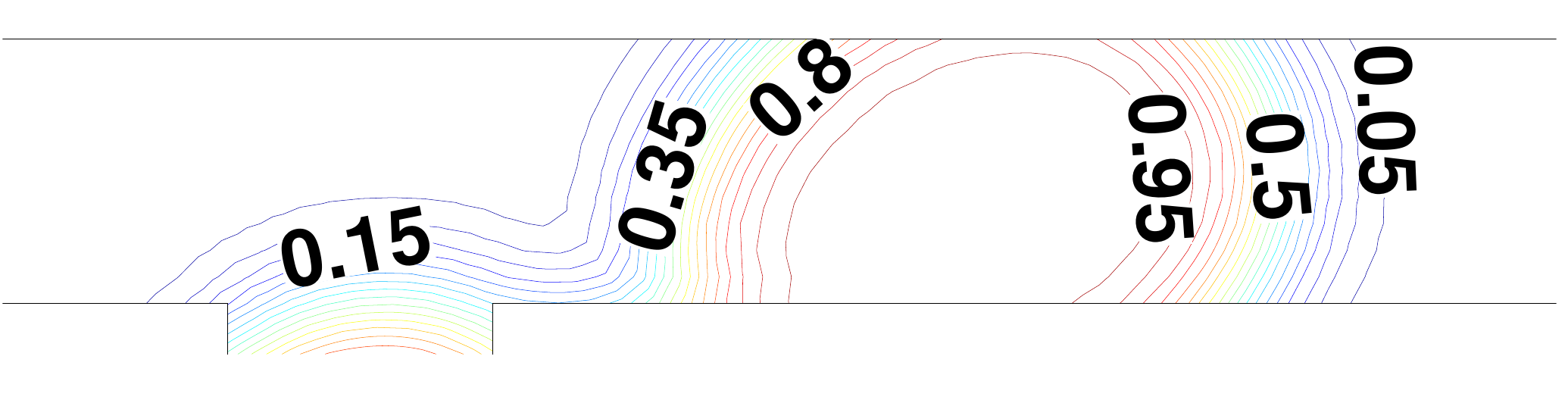}}\\
			TM3 & 11 &  11615 &44609  &163.66    &21.0  & \raisebox{-0.4\totalheight}{\includegraphics[scale = 0.25]{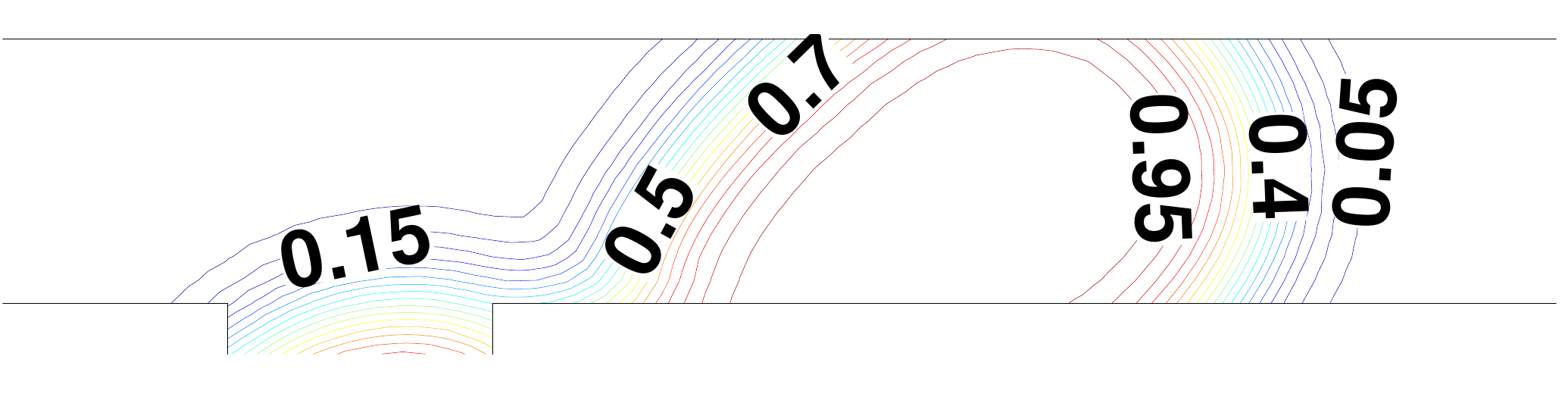}}\\ 
			TM4 & 10.5 & 12254 &47071  &163.56    &20.9  & \raisebox{-0.4\totalheight}{\includegraphics[scale = 0.25]{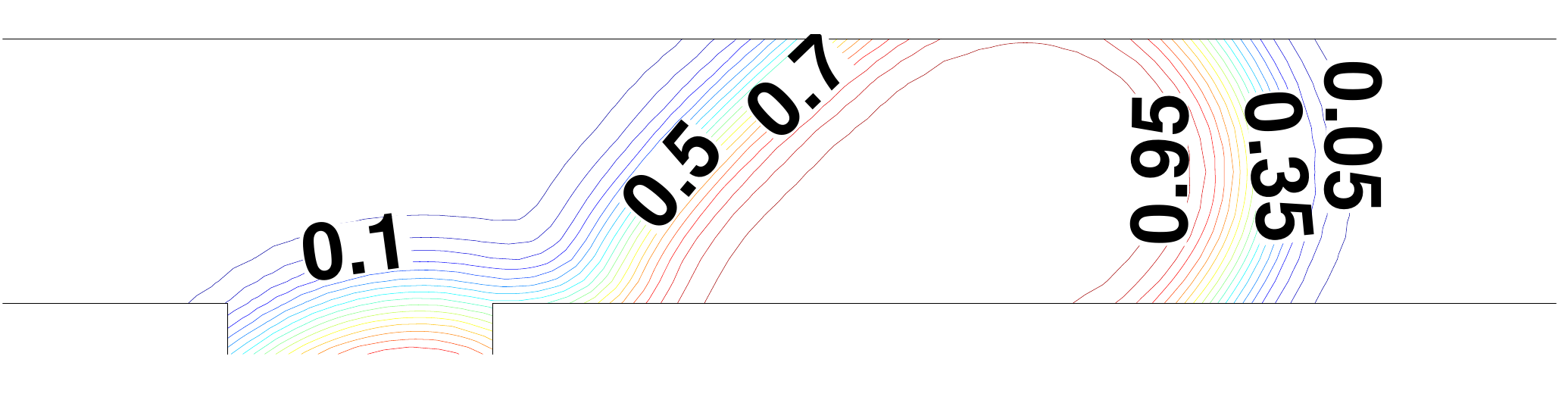}} \\ 
			TM5 & 10 &  13766 &53029  &163.26     &20.8  & \raisebox{-0.4\totalheight}{\includegraphics[scale = 0.25]{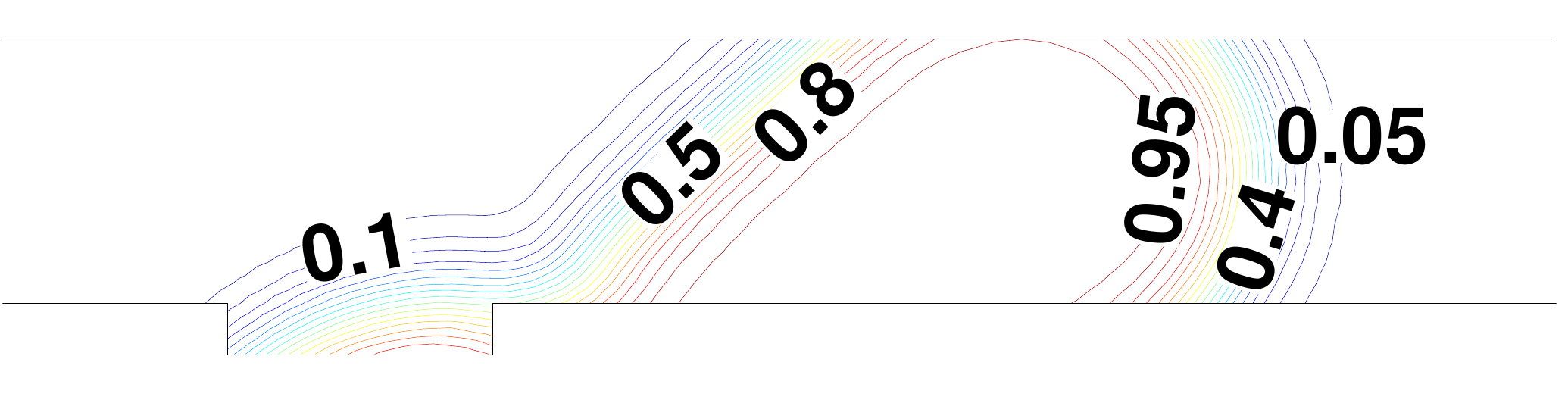}} \\
			TM6 & 9 &  16409 &63373  &163.21    &20.8  & \raisebox{-0.4\totalheight}{\includegraphics[scale = 0.25]{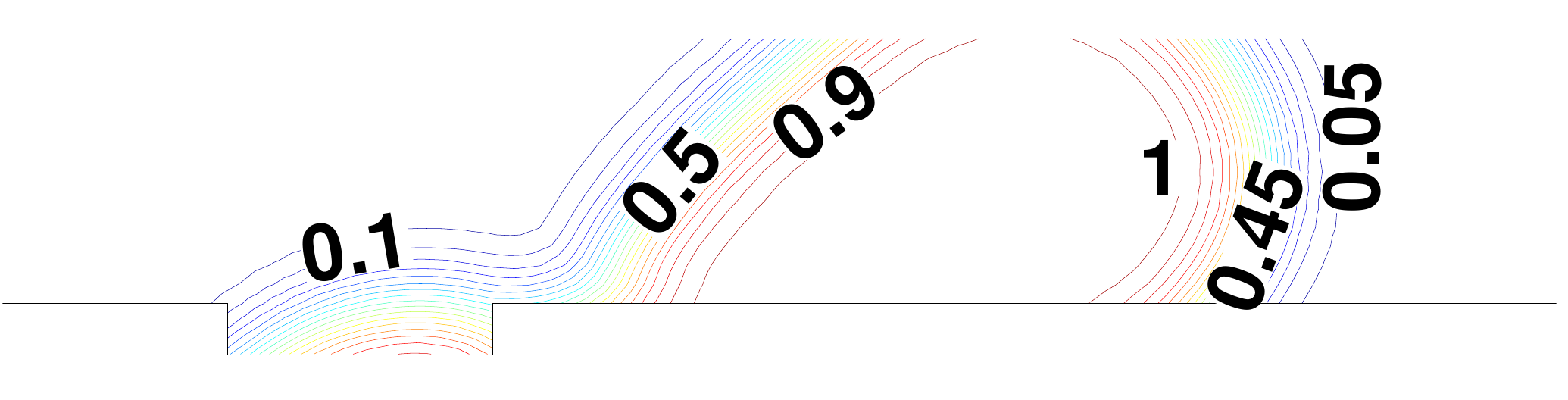}} \\
			TM7 & 8 &  20963 &71697  &163.20    &20.8  & \raisebox{-0.4\totalheight}{\includegraphics[scale = 0.25]{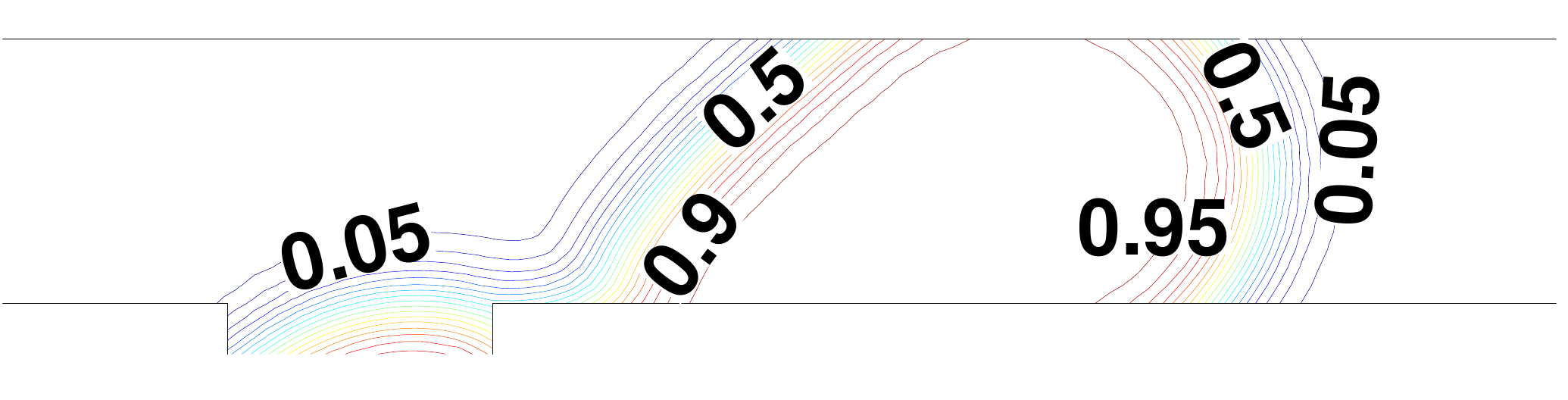}}\\  \hline
\multicolumn{7}{l}{\rev{(b) quadrilateral mesh}}\\
			QM1 &13 & 3814  &19745  &173.73    &22.8   & \raisebox{-0.4\totalheight}{\includegraphics[scale = 0.25]{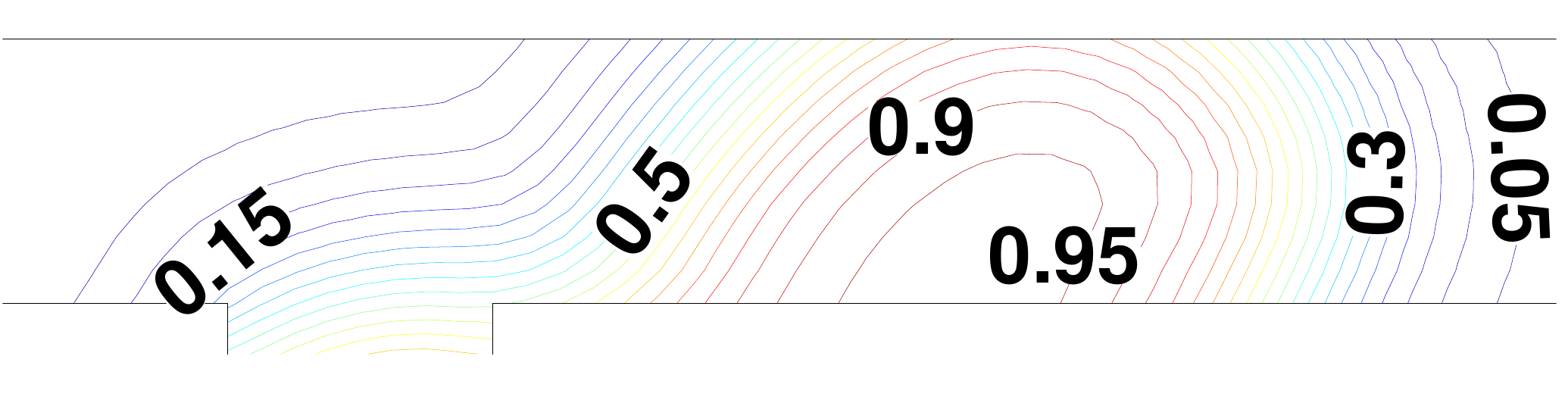}}\\
			QM2 &12.5 &  3944  &20421  &172.75   &22.7  & \raisebox{-0.4\totalheight}{\includegraphics[scale = 0.25]{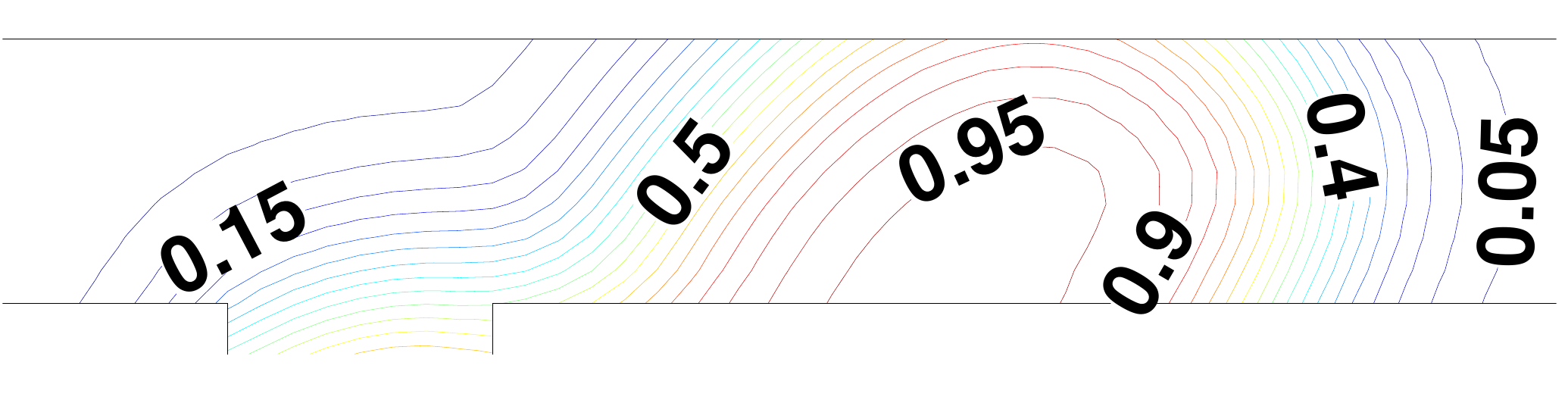}} \\
			QM3 & 11 & 4933 &25909  &170.98   &22.5  & \raisebox{-0.4\totalheight}{\includegraphics[scale = 0.25]{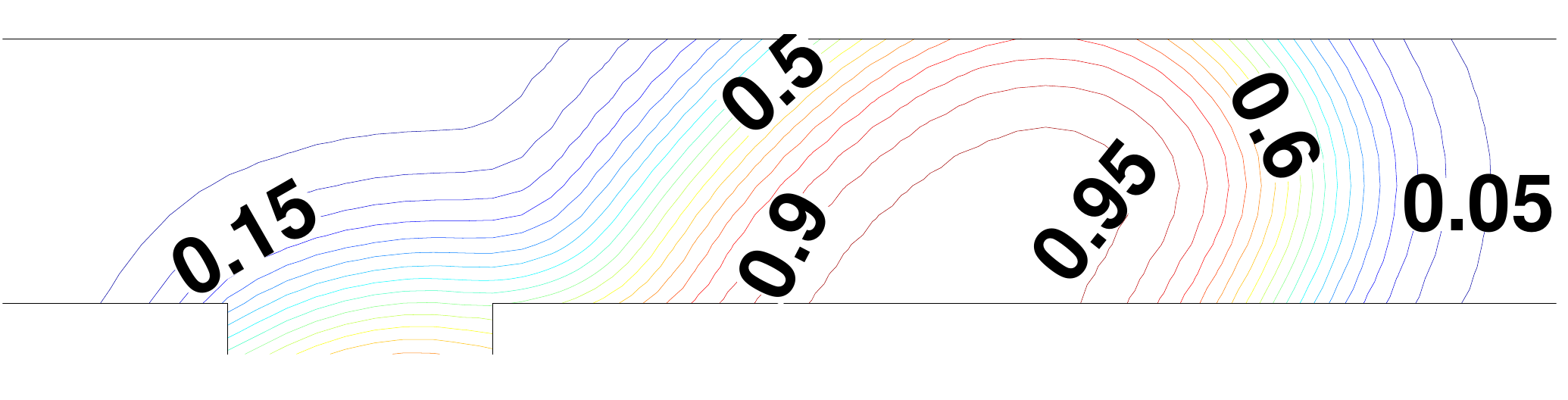}}\\ 
			QM4 &10.5 & 5646 &29997  &170.36    &22.5  & \raisebox{-0.4\totalheight}{\includegraphics[scale = 0.25]{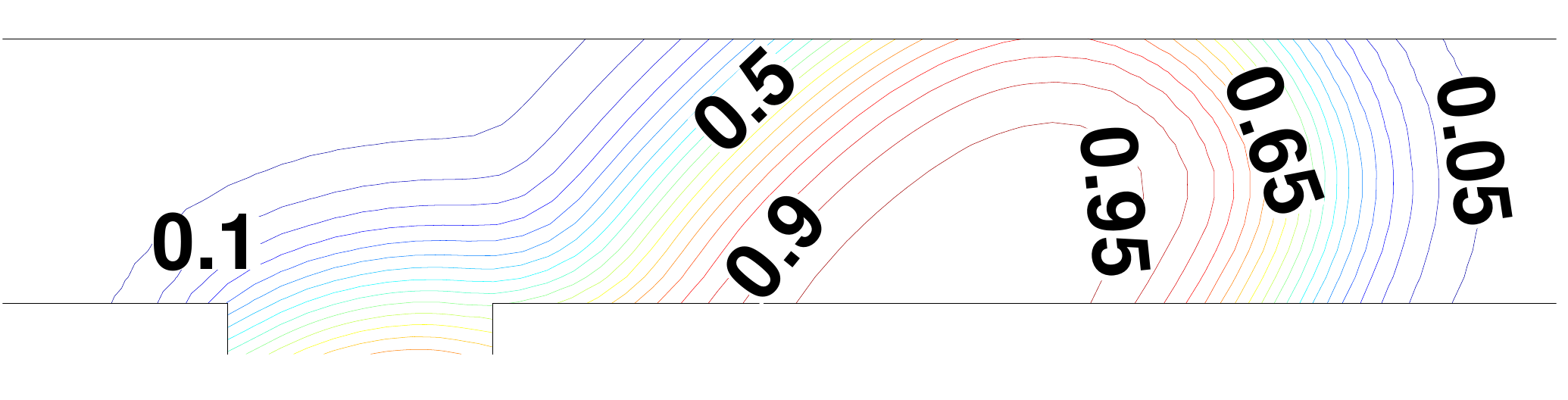}}\\ 
			QM5 & 10 & 5910 &31405  &170.26    &22.4  & \raisebox{-0.4\totalheight}{\includegraphics[scale = 0.25]{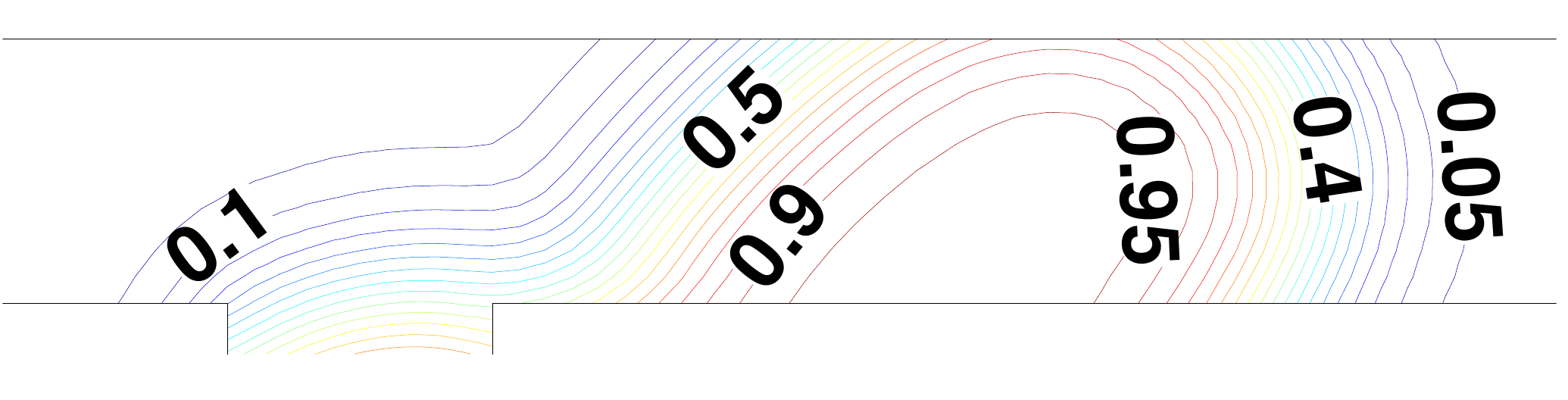}}\\
			QM6 & 9 & 6975 &38199  &170.13    &22.3  & \raisebox{-0.4\totalheight}{\includegraphics[scale = 0.25]{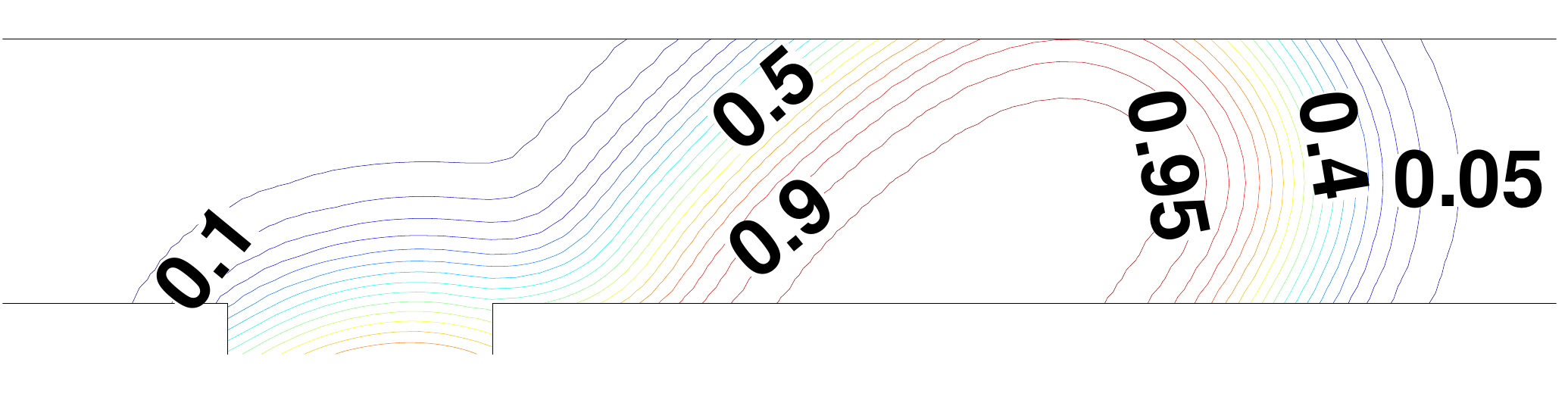}}\\
			QM7 &8 &  8114 &43905  &170.52    &22.1  & \raisebox{-0.4\totalheight}{\includegraphics[scale = 0.25]{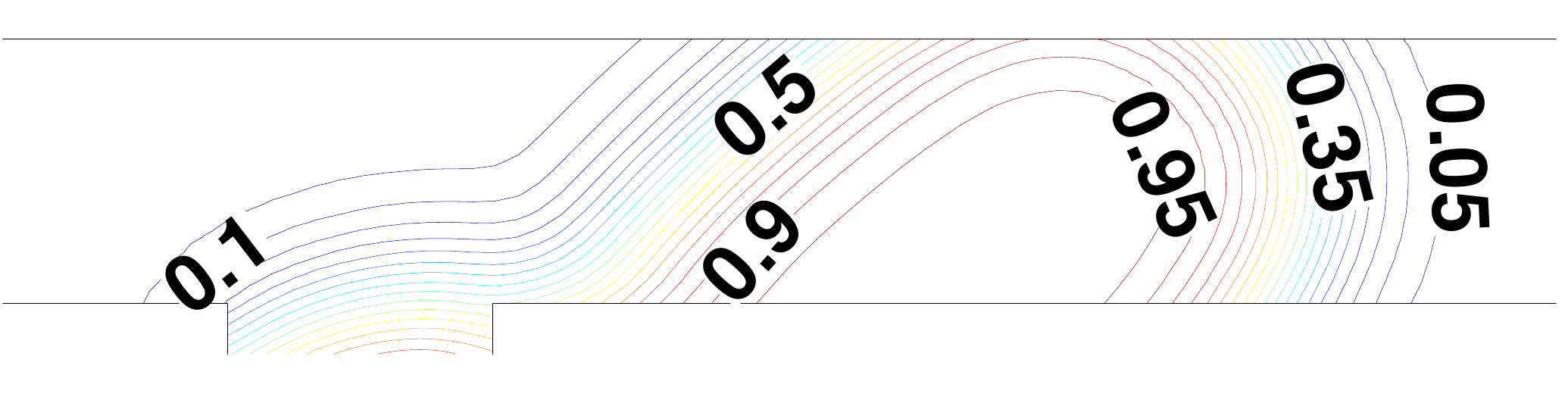}} \\  \hline
		\end{tabular}
}
	\end{center}
\end{table}
\\\noindent
The mesh convergence analysis in Table \protect\ref{tab:2} depicts the influence of triangular mesh \rev{(TM1 to TM7)}  size (i.e., number of mesh elements, N$_{\text{e}}$ \rev{and maximum size of element, $\Delta_\text{{max}}$ \micro m}) on the effective droplet diameter ($d_{\text{eff}}$, \micro m), droplet detachment time ($t_{\text{dd}}$, ms) and evolution of the interface at a fixed flow rate ($Q_r=0.5$). 
\rev{Further, the structured quadrilateral meshes (QM1 to QM7) are also tested with a similar refinement in the maximum element size ($\Delta_\text{{max}}$, \micro m) from 13 to 8. Both triangular and quadrilateral meshes have shown similar consistent patterns of droplet size and breakup time, i.e., both decrease with an increasing number of elements. 
\\
Undoubtedly, the number of elements (N$_{\text{e}}$) and degree of freedom (DoF) are higher for triangular mesh in comparison to that for the quadrilateral mesh structure. For instance, N$_{\text{e}}$ = 13766 with DoF = 53029  are in TM5 mesh whereas N$_{\text{e}}$ = 5910 and DoF =  31045 are in QM5 mesh for $\Delta_\text{{max}} = 10  \micro m$. Hence, the effective droplet diameter is a little bigger (and less accurate) for the quadrilateral and smaller (and more accurate) for the triangular mesh. The interface evolution and demarcation are also smooth and clear with triangular meshes compared to the quadrilateral meshes.}
\\
The analysis shows that the droplet behaviour (evolution of the interface, diameter, and detachment time) becomes strongly stable with increasing  N$_{\text{e}} > 12254$ \rev{(i.e., refinement after TM4)}. The changes in the results with \rev{further} mesh refinement are insignificant for $\Delta_\text{{max}} < 10$, hence, \rev{TM5 mesh} is enough to carry out the simulations. Further, DoF enhances proportionally with N$_{\text{e}}$, and so the computational efforts. Keeping in mind a trade-off between the accuracy and computational efforts, the mesh \rev{TM5} consisting of 13766 triangular \rev{non-uniform} elements is considered to be sufficiently refined to resolve the gradients and interface in most accurate manner. The new results thus presented hereafter are obtained by using \rev{TM5 mesh} consisting of N$_{\text{e}} = 13766$ \rev{non-uniform triangular} elements.
\section{Results and discussion}
\noindent
In this section, the {hydro}dynamics of the \rev{two-phase flow and} droplet \rev{generation} in T-junction microfluidic device \rev{for} the wide ranges of flow governing parameters (see section \ref{sec:sanp}) are presented and discussed.
Before presenting the new results obtained herein, the reliability and accuracy of the solution approach are established through validation of present results with the available literature. 
\begin{figure}[!b]
	\centering
	\subfloat[$L/w_{\text{c}}$ vs. $Q_{\text{r}}$ and $d_{\text{eff}}/w_{\text{c}}$ vs. $Q_{\text{r}}$]{\includegraphics[width=1\linewidth]{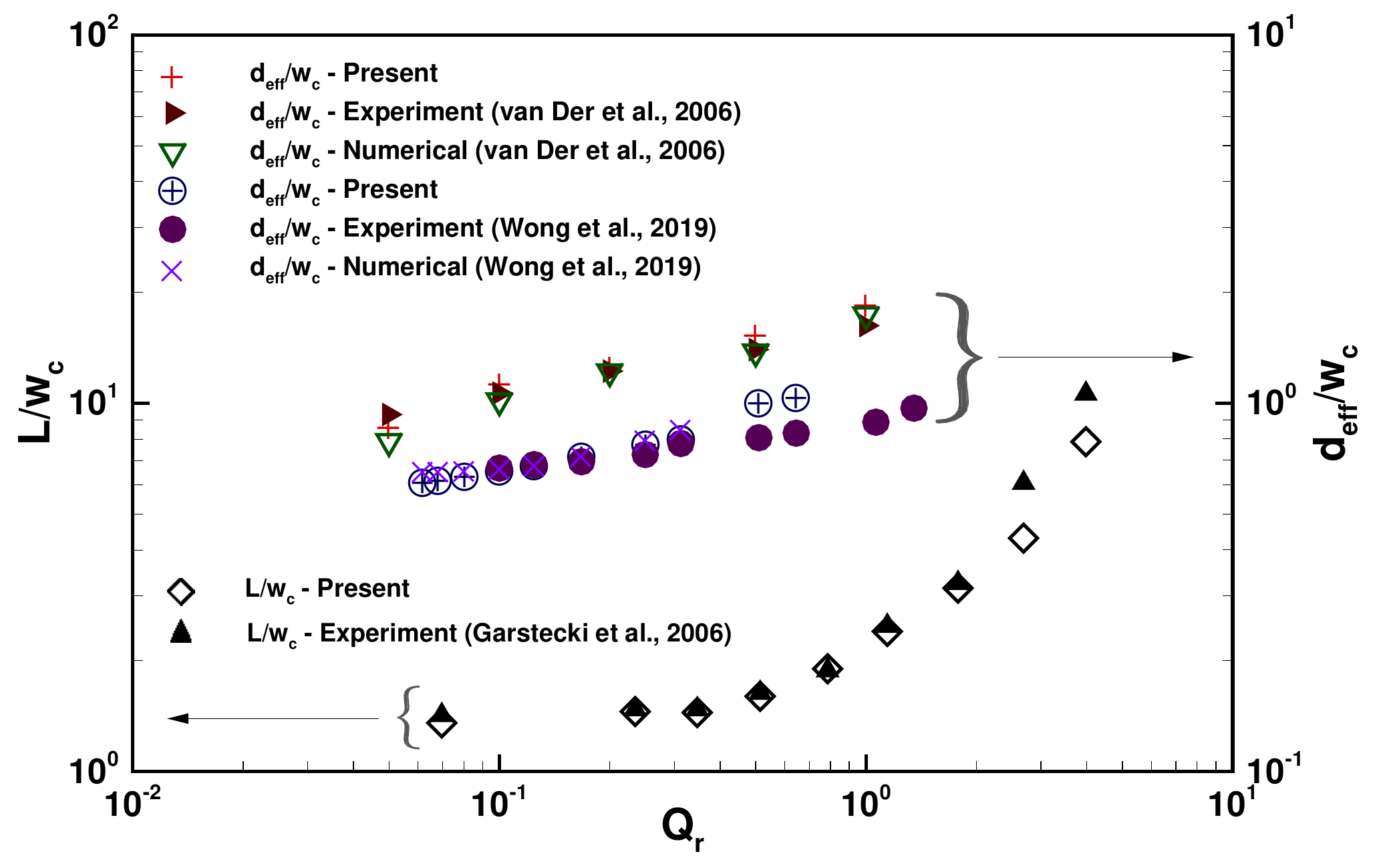}}\\
	\subfloat[$L/w_{\text{c}}$ vs. $Q_{\text{c}}$]{\includegraphics[width=0.5\linewidth]{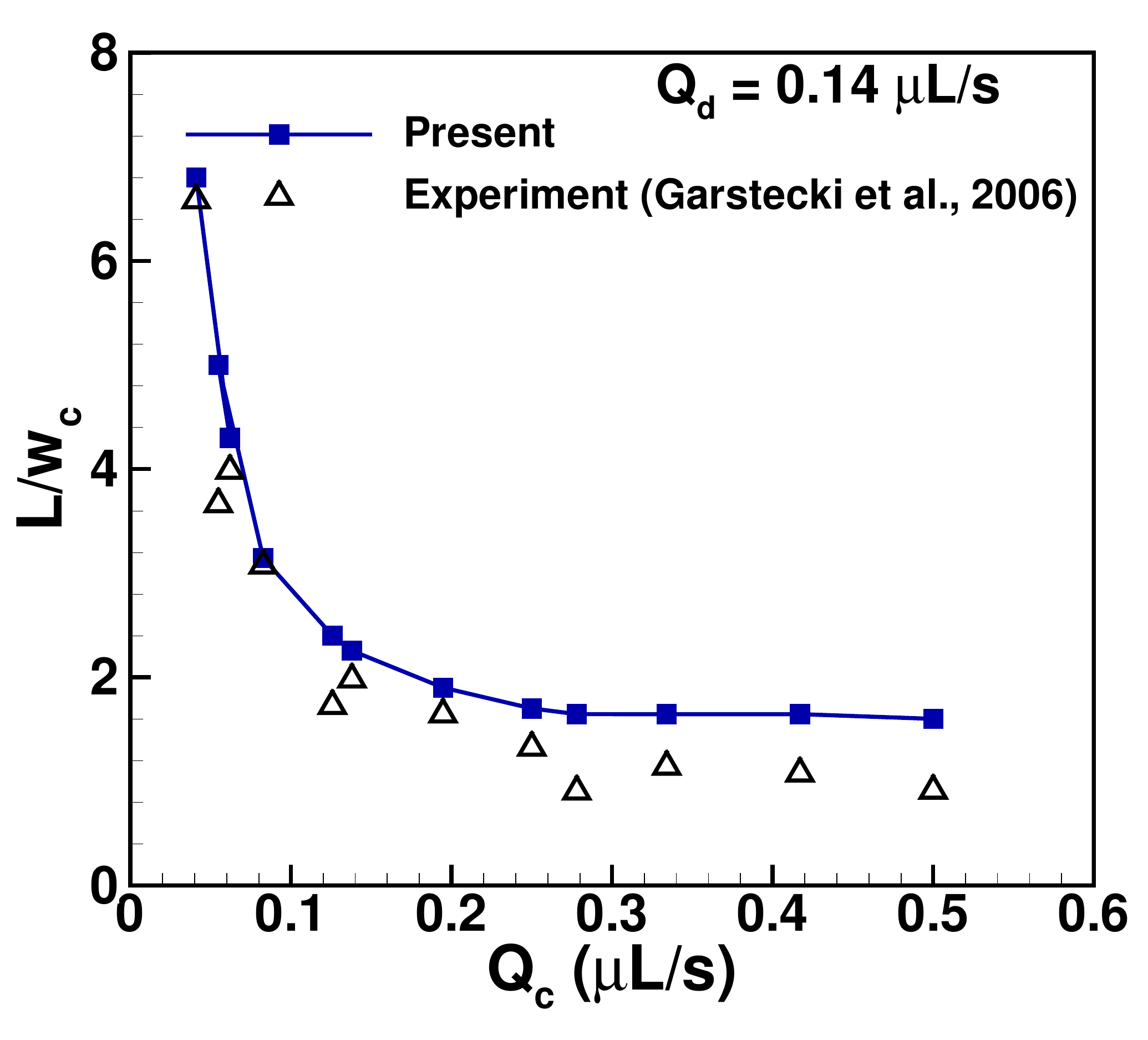}\label{fig:2b}}	
	\subfloat[$L/w_{\text{c}}$ vs. $Q_{\text{d}}$]{\includegraphics[width=0.5\linewidth]{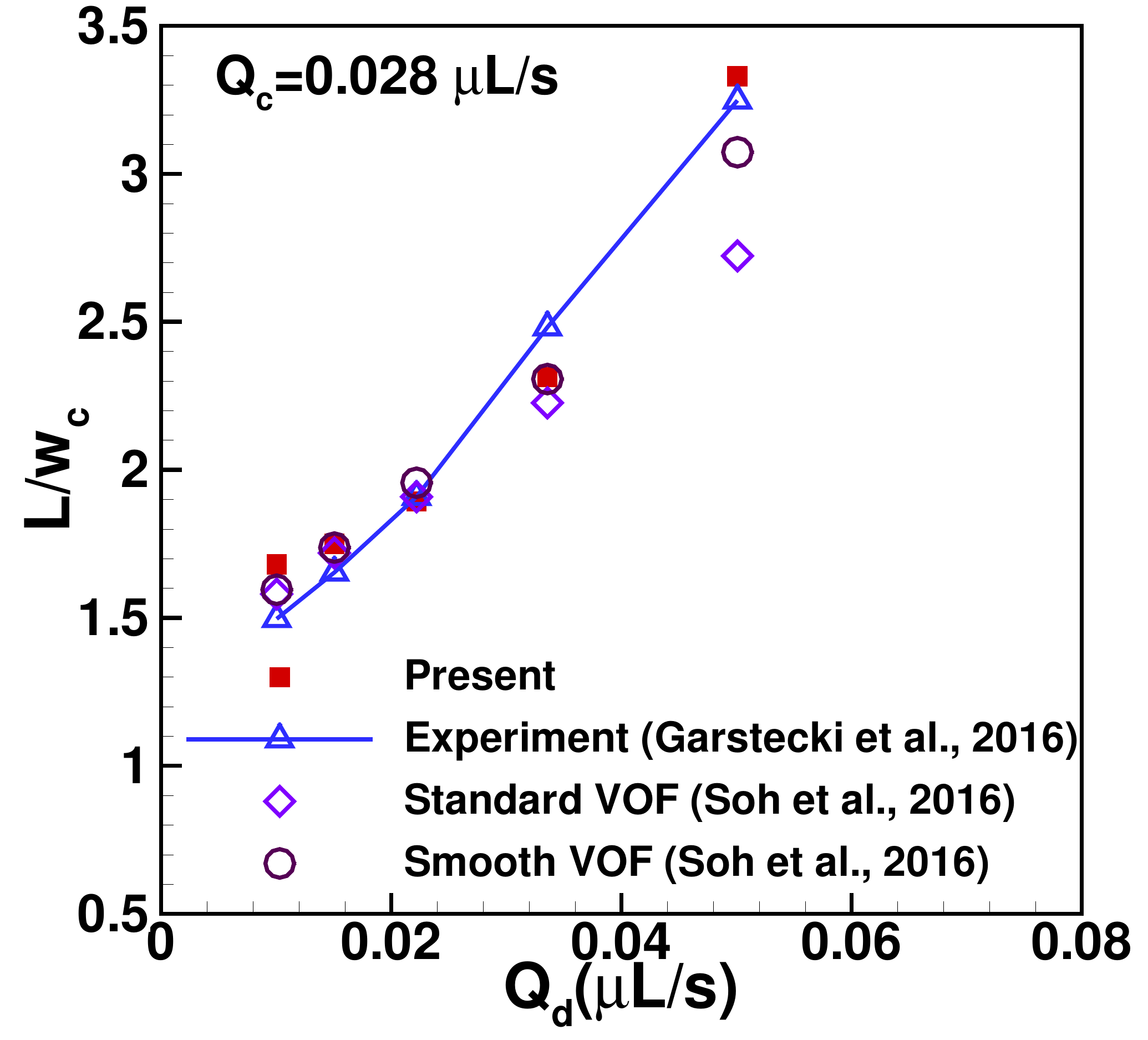}\label{fig:2c}}	
	\caption{Comparison of the present results with the literature for the wide range of $2\times10^{-3} \leq \cac\leq 4\times10^{-2}$ \citep{VanderGraaf2006}, $\cac=10^{-1}$ \citep{Wong2019}, and $\cac\leq 10^{-2}$ \citep{Garstecki2006,Soh2016} and flow rate ($10^{-2}\le \qr\le 10$).}
	\label{fig:3}
\end{figure}
\subsection{Validation of the results}
\noindent 
The present results in terms of the effective droplet diameter ($d_{\text{eff}}$) and droplet length ($L$) have been compared with experimental \citep{Garstecki2006,VanderGraaf2006} and numerical   \citep{VanderGraaf2006,Soh2016,Wong2019} works in \fig\ref{fig:3}.  
\rev{The flow and geometrical conditions are maintained consistent with the respective literature for the comparison purpose.} The present values are closely consistent with the literature  \citep{Garstecki2006, VanderGraaf2006,Soh2016,Wong2019} results for   wide range of flow conditions ($\cac$, $\qr$, $Q_{\text{c}}$, and $Q_{\text{d}}$). 
\\\noindent
\fig\ref{fig:3} shows that the effective droplet diameter ($d_{\text{eff}}$) and droplet length ($L$) are decreasing with increasing flow rate of the continuous phase ($Q_{c}$) at a constant flow rate of the dispersed phase ($Q_{d}$). \rev{Vice versa, $L$ increases with increasing $Q_{d}$ for fixed $Q_{c}$.} Broadly, the present results are much closer to the experimental results than the numerical results obtained by using the lattice Boltzmann method (LBM) with adopted interface tracking method \citep{VanderGraaf2006}, the level set method \citep{Wong2019} and VOF \citep{Soh2016}. It is because of the comparatively finer mesh used in the present work than that used by other numerical studies \citep{VanderGraaf2006,Soh2016,Wong2019}. 
\\\noindent
Based on our previous experience of CFD simulations of various problems \citep{Bharti2006,Sivakumar2006,Bharti2007a,Bharti2007b,Patnana2009,Patnana2010,Tian2014,Pravesh2016,Gangawane2018,Kumar2021,Vishal2021}, a slight deviation is quite common in simulation studies due to inherent characteristics of numerical solvers and methodologies used in related studies. 
Based on the above excellent agreement of present and literature values of $d_{\text{eff}}$ and $L$ for the broader range of $\cac$ and $\qr$, the present results are believed to have an excellent level ($\pm 1\%$) of accuracy.
%
\subsection{Instantaneous phase flow profiles}
\begin{figure}[!b]
	%
	{\includegraphics[width=0.95\linewidth]{Figures/Phase_compositions/Untitled}}
	\\	\subfloat[$\cac=10^{-4}$]{\includegraphics[width=0.95\linewidth]{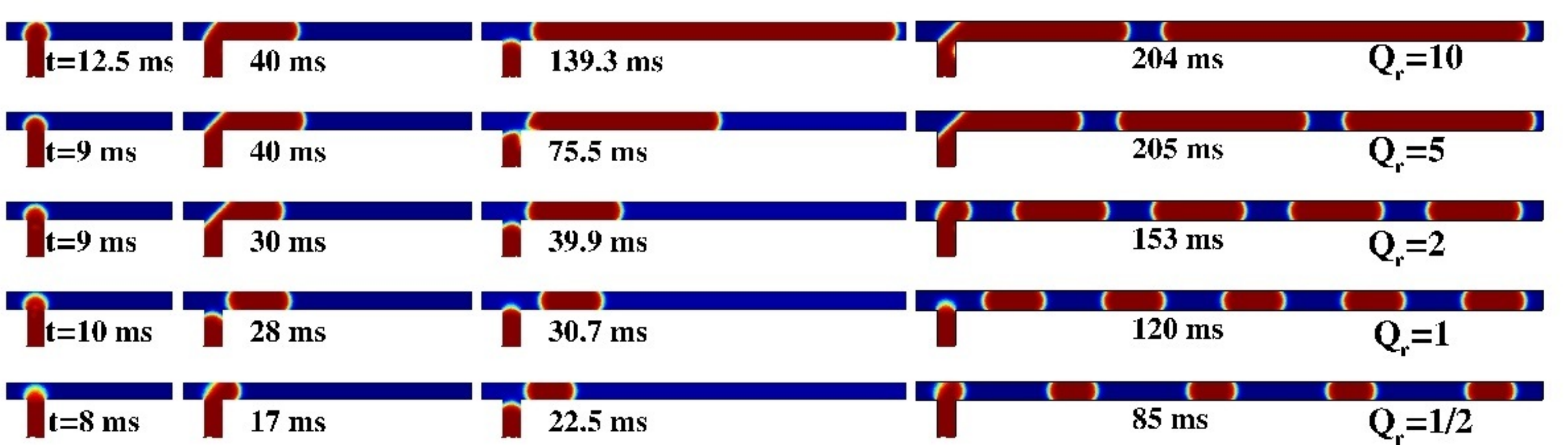}\label{fig:5a}}
	\\	\subfloat[$\cac=2\times 10^{-2}$]{\includegraphics[width=0.95\linewidth]{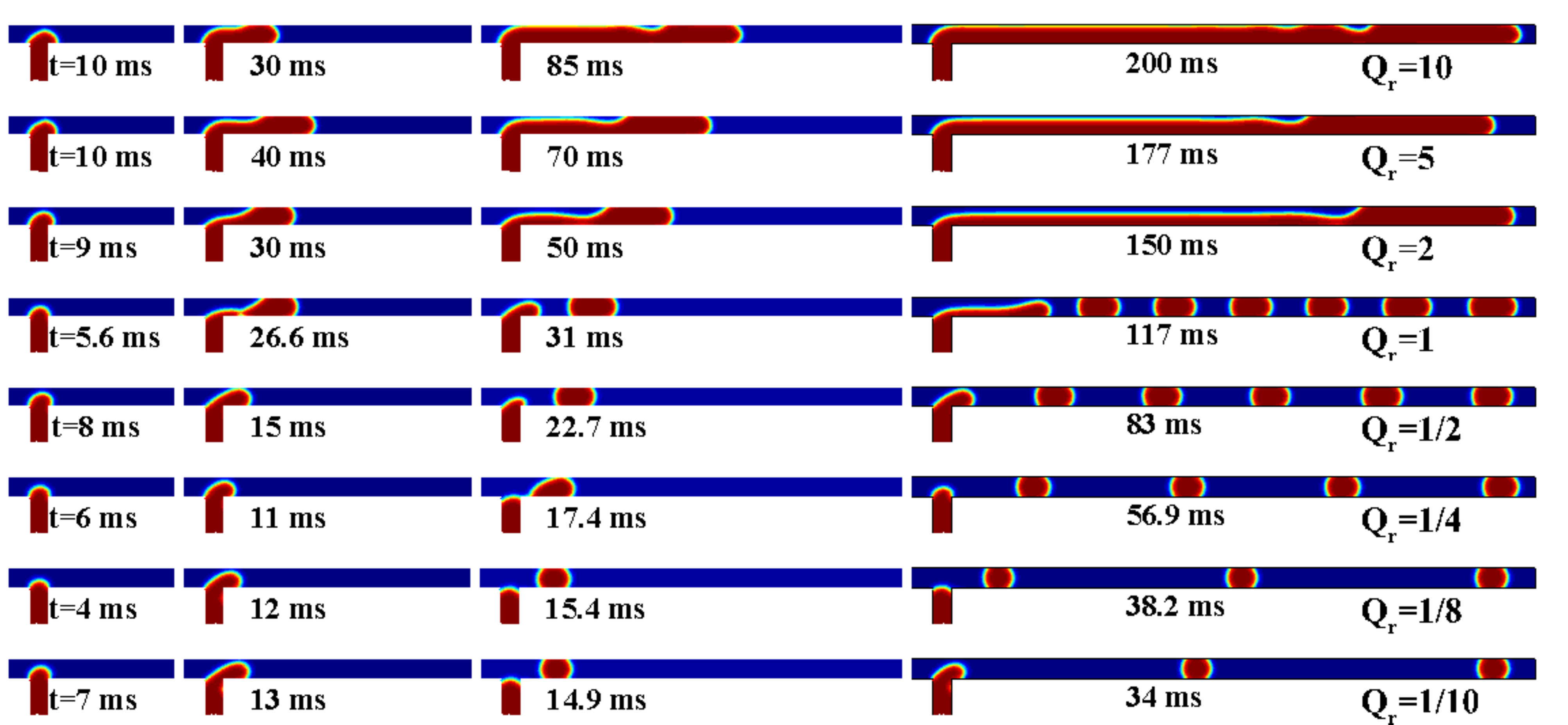}\label{fig:5b}}	%
	\\	\subfloat[$\cac=1$]{\centering\includegraphics[width=0.95\linewidth]{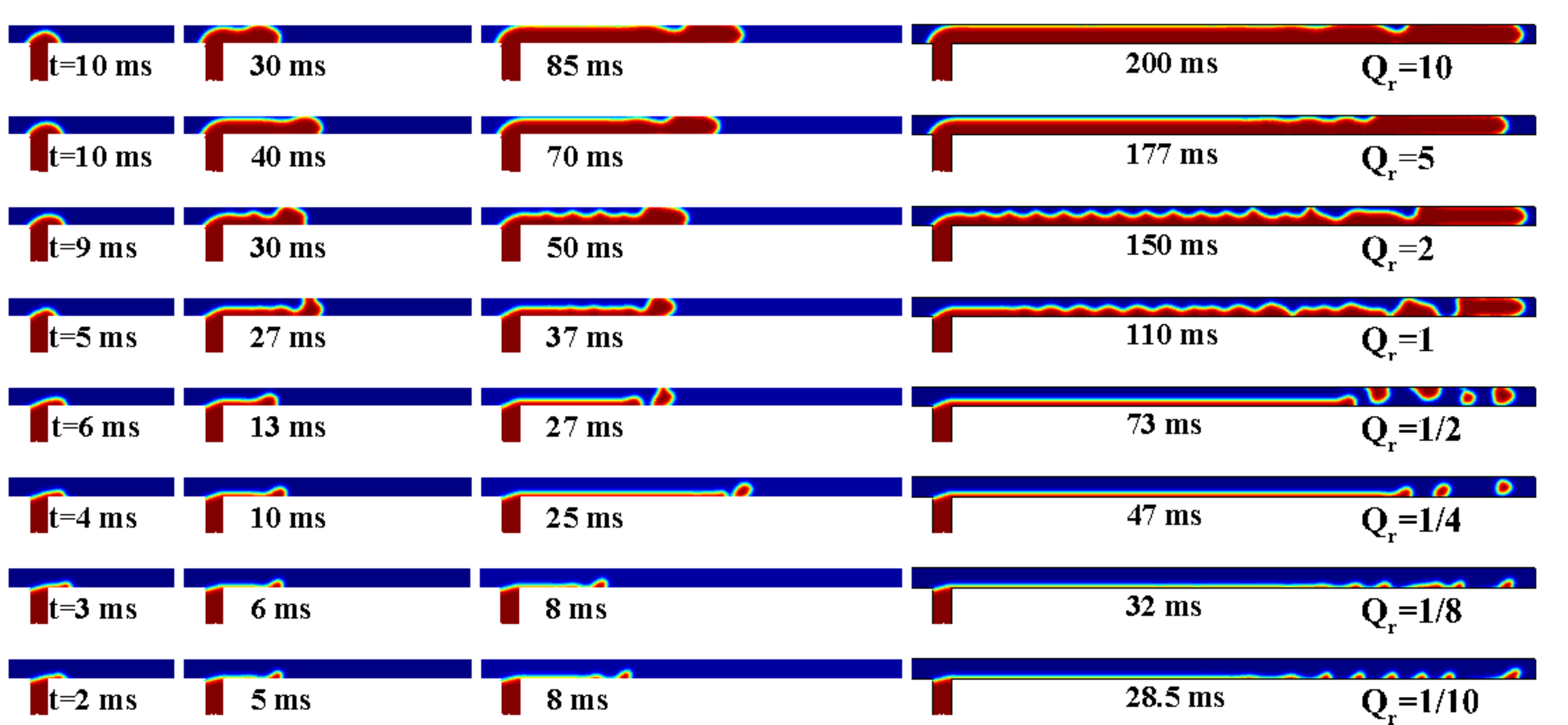}\label{fig:5c}}
	\caption{Instantaneous phase ($\phi$) flow profiles for $10^{-1}\le \qr\le 10$. (i) evolution of the dispersed phase,  (ii) droplet breakup stage, (iii) stable droplet formation, and (iv) channel filled with the \rev{hydrodynamically developed} droplets/dispersed phase.}
	\label{fig:5}
\end{figure}

%
\noindent
The \rev{hydrodynamics of two-phase flow and} droplet generation is shown in  \figs \ref{fig:4} and \ref{fig:5} through instantaneous phase ($\phi$) flow profiles in a microfluidic device for a \rev{wide} range of \rev{flow rate ratio} ($10^{-1}\le \qr\le 10$) and \rev{capillary number} ($10^{-4}\le \cac\le 1$). 
\begin{figure}[!b]
	{\includegraphics[width=1\linewidth]{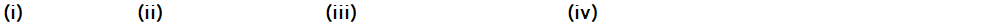}}
	\\	
	\subfloat[$\qr=1$]{\centering\includegraphics[width=1\linewidth]{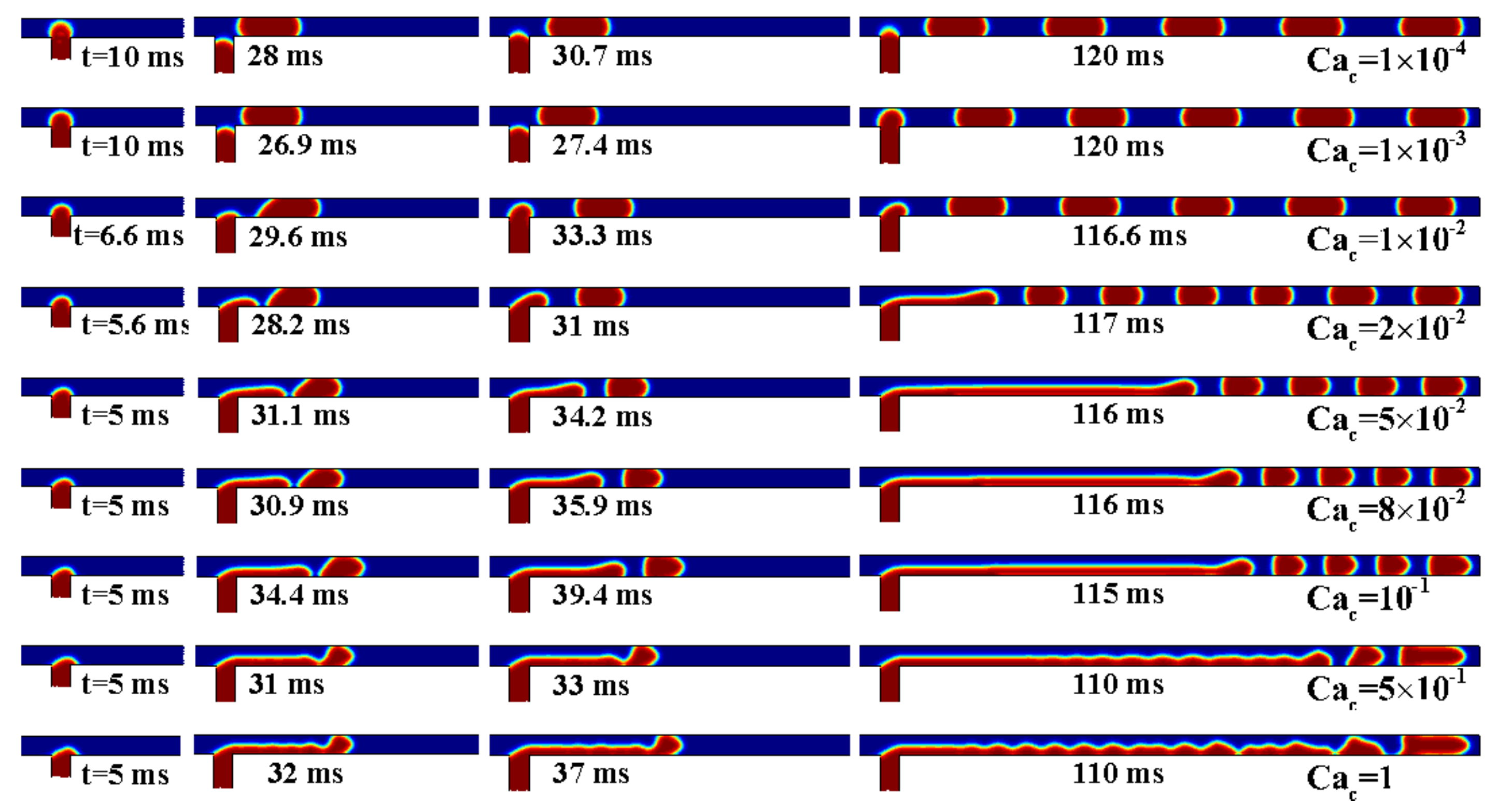}\label{fig:4a}}	
	\caption{Instantaneous phase ($\phi$) flow profiles for $10^{-4}\le \cac\le 1$. (i) evolution of the dispersed phase,  (ii) droplet breakup stage, (iii) stable droplet formation, and (iv) channel filled with the \rev{hydrodynamically developed} droplets/dispersed phase.}
	\label{fig:4}
\end{figure}
Each flow conditions show the four stages of the flow and droplet generation as follows: (i) evolution of the dispersed phase,  (ii) droplet breakup stage, (iii) stable droplet formation, and (iv) channel filled with the \rev{hydrodynamically developed} droplets/dispersed phase.
\begin{figure}[!b]
	\ContinuedFloat\
	{\includegraphics[width=1\linewidth]{Figures/Phase_compositions/Untitled}}
	\\	
	\subfloat[$\qr=10$]{\centering\includegraphics[width=1\linewidth]{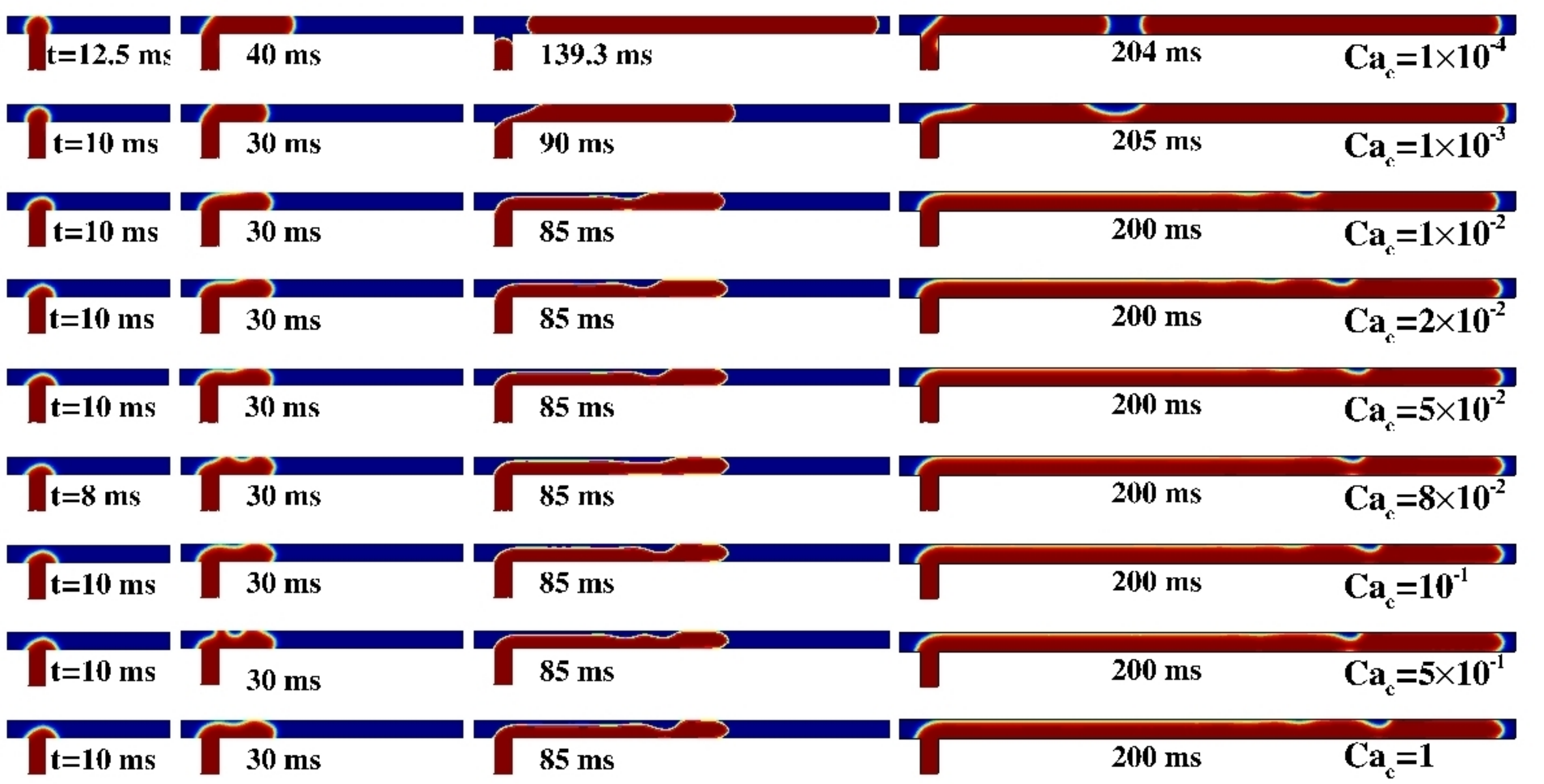}\label{fig:4b}}\\
	\subfloat[$\qr=1/10$]{\centering\includegraphics[width=1\linewidth]{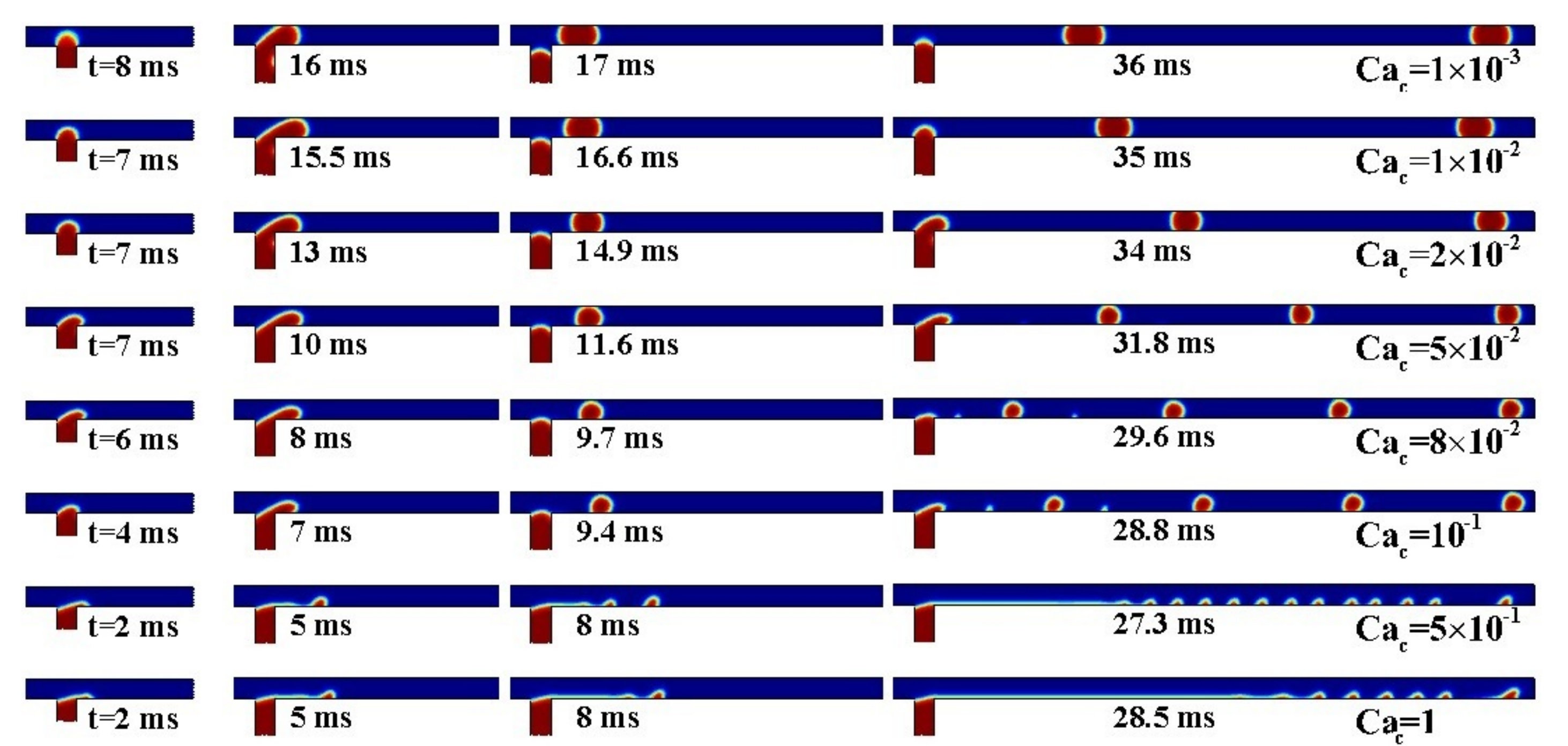}\label{fig:4c}}
	\caption{Continued.}
	\label{fig:4}
\end{figure}
To highlight the instantaneous evolution of the dispersed phase, the length of channel shown in the figures is varied according to the movement of the dispersed phase in the downstream of the main channel. 
\\\noindent 
\fig\ref{fig:4} depicts the effect of interfacial tension ($\cac$)  on the phase flow behaviour for a fixed flow rate ratio ($\qr$). It can be noted that, in the initial stage, when the dispersed phase enters into the main channel, the viscous and pressure forces that rise in the upstream region are sufficient to overcome the interfacial force to form a droplet. Once the hydrodynamic development \rev{of liquid phases} in the channel \rev{is established}, the flow becomes stratified. 
\\\noindent 
\fig \ref{fig:4a} presents a special case of equal flow rate of both the phases ($\qr=1$, i.e., $Q_{\text{d}} = Q_{\text{c}}$). At lower values of $ \cac (< 10^{-2})$, the elongated droplets ($L/w_{\text{c}} \gg 1$) generate immediate after the interaction of two phase. 
The fluid flow in such cases is primarily attributed to the dominance of the interfacial force balanced by the pressure rise in the upstream region. 
At $\cac =2\times10^{-2}$, the location of droplet detachment shifts towards the downstream with an increase in time, and both the immiscible fluids flow parallel to each other. The size of subsequently generated elongated droplets  ($L/w_{\text{c}} > 1$) also reduces.    
Therefore, there is a transition occurring at $Ca_{\text{c,trans}} \approx 2\times10^{-2}$. For $Ca_{c} > Ca_{\text{c,trans}}$, the strengthening of the viscous (i.e., shear) stress results in the formation of a stratified/parallel flow. Further, \rev{an increasing portion of} the downstream region is occupied by the stratified flow with increasing $Ca_{c}$.  It results in the reduction of the downstream region available for the droplet formation.  Therefore, it is evident that the interplay between the viscous and interfacial forces has a significant impact on the droplet formation \rev{even} for the simplest limiting case of $\qr=1$. 
Such flow behaviors, however, change with the flow rate ratio ($\qr$), as seen in \figs \ref{fig:4b} and \ref{fig:4c}. 
\\\noindent 
\fig \ref{fig:4b} presents the phase ($\phi$)  flow profiles  for  $\qr=10$ wherein the droplet formation can be witnessed at much lower values of $\cac(\approx 10^{-4})$. However, two-layered stratified flow is visible for $\cac > 10^{-3}$. It is reported \rev{in literature} \citep{Garstecki2006}  for $\cac < 10^{-2}$ that majority of the downstream region is filled with the dispersed phase.  It thereby causes the pressure rise in the upstream region, which is the primary contributor to provide the necessary mechanism for droplet formation. This \rev{flow nature} corresponds to the `squeezing regime' of two-phase flow. When $\cac \gtrsim10^{-2}$, the shear stress becomes more important to initiate the droplet breakup, and the flow follows `dripping like regime'.
Surprisingly, the present study has observed that the `squeezing regime' continues even for $\cac > 10^{-2}$ for a fixed $\qr~(\geq2)$. This effect is attributed solely to the variation of interfacial tension, i.e., $Ca_{\text{c,trans}}$ value is dependent only on the surface tension for a fixed $\qr$.
\\\noindent 
At the lower $\qr~(<0.5)$, the droplets being formed are smaller in size compared to that at the higher $\qr$, as shown in  the phase ($\phi$)  flow profiles in \fig \ref{fig:4c} for $\qr=0.1$.  It is also noted from \fig \ref{fig:4c} that the distance between the subsequent droplets has increased significantly. It implies that as $\qr$ is decreased (i.e., $Q_{\text{c}}$ increased while keeping $Q_{\text{d}}$ constant), the generated droplets are rapidly swept away with continuous fluid phase. 
Moreover, for lower $\qr=0.1$, the droplets are almost circular in shape with a diameter of droplet in the order of the channel width (i.e., $d_{\text{eff}} \le w_{\text{c}}$).
\\\noindent
Subsequently, combined effects of the interfacial tension and fluid viscosity on the droplet formation are analyzed through the phase ($\phi$) flow profiles in \fig \ref{fig:5} in the range $10^{-1}\le \qr\le 10$. The variation of $\qr$ for fixed $\cac$ implies that the interfacial tension force and viscous force of the continuous phase responsible for the droplet formation are balanced, in order to maintain a constant $\cac$.
\fig \ref{fig:5a} displays the phase ($\phi$) flow profiles for $5\times10^{-1}\le \qr\le 10$ at lowest $\cac=10^{-4}$. At higher $\qr \ge 5$, the plug type \rev{(i.e., axially elongated)} droplets are formed. The length of droplet \rev{is generally} at least ten times higher than the channel width ($L/w_{\text{c}} \gtrapprox 10$). 
On further decrease in $\qr (\le 2)$, the length of the droplets decreases, and eventually resulting in droplets of lengths comparable to the channel width \rev{($L/w_{\text{c}} \approx 1$)} for significantly low $\qr$.  The generated droplets are clearly distinguishable and mono-dispersed in nature.  The distance between the subsequent droplets is also uniform in this region. Also, the time taken to form the droplets is decreased, and droplets generation frequency is increased on varying $\qr$ from $10$ to $1/2$. 
\\\noindent 
\fig \ref{fig:5b} presents the observations of the formation of the droplet at a relatively larger $\cac$ ($=2\times10^{-2}$).  At higher $\qr(\geq 2)$, droplets are not formed as the dispersed phase is flowing and occupying more volume of the downstream channel until it reaches the outlet of the channel.  It is also evident that the pressure rise in the upstream and shear stress exerted by the outer fluid is not sufficient to break the interface between the two fluid phases. On varying $\qr$ from $1$ to $1/10$, the distance between the subsequent droplets is increasing as the viscosity ratio ($\mu_{\text{r}}$) is decreasing, and this trend is consistent with the observations reported in literature \citep{Liu2011a}. 
Furthermore, the phase flow behaviours shown in \fig \ref{fig:5c} are observed for a spacial case of $\cac=1$ wherein both viscous and interfacial forces contribute equally. In this case, droplet formation is not seen irrespective of $\qr$. The flow is stratified and jet-type at $\qr > 1$ and $\qr<1$, respectively.
\\\noindent 
\rev{In general,} the two-phase flow patterns have shown a complex interplay of inertial, viscous, and interfacial forces in governing the \rev{the hydrodynamics of} droplet generation. The flow patterns are categorized in the subsequent section.
%
\subsection{Classification of flow regimes}
\begin{figure}[!b]
	\centering
	\subfloat[Schematics of flow regimes characterization]{\includegraphics[width=0.45\linewidth]{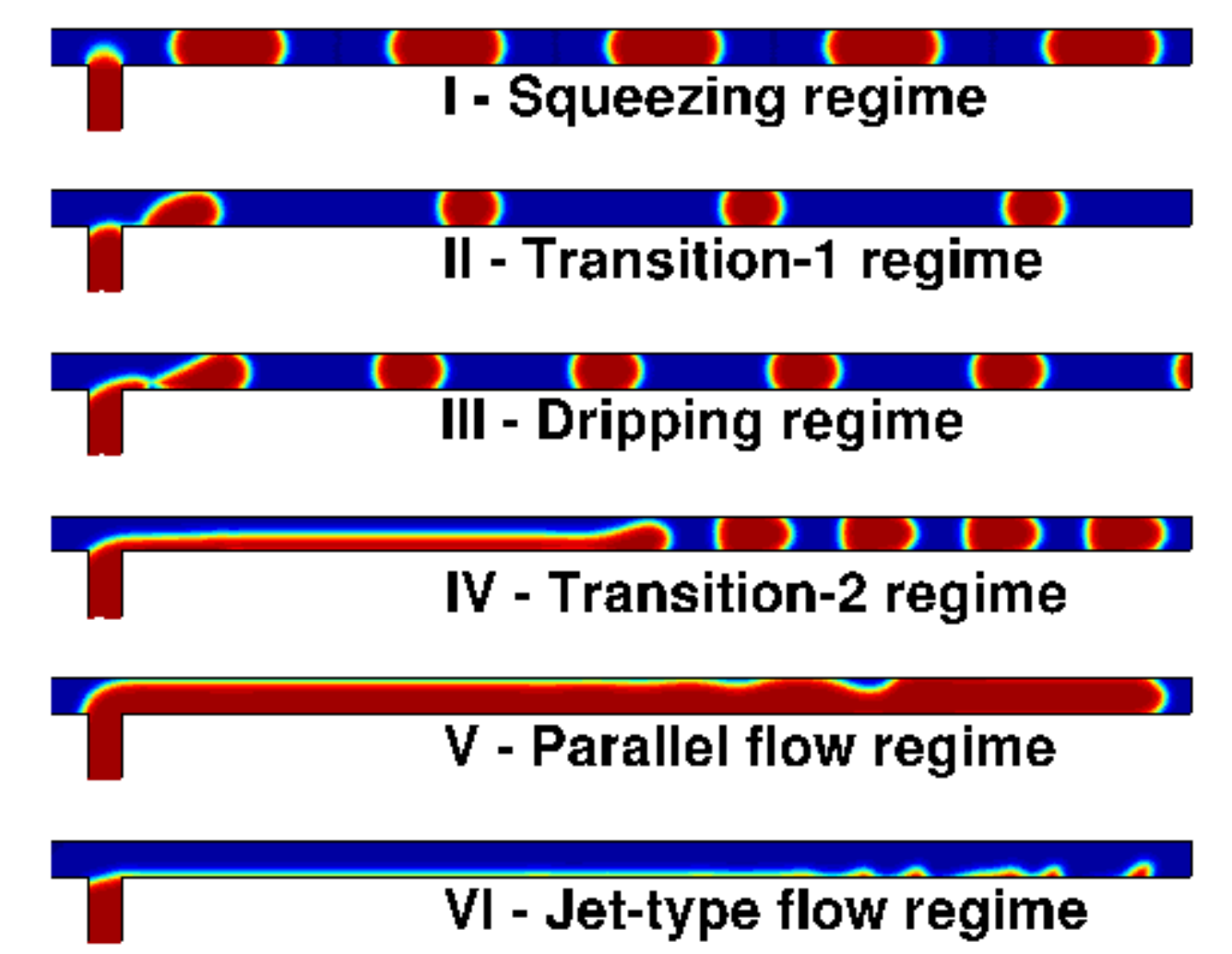}\label{fig:14b}}
	\subfloat[Ranges of conditions]{\scalebox{0.75}{
			\begin{tabular}[b]{c|c|c|c}
			\hline Regime & $\cac$ & $\qr$ & Droplets\\\hline
			I  &  $< 10^{-2}$      & $0.1\le \qr\le 10$  & Yes\\\cline{1-3}
			II  & $\approx10^{-2}$  & $\leq 1$ & \\\cline{1-3}
			III & $2\times10^{-2}<\cac<10^{-1}$ &  $0.5<\qr<0.1$  & \\\hline
			IV  &$2\times10^{-2} < \cac < 10^{-2}$  &$0.25\le \qr\le 1$  & No\\\cline{1-3}
			V   &$\leq 10^{-2}$     &$>1$  &  \\\cline{1-3}
			VI  &$>10^{-2}$         &$0.1\leq \qr \leq 0.125$  &   \\\cline{3-3}
			&         &$0.5\leq \qr \leq 1$   &  \\\hline
		\end{tabular}\label{fig:14d}}}
\\
	\subfloat[Flow regime map]{\includegraphics[width=0.5\linewidth]{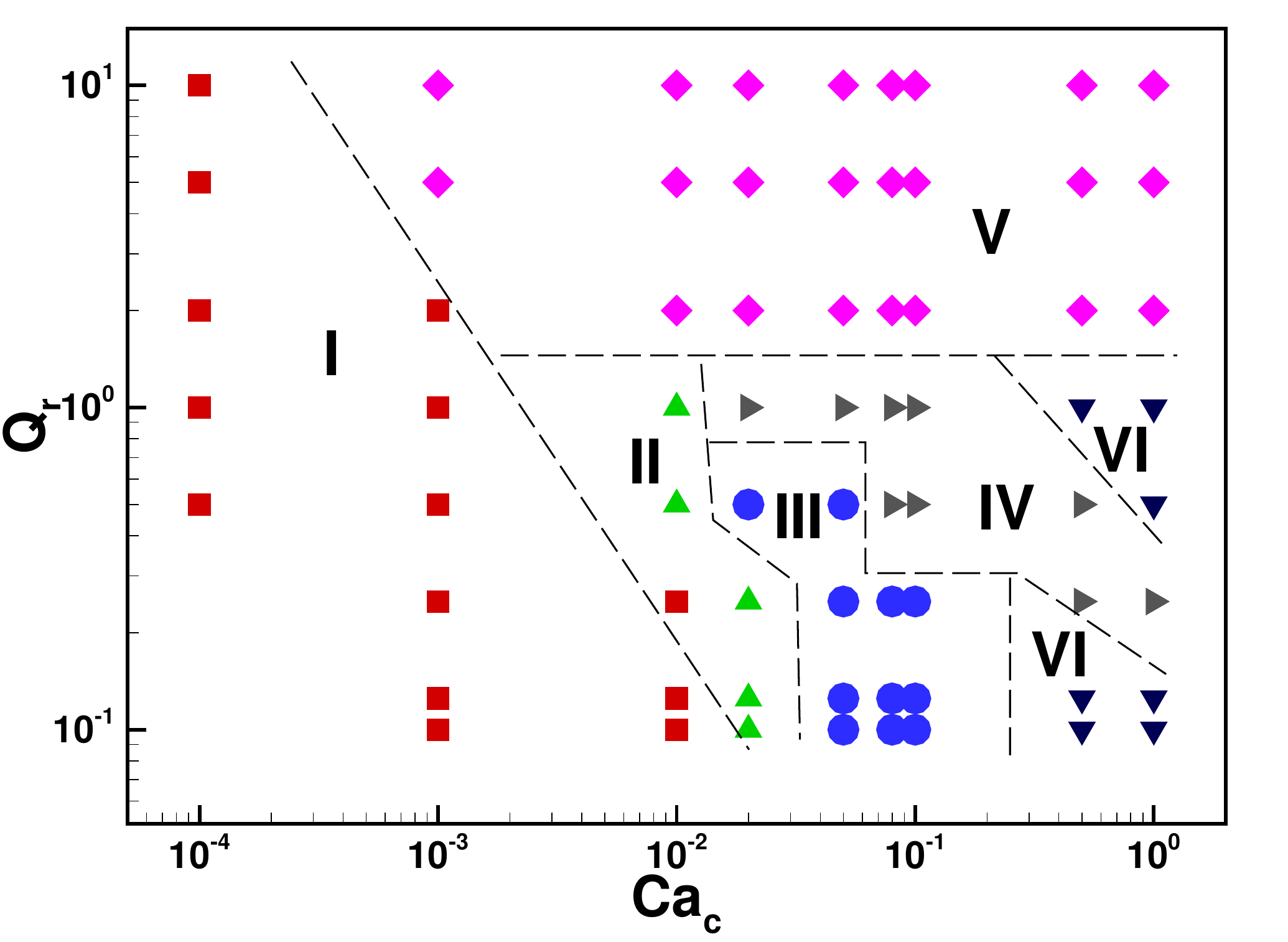}\label{fig:14c}}
	\subfloat[Droplet and non-droplet zones]{\includegraphics[width=0.5\linewidth]{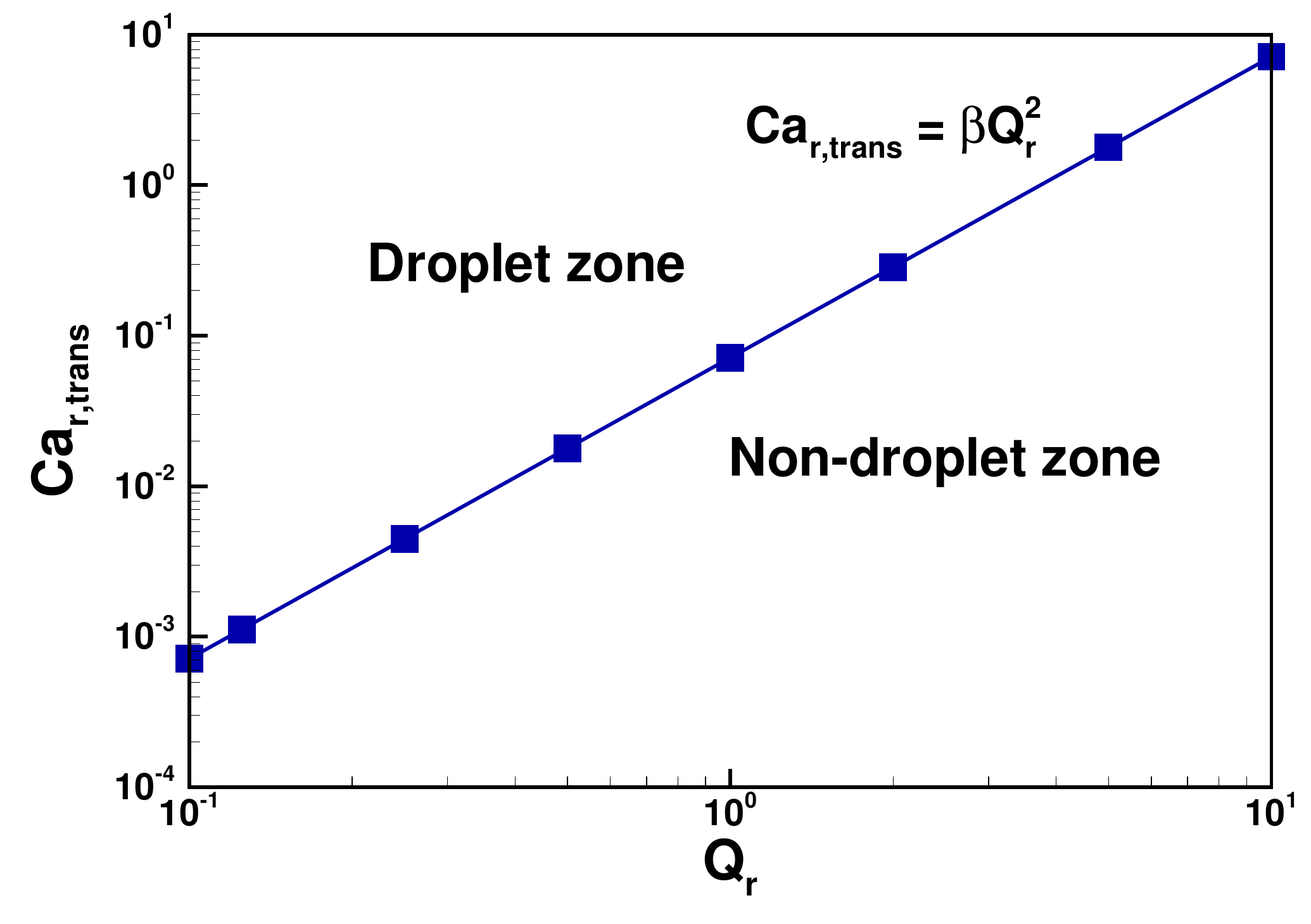}\label{fig:14a}}
	\caption{Classification of regimes for two-phase flow through T-junction microfluidic device.}
	\label{fig:14}
\end{figure}

%
\noindent The phase profiles (\figs \ref{fig:4} and \ref{fig:5}) have displayed that the flow patterns are strongly influenced by the flow rate ($\qr$) and capillary number ($\cac$). Various transitional features are also noted under otherwise identical flow governing conditions.  \fig\ref{fig:14b} has schematically represented the nature of flow during various transitions observed for the ranges of conditions studied herein.  
In culmination, essentially, six types of flow regimes, namely, ({I}) squeezing, (II) first transition, (III) dripping, (IV) second transition, (V) parallel, and (VI) jet-type flow, are observed in the present study. 
\rev{The ranges of conditions for each flow regime are summarized in  \fig \ref{fig:14d} and also sketched through the flow map in \fig \ref{fig:14c}.}
\\\noindent  
The `squeezing regime' is observed for the low values of $\cac\ll10^{-2}$ wherein the highly elongated droplets ($L\ggg w_{\text{c}}$) \rev{are generated}. 
The `dripping regime', wherein the low-to-negligible elongated droplets ($L\ge w_{\text{c}}$), is seen for $2\times10^{-2}<\cac<10^{-1}$ and $1/2<\qr<1/10$. \rev{The droplets generated in this regime are generally mono-dispersed in nature.}
The `first transition' between the `squeezing' and `dripping' regimes is occurring at $\cac\approx10^{-2}$ for $\qr\leq 1$ wherein the droplet size reduces in large extent. 
The `second transitional' regime \rev{appears in} between the `dripping' and `parallel flow' regimes for  $2\times10^{-2} < \cac < 10^{-2}$ and $1/4\le \qr\le 1$. In this regime, few droplets formed in the initial stage, immediately after the interaction of both phases in the downstream channel. Subsequently, the dispersed phase is flowing parallel to the continuous phase. The `parallel flow' regime appears after the second transition for $\cac\leq 10^{-2}$ and $\qr>1$. In this type of flow, the dispersed phase enters the channel and fills the entire downstream region. 
\rev{After some finite time, a moving interface is eventually created between the two immiscible phases due to the strong resistance imposed by the dispersed phase on the continuous phase. It, consequently, prevents droplet formation.}
Further, a `jet-type flow' regime appears for $\cac>10^{-2}$ and $\qr<1$. In this regime, the dispersed phase continues to flow, until the outlet of the main channel, as a single thread-like jet without droplet formation, \rev{due to the dominance of inertia imposed by the continuous phase}.  The droplet formation in the \rev{second transition,} parallel and jet-type flow regimes is not affirmative.
\\\noindent 
It has been established above that the droplets are certainly formed in the first three (i.e., squeezing, the first transition, and dripping) flow regimes. In contrast, droplet formation is not evident in the other three (second transition, parallel, and jet-type) flow regimes.  The two-phase flow behaviour in microfluidic geometry, thus, can also be categorized into the `droplet' and `non-droplet' zones\rev{, as depicted in \fig\ref{fig:14a}}.  
These two zones are distinguished through the ratio of the threshold capillary numbers of the dispersed and continuous phases ($Ca_{\text{r,trans}}=Ca_{\text{d,trans}}/Ca_{\text{c,trans}}$). 
For example, the transition from the `droplet' to `non-droplet' zone for $\qr=1$ takes place (see \fig \ref{fig:4a}) at $\cac=2\times10^{-2}$ and correspondingly at $Ca_{\text{d}}=1.43\times10^{-3}$, and therefore, $Ca_{\text{r,trans}}=7.15\times10^{-2}$. 
\\\noindent 
To depict the `droplet' and `non-droplet' zones in two-phase microfluidic flows, \fig \ref{fig:14a} displays the variation of $Ca_{\text{r,trans}}$  with $\qr$ over the ranges of conditions explored in this work.  Markedly, \rev{a log-log curve between $Ca_{\text{r,trans}}$  and $\qr$ separates the `droplet' and `non-droplet' zones by a straight line characterized by} $Ca_{\text{r,trans}}$ proportional to square of $\qr$\rev{, as expressed by \eqn\eqref{eq:10}.} 
\begin{equation}
Ca_{\text{r,trans}} =\beta \qr^2\qquad\text{where}\qquad Ca_{\text{r,trans}}=\frac{Ca_{\text{d,trans}}}{Ca_{\text{c,trans}}}
\label{eq:10}
\end{equation}
where $\beta = 0.07143$ is obtained by statistical analysis of the \rev{numerical} data. 
\rev{In other words, the straight line (\eqn\ref{eq:10}) flow map represents a boundary between zones (III) dripping and (IV) second transition.} 
It is noted that there is a precise formation of the droplets above the curve (\fig \ref{fig:14a}), and hence, called the `droplet zone'. 
\rev{For the combination of $Ca_{\text{r,trans}}$ and $\qr$} under the curve, the viscous force exerted by the continuous phase at the interface is not sufficient to overcome the interfacial tension force. 
Due to this, both the continuous and dispersed phases are flowing parallel to each other without any droplet formation. 
At very high $Ca_{\text{r,trans}}$, the dispersed phase \rev{attains} a jet-type flow \rev{as} the continuous phase exerts a considerable viscous force on the \rev{interface} of the dispersed phase. It is thus \rev{unable to produce droplets}, and called the `non-droplet' zone. \rev{Further, in addition to $Ca_{\text{r}}$ and $\qr$, the viscosity and density of fluid phases significantly influence the hydrodynamic nature. The proportionality factor ($\beta$) demarcating the boundary between the `droplet' and `non-droplet' zones thus may be related to Reynolds number ratio ($Re_{\text{r}}$) of the two phases. It, however, cannot be affirmed at this stage, and more experimentation is required with varying $Re_{\text{r}}$ to affirm the functional relationship between $\beta$ and $Re_{\text{r}}$.}
\\\noindent 
The above discussed phase ($\phi$) flow profiles (\figs\ref{fig:4} to \ref{fig:14}) show a complex interplay of inertial, viscous, and interfacial forces governing the \rev{hydrodynamics of two-phase flow and} droplet generation in microfluidic systems. The droplet behaviour as function of $\cac$ and $\qr$ is further analyzed in subsequent sections.
%
\subsection{Droplet size}
\begin{figure}[!b]
	\centering
	\subfloat[$\qr=10$]{\includegraphics[width=0.48\linewidth]{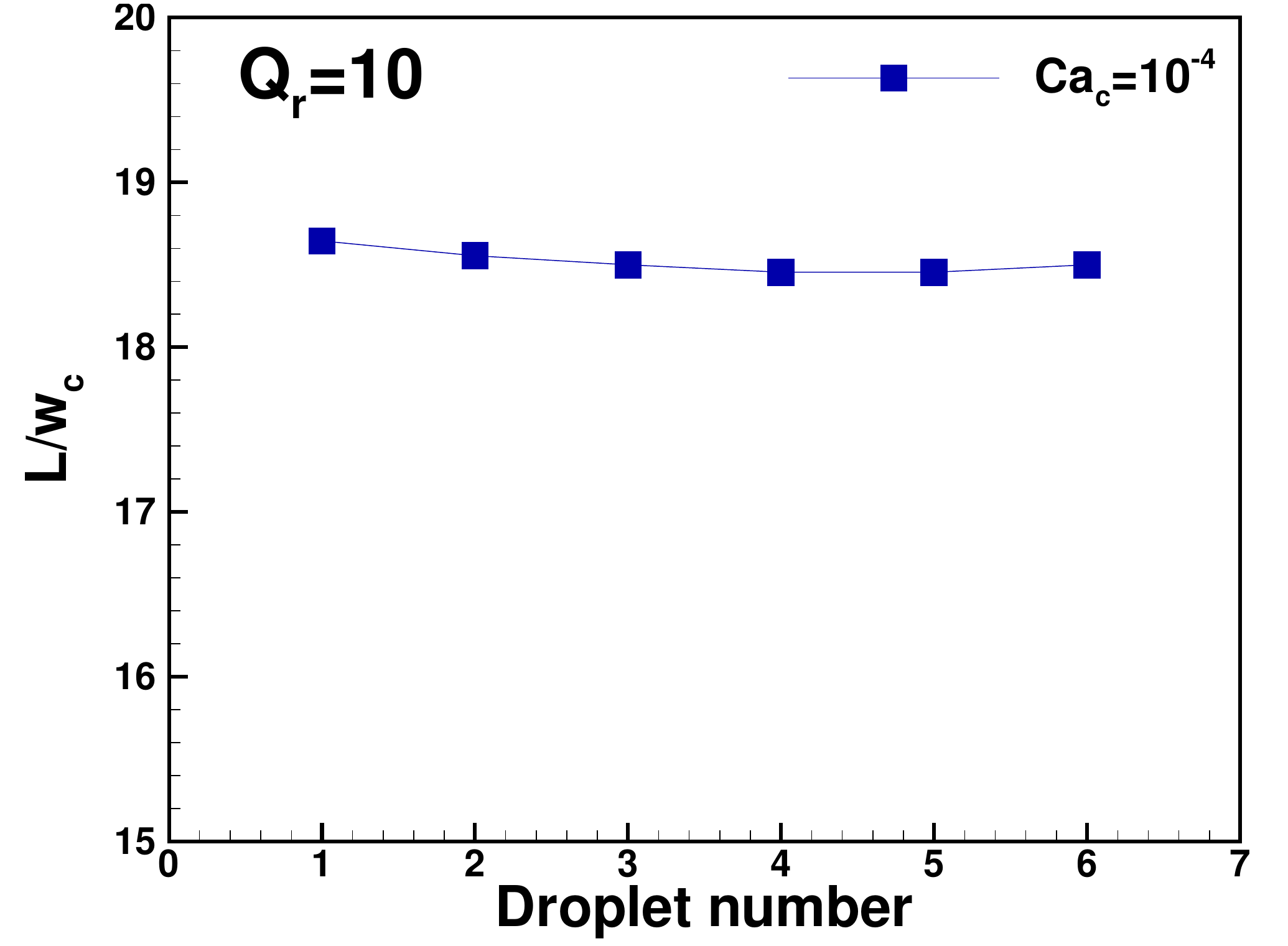}\label{fig:6a}}
	\subfloat[$\qr=2$]{\includegraphics[width=0.48\linewidth]{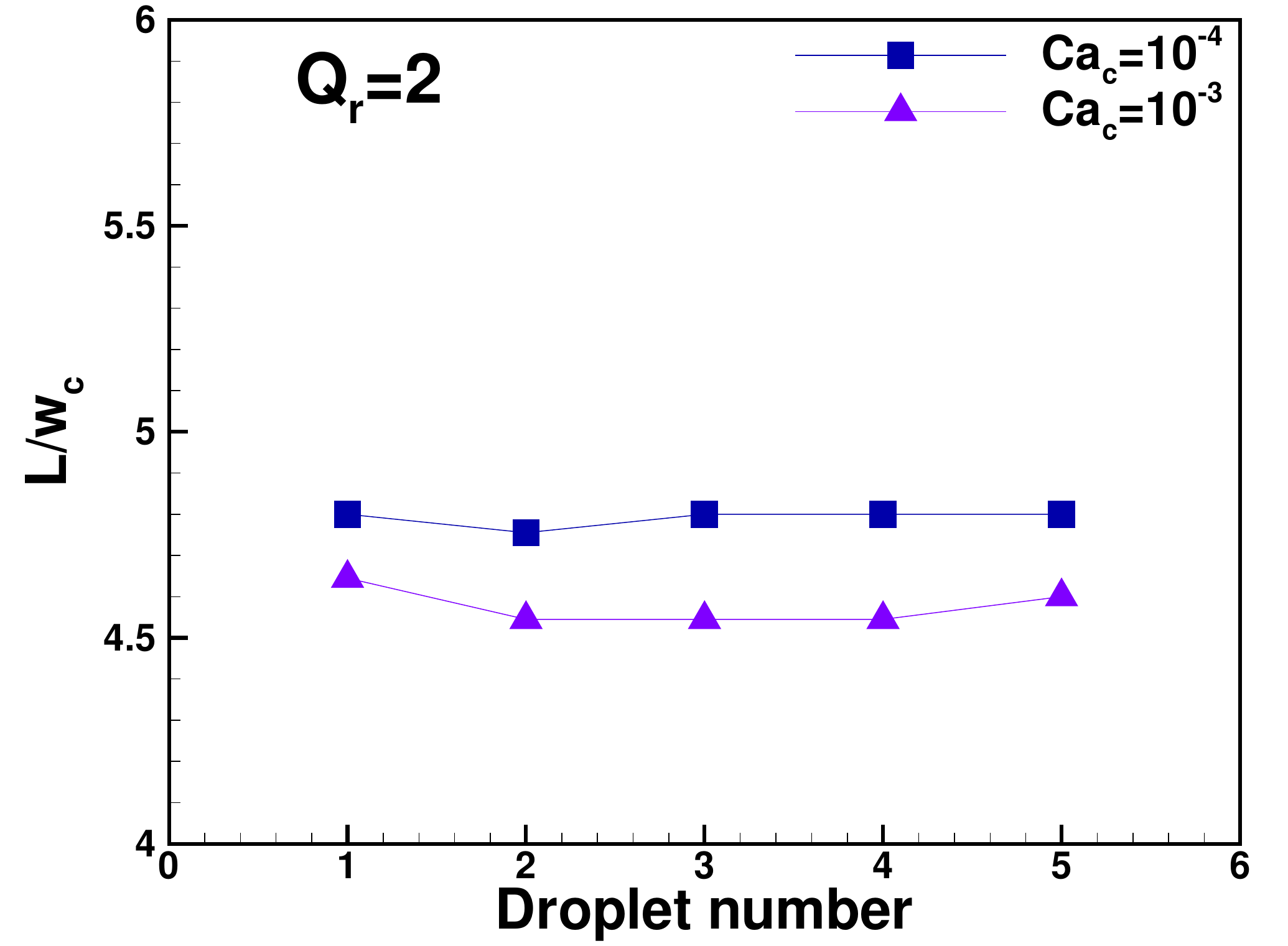}\label{fig:6b}}
	\\
	\subfloat[$\qr=1$]{\includegraphics[width=0.48\linewidth]{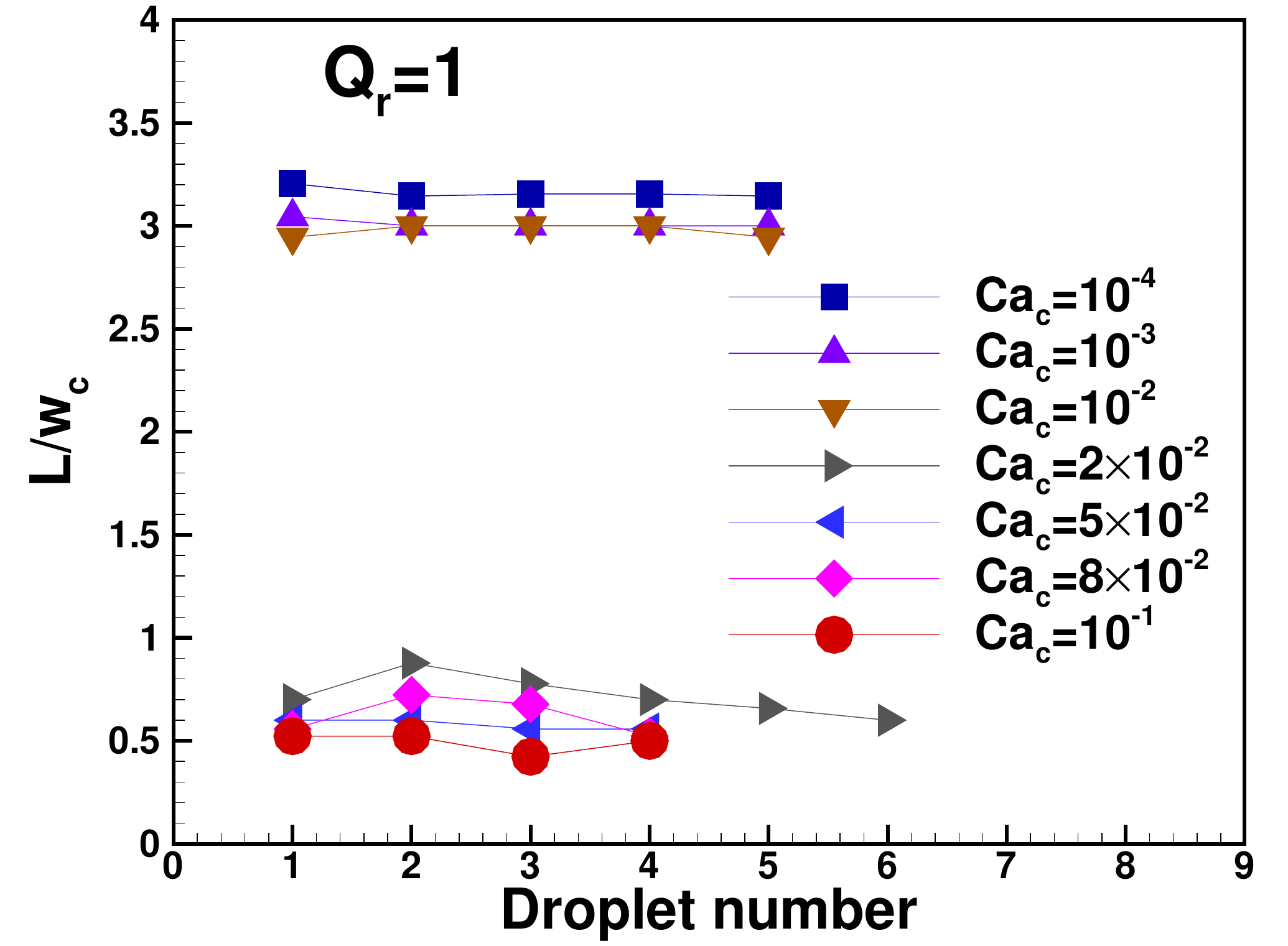}\label{fig:6c}}
	\subfloat[$\qr=1/10$]{\includegraphics[width=0.48\linewidth]{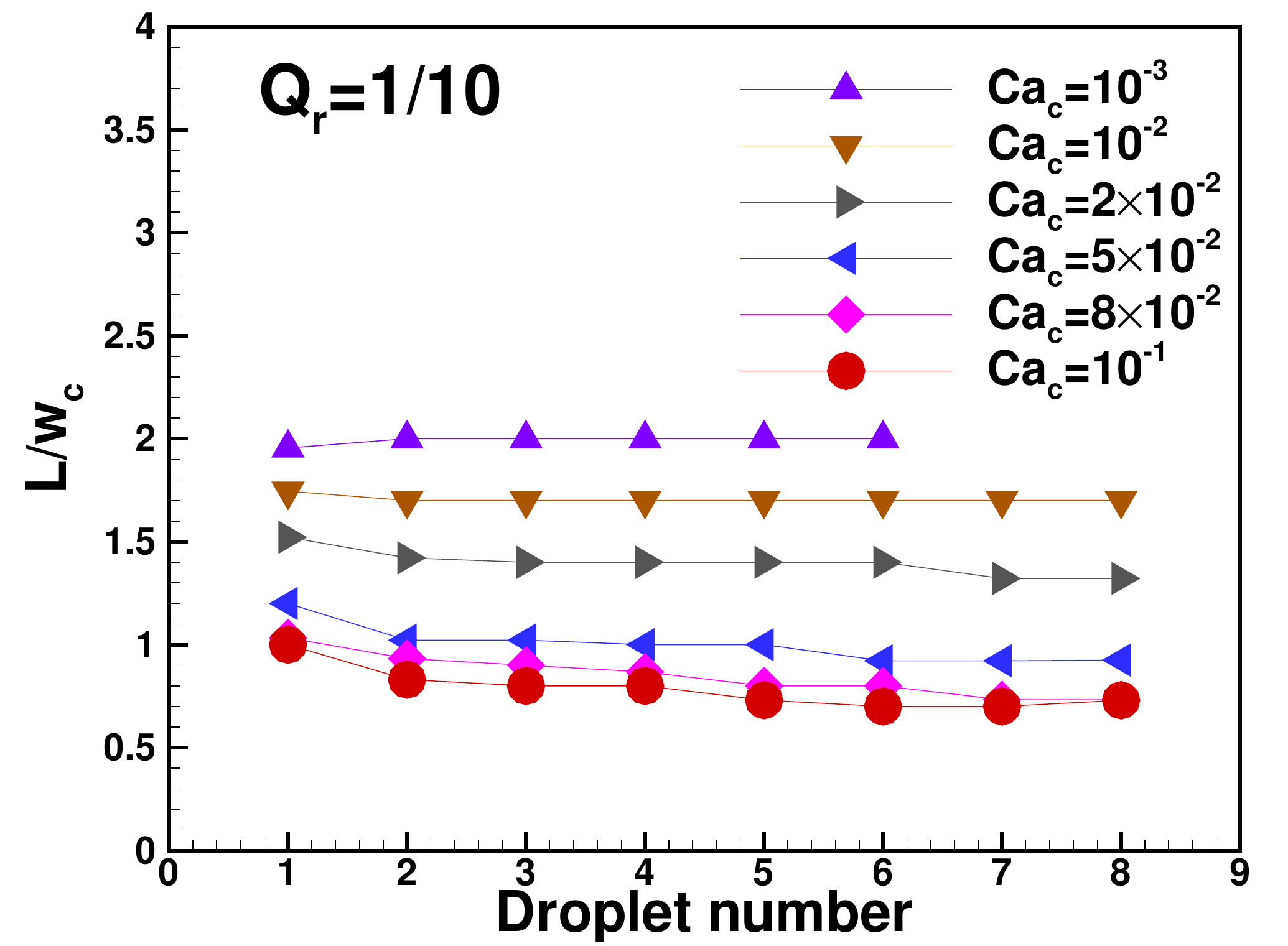}\label{fig:6d}}
	\caption{Dependence of droplet length ($L/w_{\text{c}}$) on $\cac$.}
	\label{fig:6}
	\vspace{-1em}
\end{figure}

\begin{figure}[!b]
	\centering
	\subfloat[$\cac=10^{-4}$]{\includegraphics[width=0.48\linewidth]{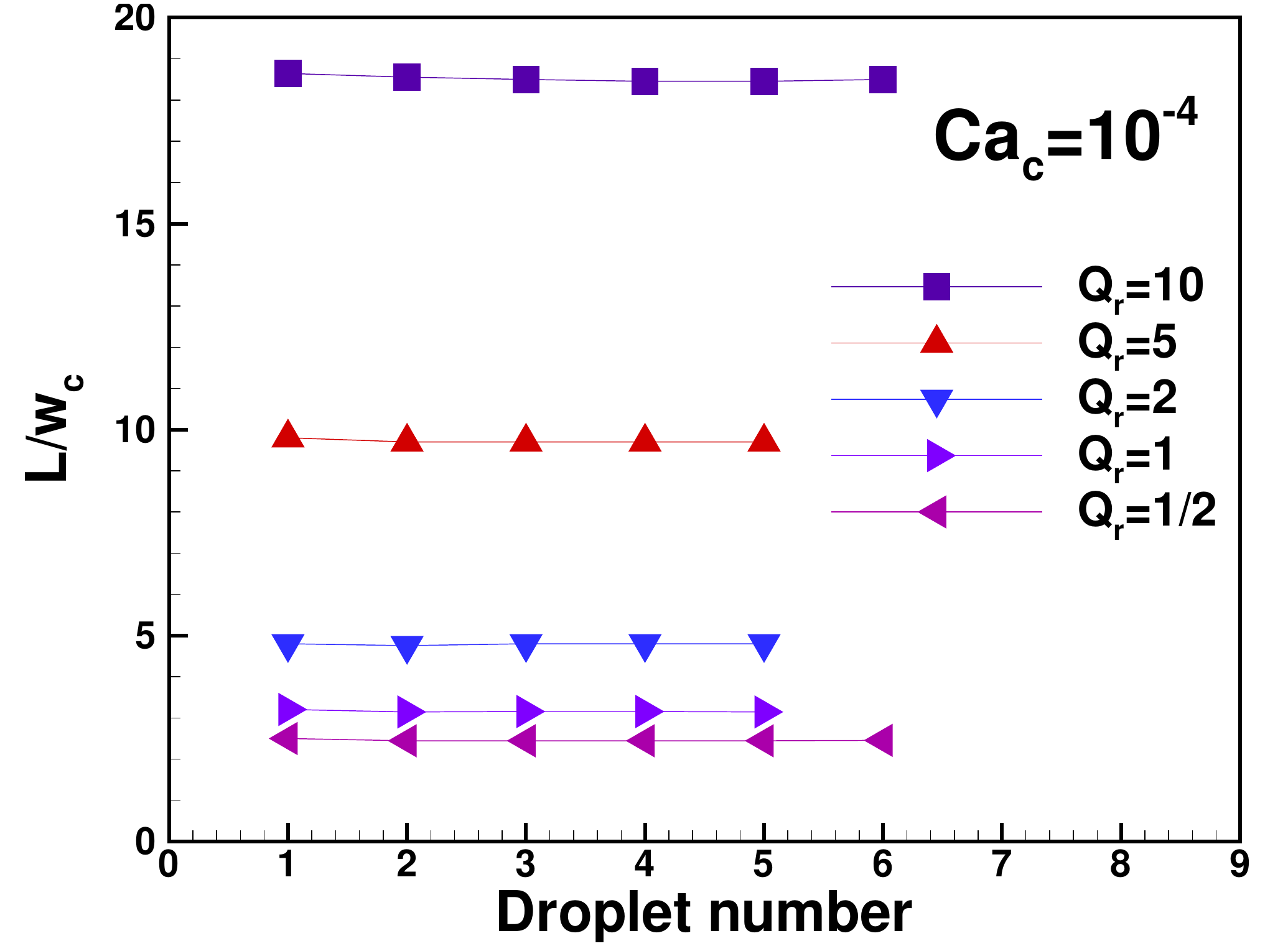}\label{fig:7a}}
	\subfloat[$\cac=10^{-3}$]{\includegraphics[width=0.48\linewidth]{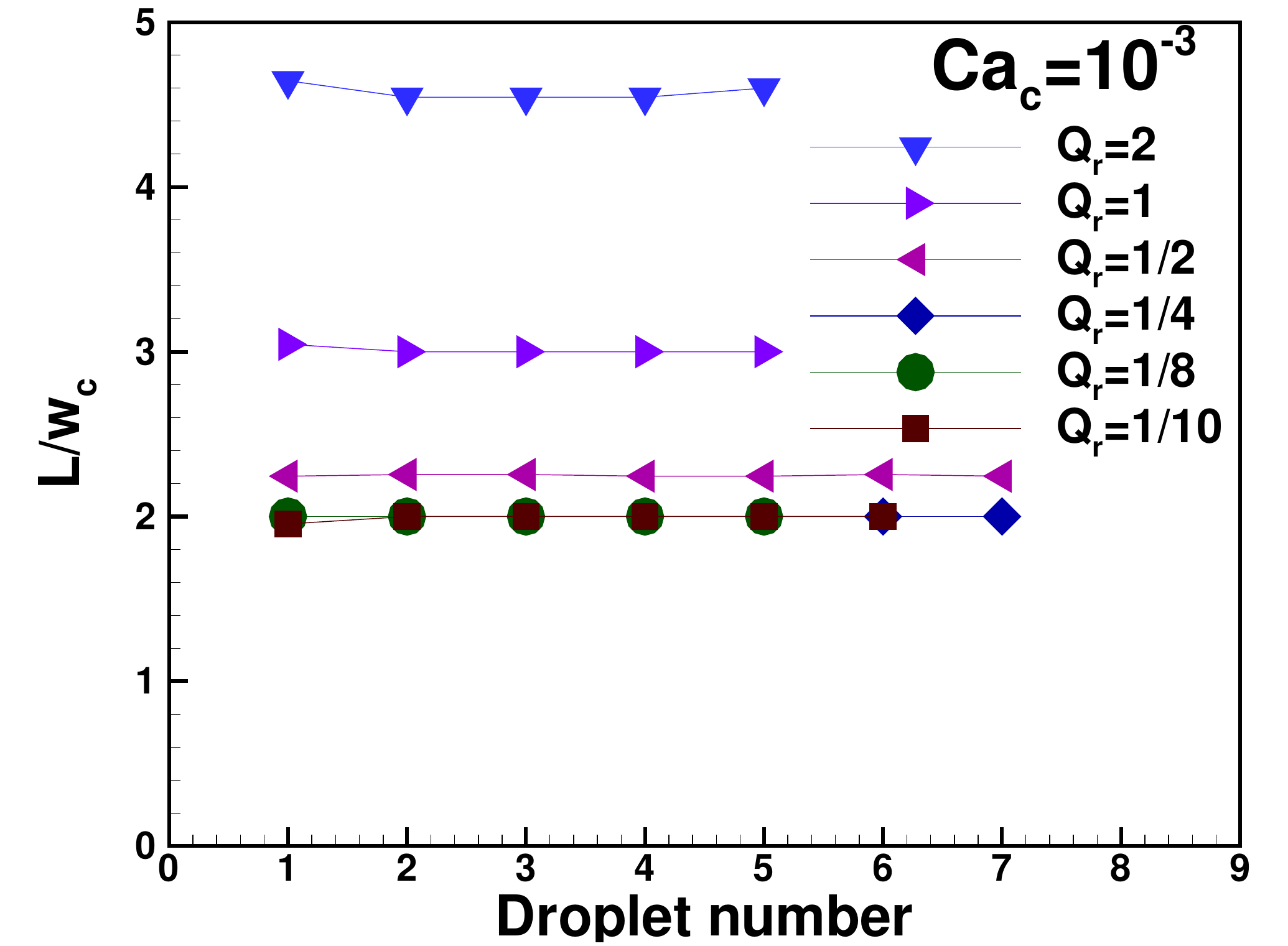}\label{fig:7b}}
	\\
	\subfloat[$\cac=10^{-2}$]{\includegraphics[width=0.48\linewidth]{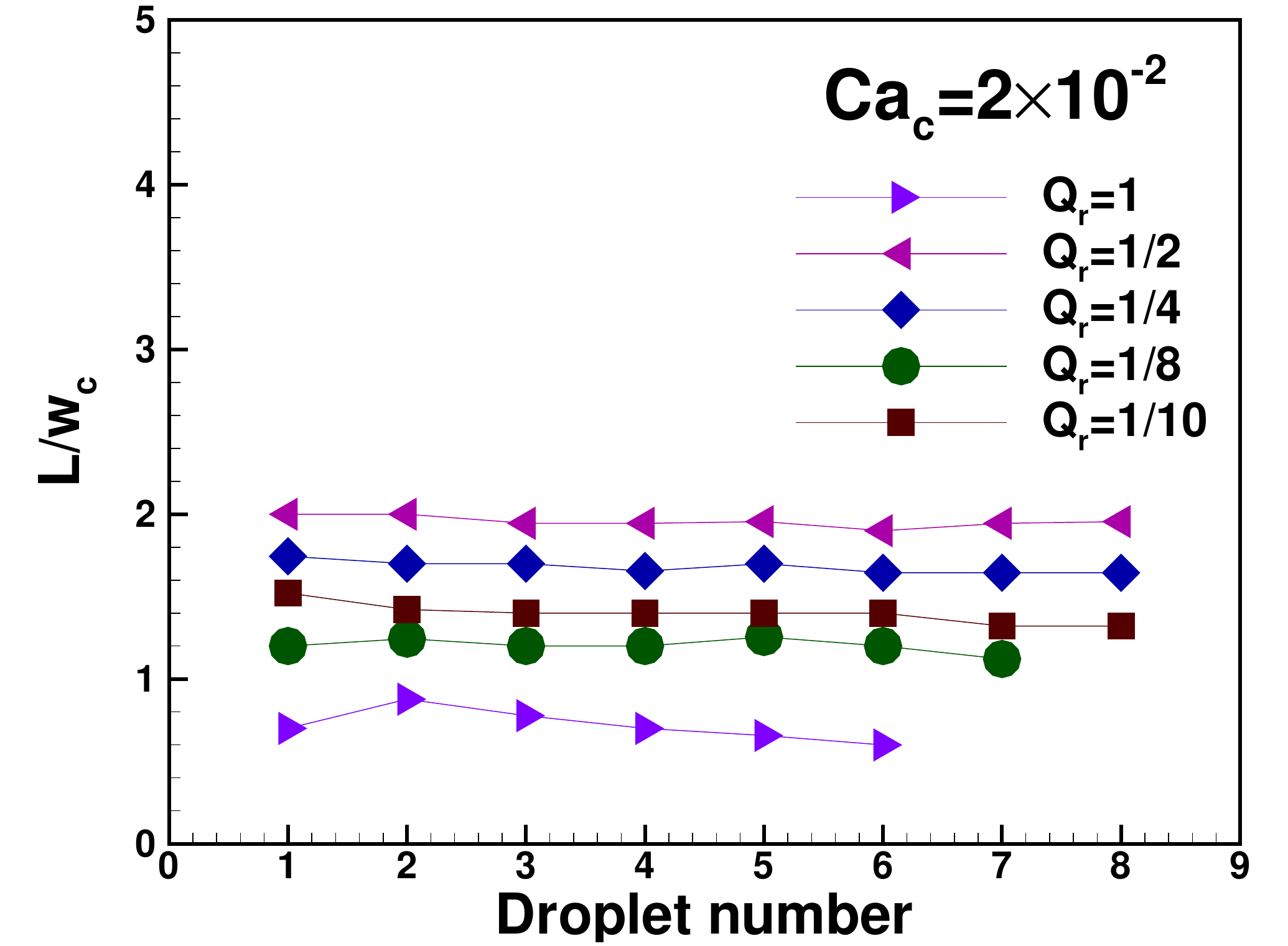}\label{fig:7c}}
	\subfloat[$\cac=10^{-1}$]{\includegraphics[width=0.48\linewidth]{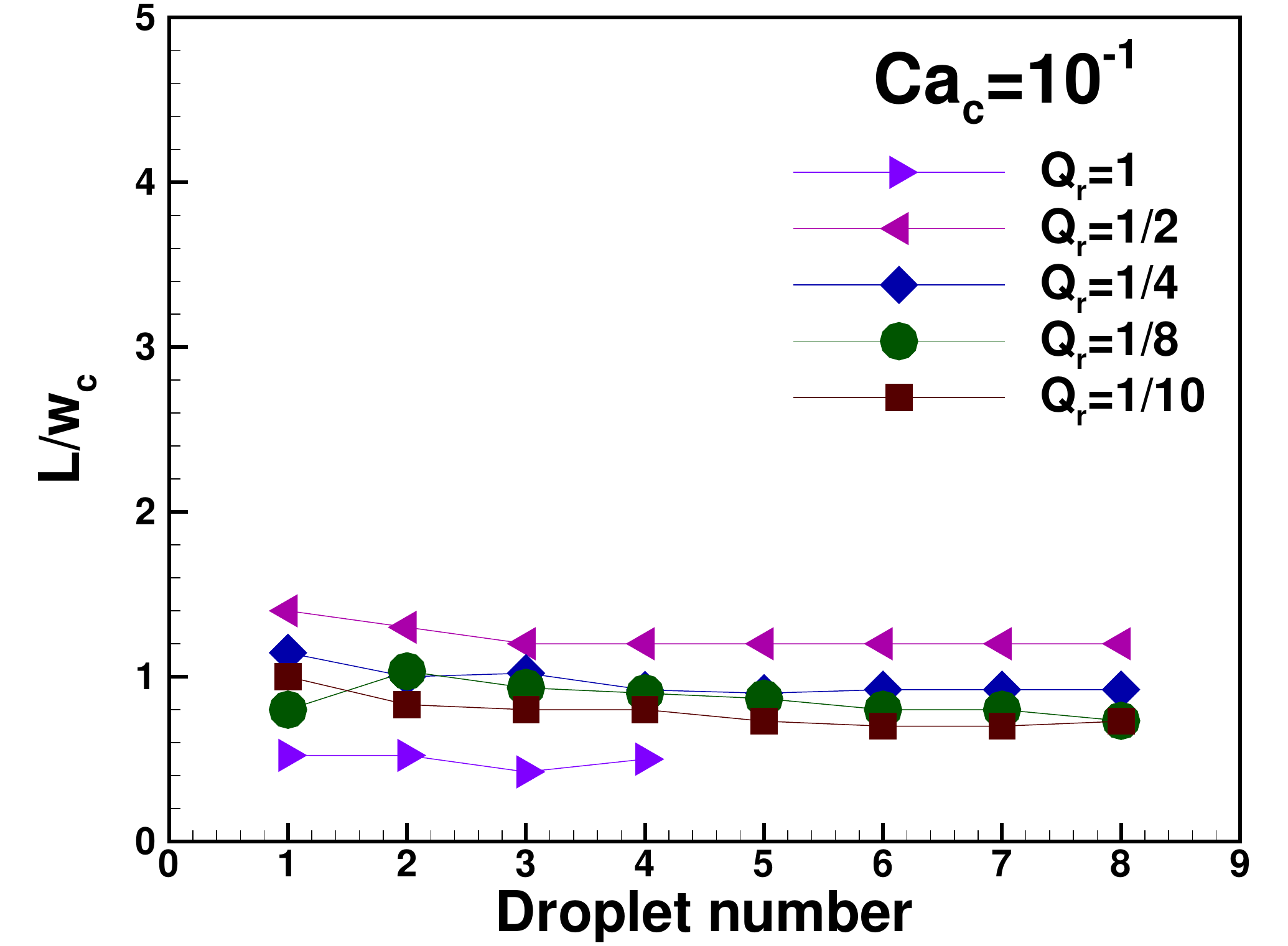}\label{fig:7d}}
	\caption{Dependence of droplet length ($L/w_{\text{c}}$) on $\qr$.}
	\label{fig:7}
\end{figure}

\noindent In order to investigate the combined influences of the interfacial tension, viscosity, and flow rate on the droplet size, \figs \ref{fig:6} and \ref{fig:7} have plotted for the length of each droplet as a function of $\cac$ and $\qr$. 
The dependence of dimensionless droplet length ($L/w_{\text{c}}$) on capillary number ($\cac$) for each sequentially generated droplets is shown in \fig \ref{fig:6} for the favourable conditions \rev{(\eqn\ref{eq:10})} of droplet generation. In each plot, the flow rate ratio ($\qr$) has been kept constant, whereas $\cac$ is varied in order to emphasize the effect of interfacial tension on the length of each droplet.  
\\\noindent 
As the phase flow profiles (\figs\ref{fig:4} to \ref{fig:5}) have displayed shrinkage of droplets with increasing $\cac$, the length of droplets ($L/w_{\text{c}}$) is inversely proportional to $\cac$. 
At higher $\qr > 1$ and lower $\cac < 10^{-2}$, the elongated droplets, like plug type, are formed wherein  $L/w_{\text{c}}\gg 1 $. All the subsequent droplets generated are found to exhibit almost the same length as that of the first droplet for a given $\qr$ and $\cac$.  This observation also holds for low $\qr$  when $\cac < 10^{-2}$, as seen in \fig \ref{fig:6c}. 
As $\cac$ is further increased, the flow is transiting into another regime at $\cac=2\times10^{-2}$.  It is \rev{denoted as} a critical or transitional capillary number ($Ca_{\text{c,trans}}$) \rev{for the continuous phase}.  For $\qr < 1$, the critical capillary number ($Ca_{\text{c,trans}}$) increases up to $\qr = 1/2$. For sufficiently low values of $\qr$ $(<1/2)$, $Ca_{\text{c,trans}}$ becomes constant.
\\
Further, the combined influences of viscosity and flow rate on the droplet length are analyzed by plotting the length of each droplet as a function of $\qr$  in \fig \ref{fig:7}  at a fixed $\cac$.  For a given $\cac$ and $\qr$, it is observed that the droplets being formed are of the same length and mono-dispersed.  However, the droplet length is decreasing with $\qr$, under otherwise identical conditions.  This observation remains valid for a wide range of $\cac$ and $\qr$, as shown in \fig \ref{fig:7}. Nevertheless, the curves corresponding to $\qr\ge 1$ in \figs \ref{fig:7c} and \ref{fig:7d} show a different trend, as discussed previously.
The downstream region fills with the droplets after an initial transient in the hydrodynamic development of two-phase flow. Hence, the droplet length is calculated as an average only after the downstream channel is completely filled. Further, the non-repeatability of exact necking instability leading to the formation of droplets results in the droplets length variation of about $2-5$\% in the dripping regime. 
It can also be observed from \fig \ref{fig:7} about the droplet length varying initially and reaching a steady yet \rev{insignificant} oscillatory behavior. 
\\\noindent 
The numerical data of the droplet length ($L/w_{\text{c}}$) in the `squeezing regime' is statistically represented by the following empirical linear relation.
\begin{equation}
L = (\alpha + \beta \qr)w_{\text{c}}\qquad \text{(squeezing regime)}
\label{eq:l_squeezing}
\end{equation}
where $\alpha=1$ and  $\beta = 1.7648$ for all values of $\qr$. 
\\\noindent
The empirical correlation (\eqn\ref{eq:l_squeezing})  suggests that the droplet length ($L$) is mainly dependent on the flow rate ratio ($\qr$), and the size of the channel ($w_{\text{c}}$) whereas independent of the physical properties like viscosity ($\mu$) and interfacial tension ($\sigma$).  Similar empirical correlations are reported for other geometrical arrangements by several experimental and numerical studies  \citep{Garstecki2006,Demenech2008,Bashir2011,Nekouei2017}.  
\\\noindent
Further, the numerical data for the droplet length ($L/w_{\text{c}}$) in the `dripping regime' is represented statistically by the following empirical power-law relation.
\begin{equation}
L=(\alpha  \qr^{\beta}  \cac^{\gamma})w_{\text{c}}\qquad \text{(dripping regime)}
\label{eq:l_dripping}
\end{equation}
where $\alpha=0.5358$, $\beta=0.2307$ and $\gamma=-0.3682$ for $\cac > 10^{-2}$. 
\\\noindent 
\rev{The droplet length in the dripping regime depends in a non-linear manner on the capillary number, flow rate, and channel dimensions.}
\begin{figure}[!b]
	\centering
	\subfloat[Squeezing regime]{\includegraphics[width=0.48\linewidth]{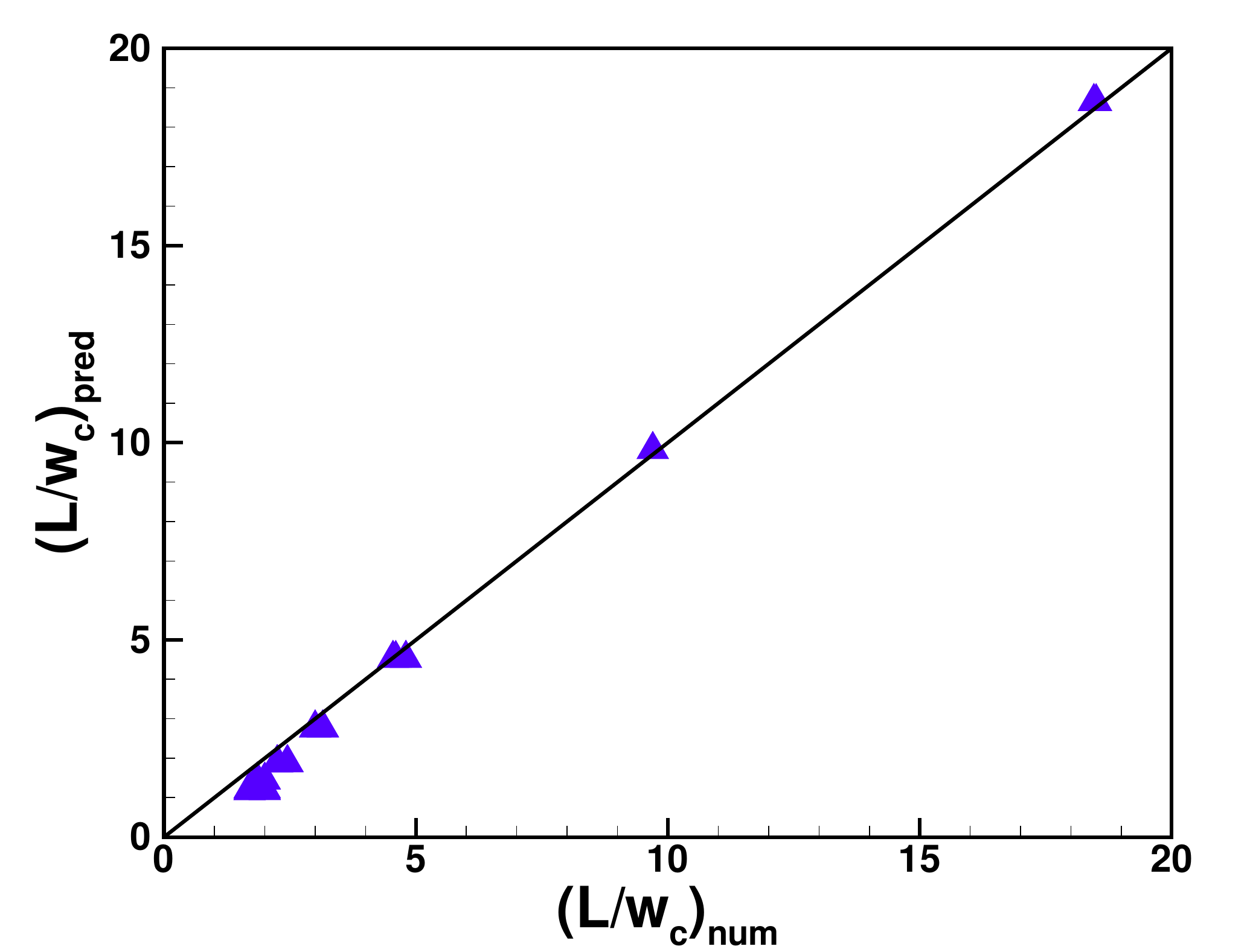}\label{fig:8a}}
	\subfloat[Dripping regime]{\includegraphics[width=0.48\linewidth]{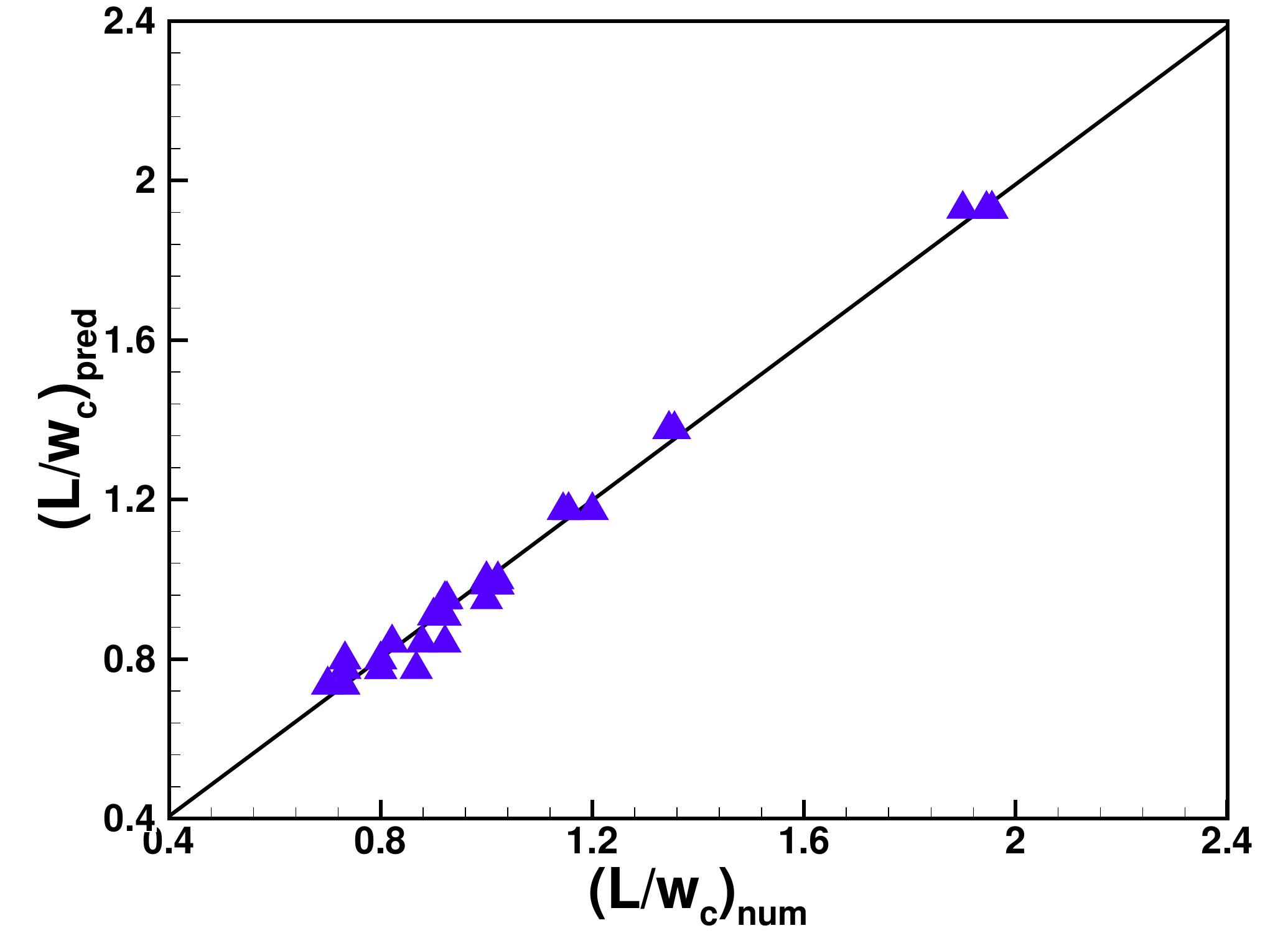}\label{fig:8b}}
	\caption{Parity plots between numerical and predicted (\eqns\ref{eq:l_squeezing} and \ref{eq:l_dripping}) values of droplet length ($L/w_{\text{c}}$).}
	\label{fig:8}
	\vspace{-1em}
\end{figure}
This relation (\eqn \ref{eq:l_dripping})  is qualitatively consistent with the literature \citep{Christopher2008,Xu2008,Gupta2010,Zhang2018} for other geometrical arrangements. 
\\\noindent
A parity plot (\figs \ref{fig:8a} and \ref{fig:8b}) shows excellent comparison ($R^{2} = 0.9899$) of the present numerical data and predictions of empirical relation (\eqns\ref{eq:l_squeezing} and \ref{eq:l_dripping}) for the droplet length ($L/w_{\text{c}}$)  in the squeezing and dripping regimes.   
%
\subsection{Droplet detachment time}
\begin{figure}[!b]
	\centering
	\subfloat[$\qr=10$]{\includegraphics[width=0.48\linewidth]{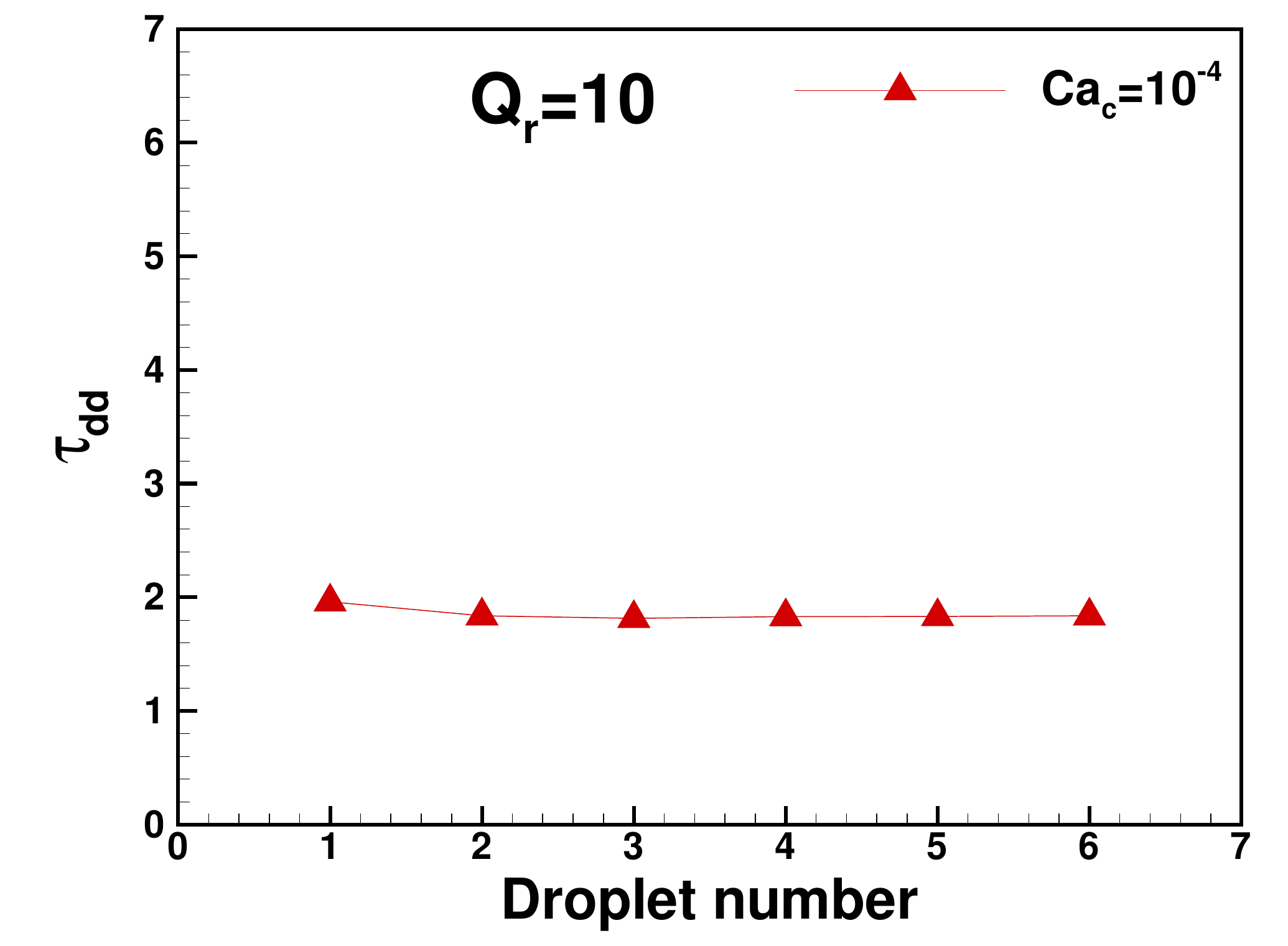}\label{fig:9a}}
	\subfloat[$\qr=2$]{\includegraphics[width=0.48\linewidth]{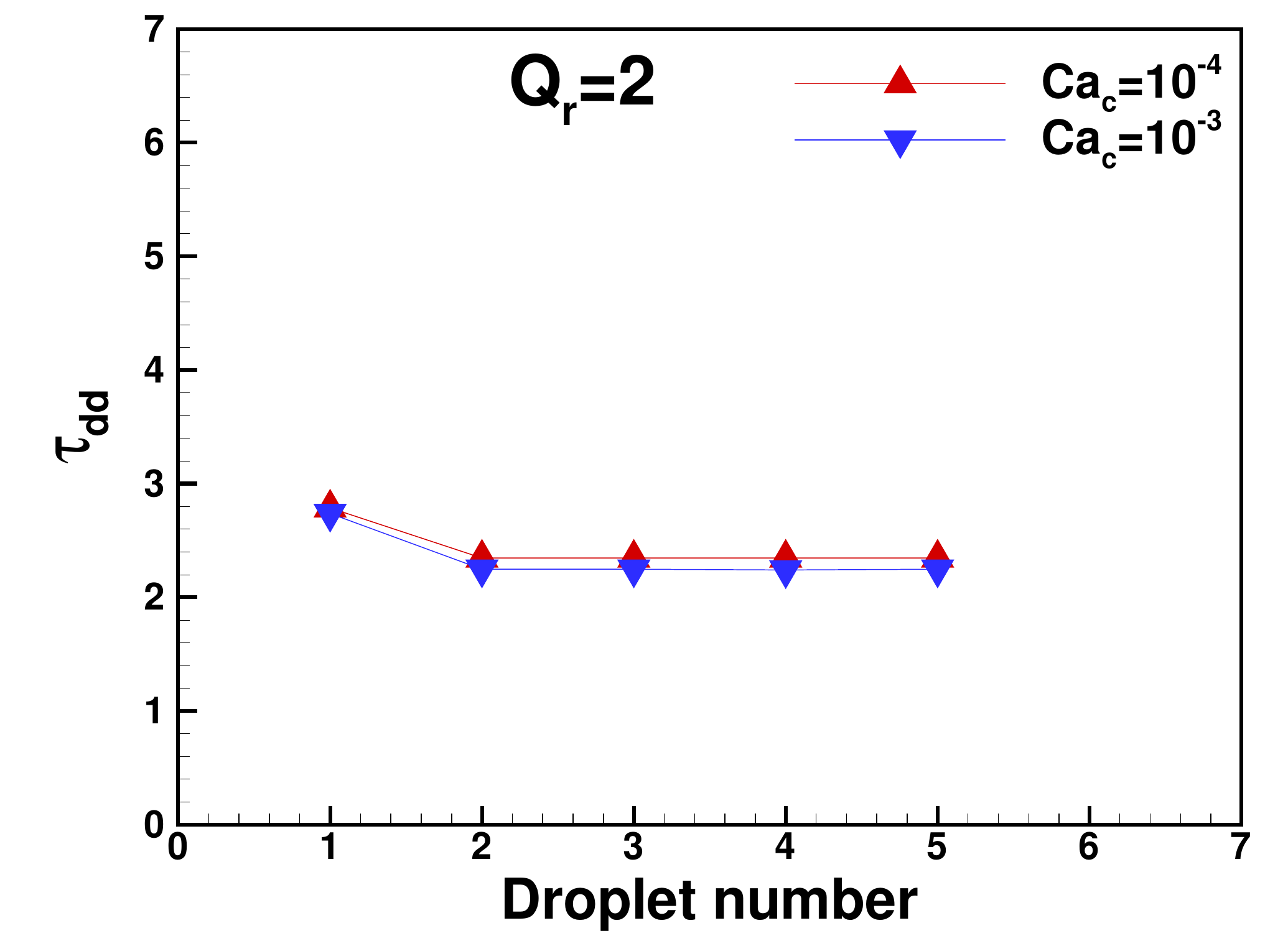}\label{fig:9b}}
	\\
	\subfloat[$\qr=1$]{\includegraphics[width=0.48\linewidth]{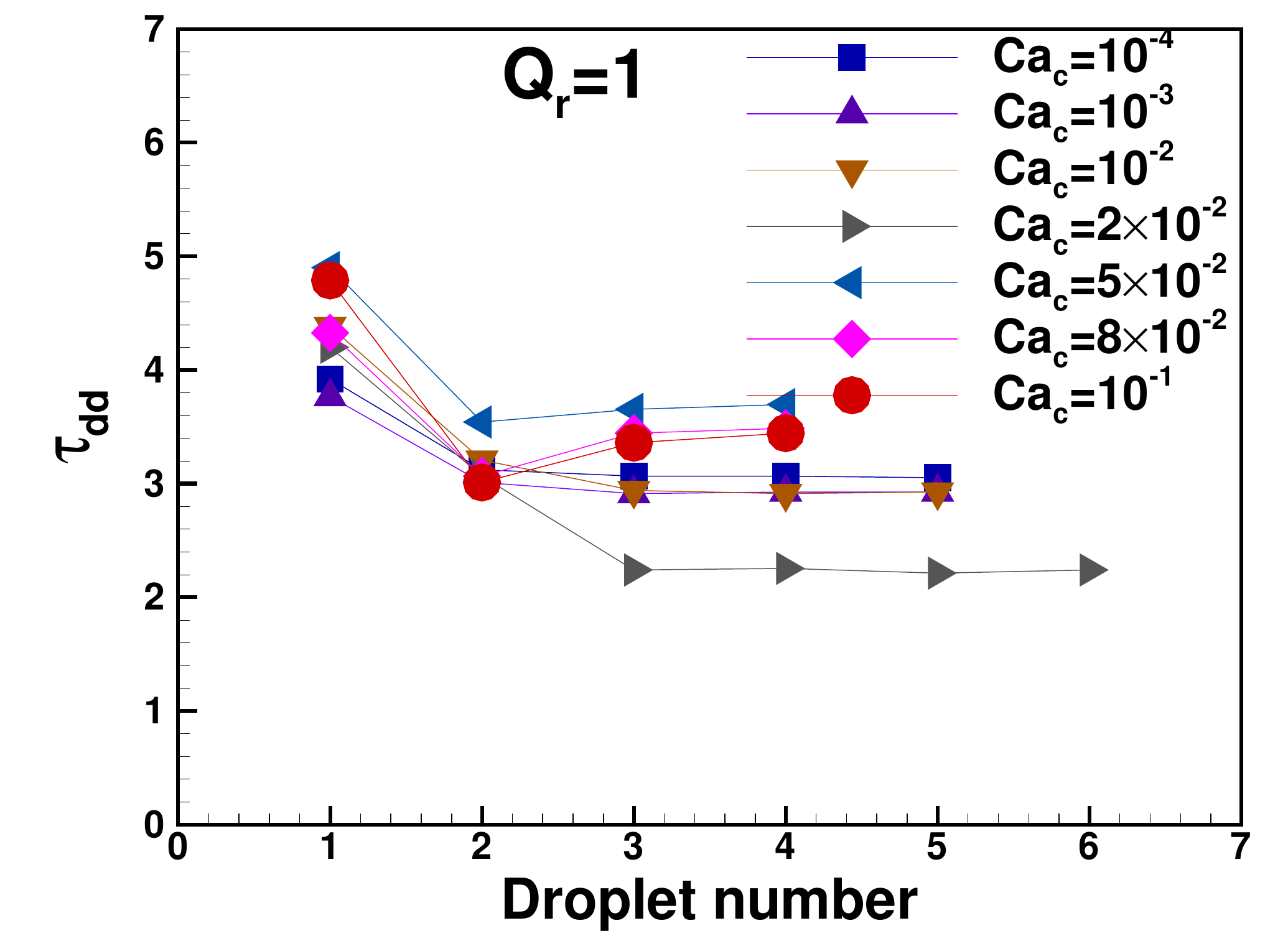}\label{fig:9c}}
	\subfloat[$\qr=1/10$]{\includegraphics[width=0.48\linewidth]{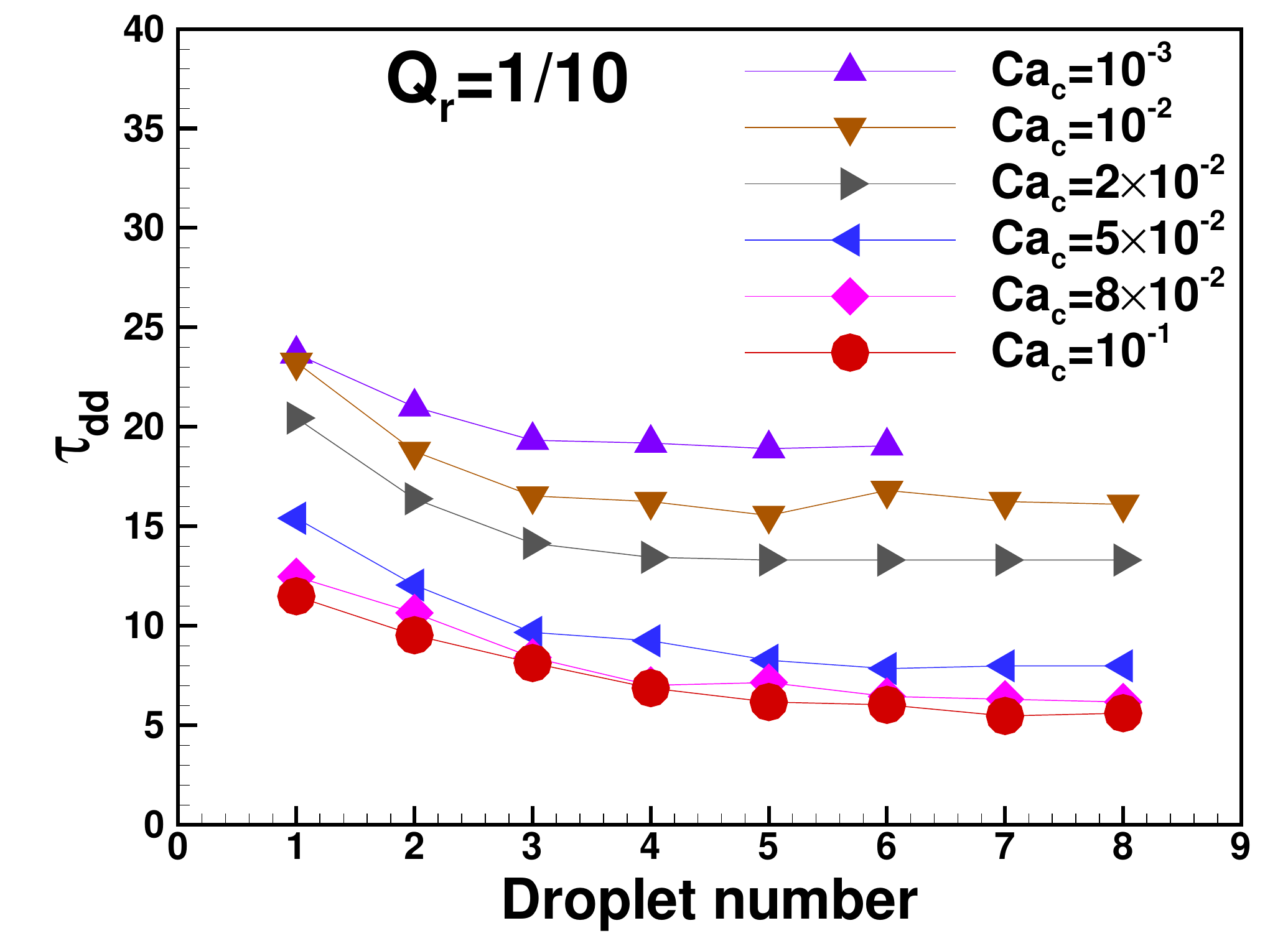}\label{fig:9d}}
	\caption{Droplet detachment time ($\tau_{\text{dd}}$) as a function of  $\cac$ for fixed $\qr$.}
	\label{fig:9}
\end{figure}

\begin{figure}[!b]
	\centering
	\subfloat[$\cac=10^{-4}$]{\includegraphics[width=0.48\linewidth]{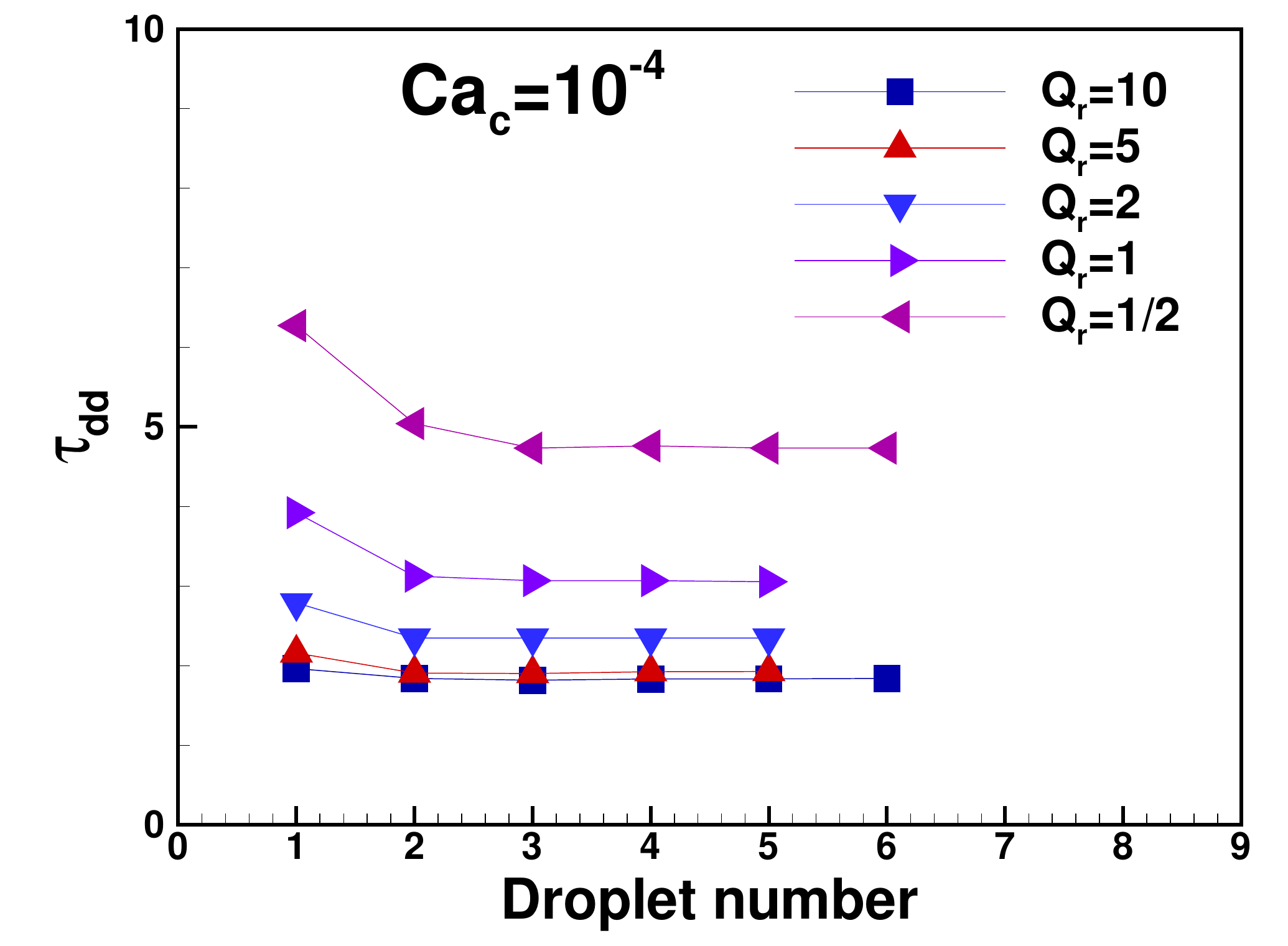}\label{fig:10a}}
	\subfloat[$\cac=10^{-3}$]{\includegraphics[width=0.48\linewidth]{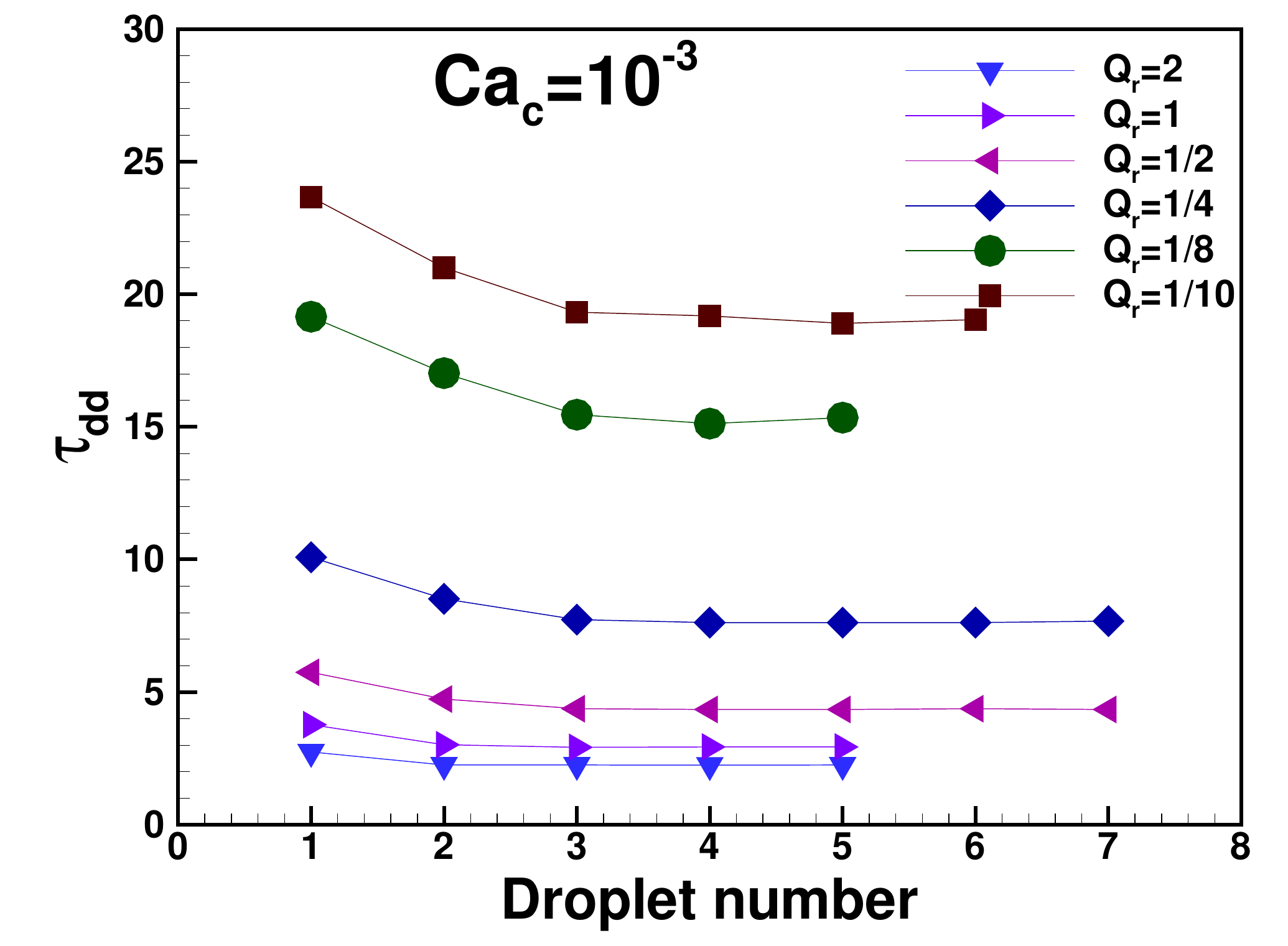}\label{fig:10b}}
	\\
	\subfloat[$\cac=10^{-2}$]{\includegraphics[width=0.48\linewidth]{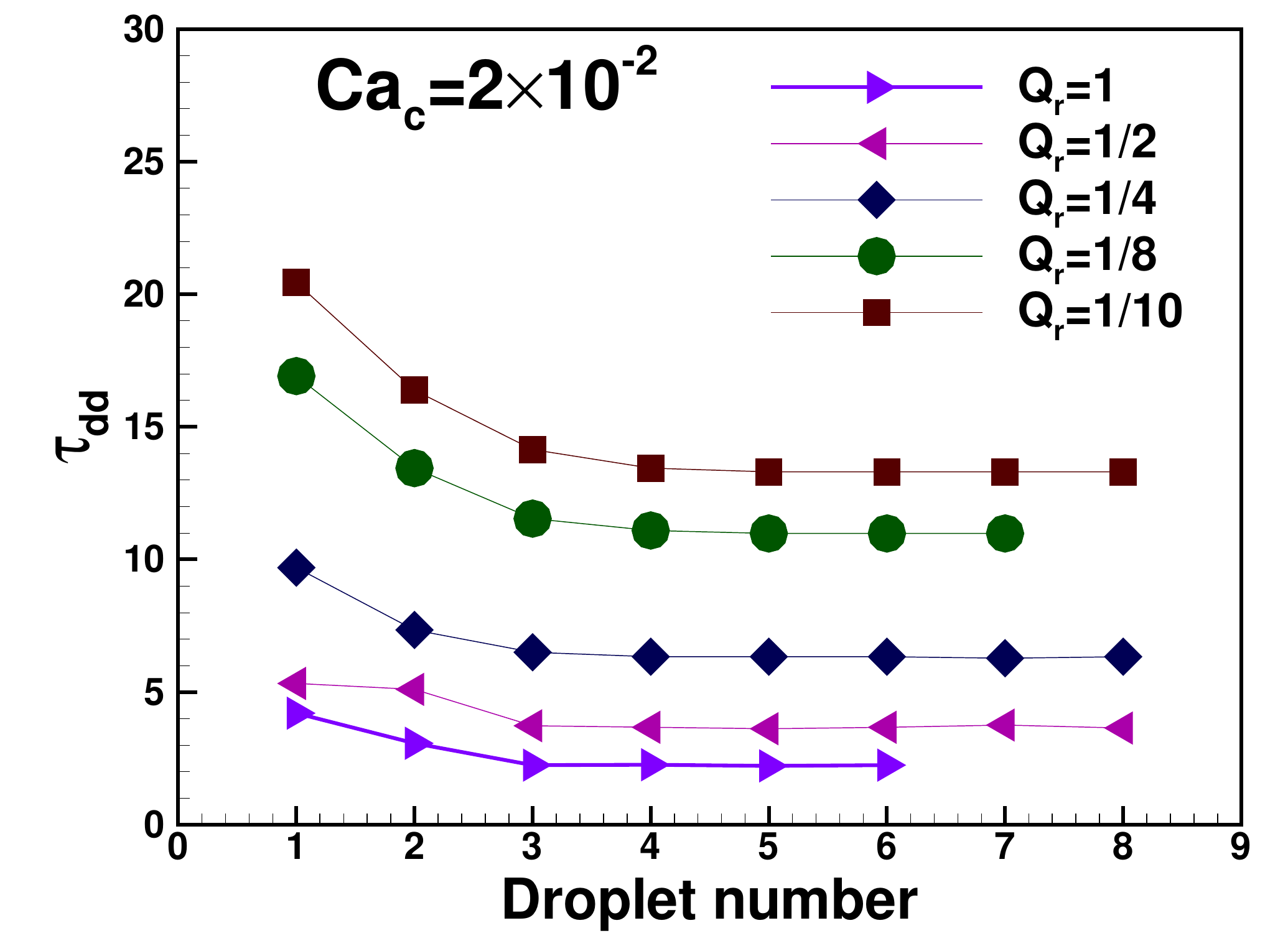}\label{fig:10c}}
	\subfloat[$\cac=10^{-1}$]{\includegraphics[width=0.48\linewidth]{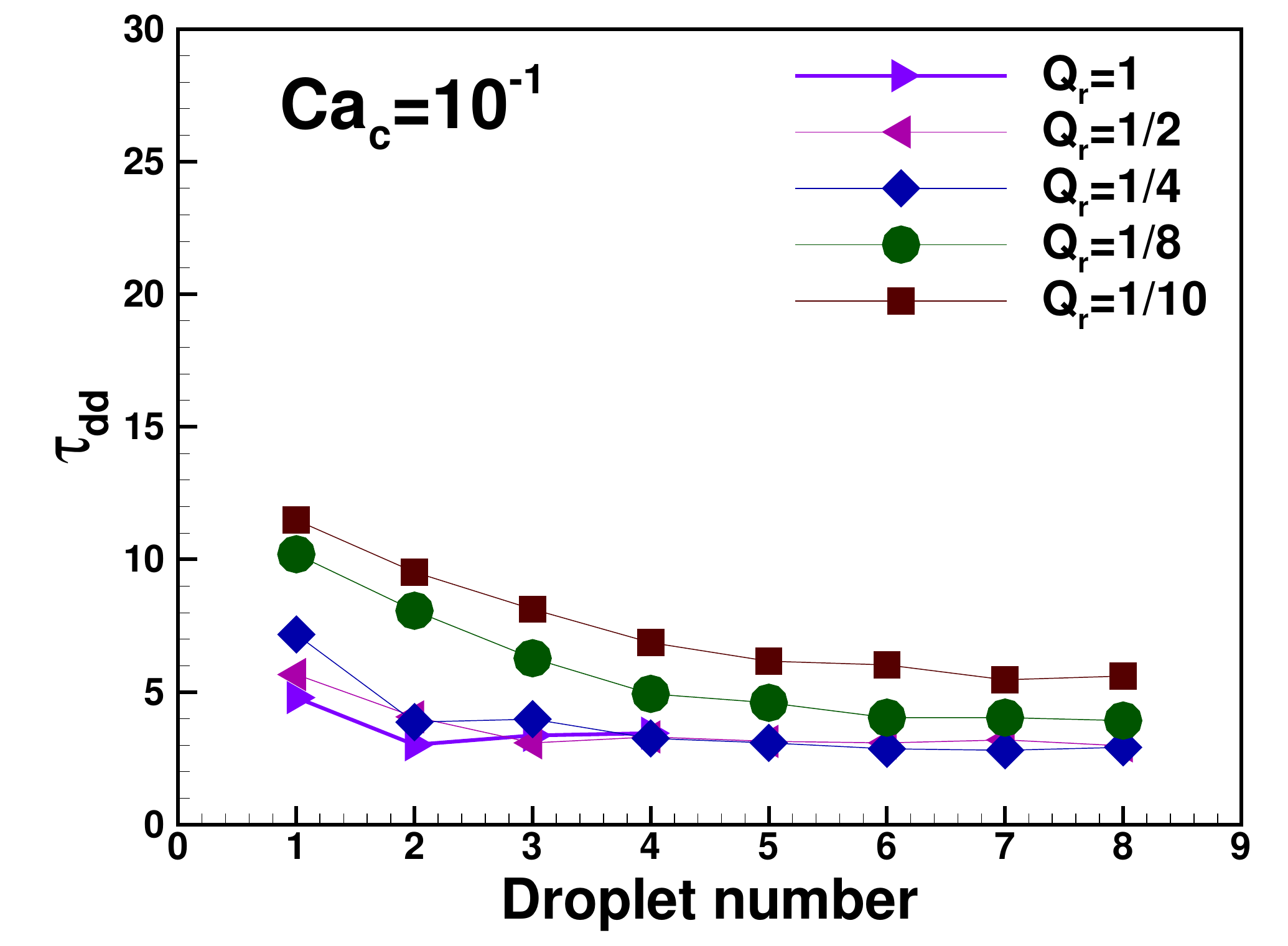}\label{fig:10d}}
	\caption{Droplet detachment time ($\tau_{\text{dd}}$) as a function of $\qr$ for fixed  $\cac$.}
	\label{fig:10}
\end{figure}

\noindent The dimensionless detachment time of the droplet is defined as $\tau_{\text{dd}}=(u_{c}t_{\text{dd}}/{w_{c}})$, where $u_{c}$ is the velocity of the continuous phase and $t_{\text{dd}}$ (\eqn\ref{eq:tfdd}) is detachment time of droplet.  \figs \ref{fig:9} and \ref{fig:10}  illustrate the variation of the droplet detachment time ($\tau_{\text{dd}}$) with $\cac$ and $\qr$ under the droplet formation \rev{regimes (\eqn\ref{eq:10})}. 
To explore the effect of interfacial tension on the droplet generation, \fig \ref{fig:9} shows variation of $\tau_{\text{dd}}$ (time interval between generation of two subsequent droplets) with $\cac$ for the fixed $\qr$.  In general, the droplets detachment time ($\tau_{\text{dd}}$) increases with decreasing $\cac$ for fixed $\qr$. It is observed that the time taken for the formation ($\tau_{\text{dd}}$)  of the first droplet is either more or equal than the subsequent droplets due to the hydrodynamic development in the main channel, depending on the values of $\cac$ and $\qr$.  
\\\noindent
Under the squeezing regime ($\qr>1$ and $\cac<10^{-2}$),  $\tau_{\text{dd}}$ is almost equal for all, including the first, droplets at very high $\qr$, see \fig \ref{fig:9a},  whereas as $\qr$ decreased, $\tau_{\text{dd}}$ for first droplets is greater than that for all other subsequent droplets formed at approximately constant $\tau_{\text{dd}}$, see \fig \ref{fig:9b}. 
For $\qr=1$, $\tau_{\text{dd}}$ is decreasing with increasing $\cac$ till the formation of the few droplets and showing some fluctuations in the initial stage because both phases are flowing with the same flow rates having equal viscosity and density, as shown in \fig \ref{fig:9c}.  For $\qr>1$, $\tau_{\text{dd}}$ is decreasing smoothly and reaching steady for all the values of $\cac$, as shown in \fig \ref{fig:9d}. 
\\\noindent
Further, to explore the effect of relative flow rates of the two-phases on the droplet generation, \fig \ref{fig:10} shows the droplet formation time ($\tau_{\text{dd}}$) of each droplet as a function of $\qr$ for the fixed $\cac$. 
Here also $\tau_{\text{dd}}$ is decreasing with a decrease in $\qr$ for a fixed $\cac$. 
The $\tau_{\text{dd}}$ is smoothly decreasing and becoming constant after the formation of a few droplets. It implies that both viscosity and interfacial tension have qualitatively similar influences on the droplet formation time ($\tau_{\text{dd}}$) in the droplet zone, i.e., squeezing and dripping flow regimes. 
The droplet detachment time ($\tau_{\text{dd}}$) shows a complex dependence on $\qr$ and $\cac$. Such behaviour of $\tau_{\text{dd}}$ is qualitatively consistent with the previous studies \citep{Husny2006,Nazari2018} for varied geometrical arrangements of the microfluidic devices. In particular, a smaller value of $\tau_{\text{dd}}$ is favorable to produce mono-dispersed droplets of uniform size. It is further analyzed in the subsequent section in terms of the frequency of the droplet generation. 
\begin{figure}[!b]
	\centering
	\subfloat[$\overline{f}_{\text{dd}}$ vs. $\cac$]{\includegraphics[width=0.48\linewidth]{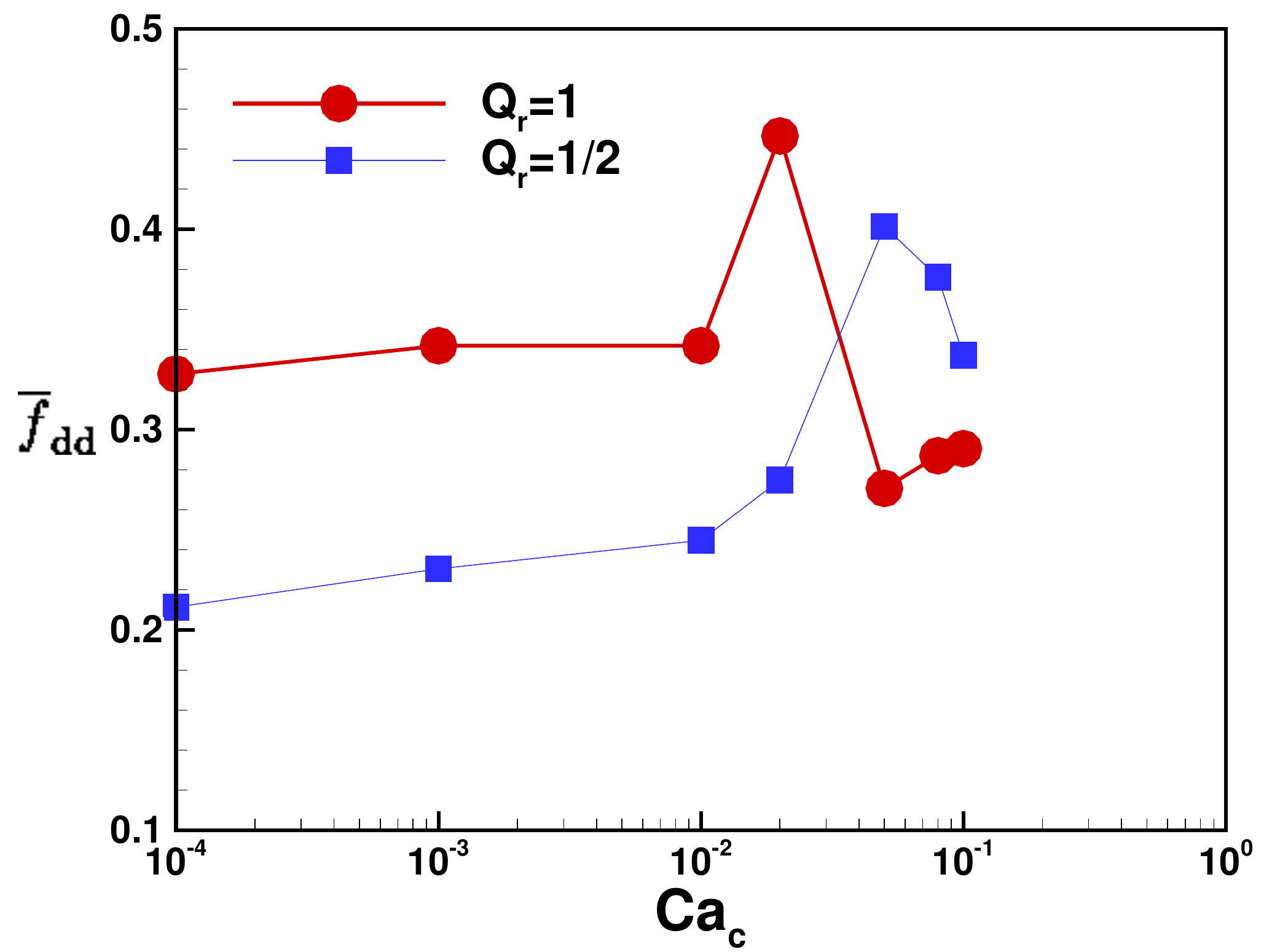}\label{fig:12}}
	\subfloat[$\overline{f}_{\text{dd}}$ vs. $\cac$]{\includegraphics[width=0.48\linewidth]{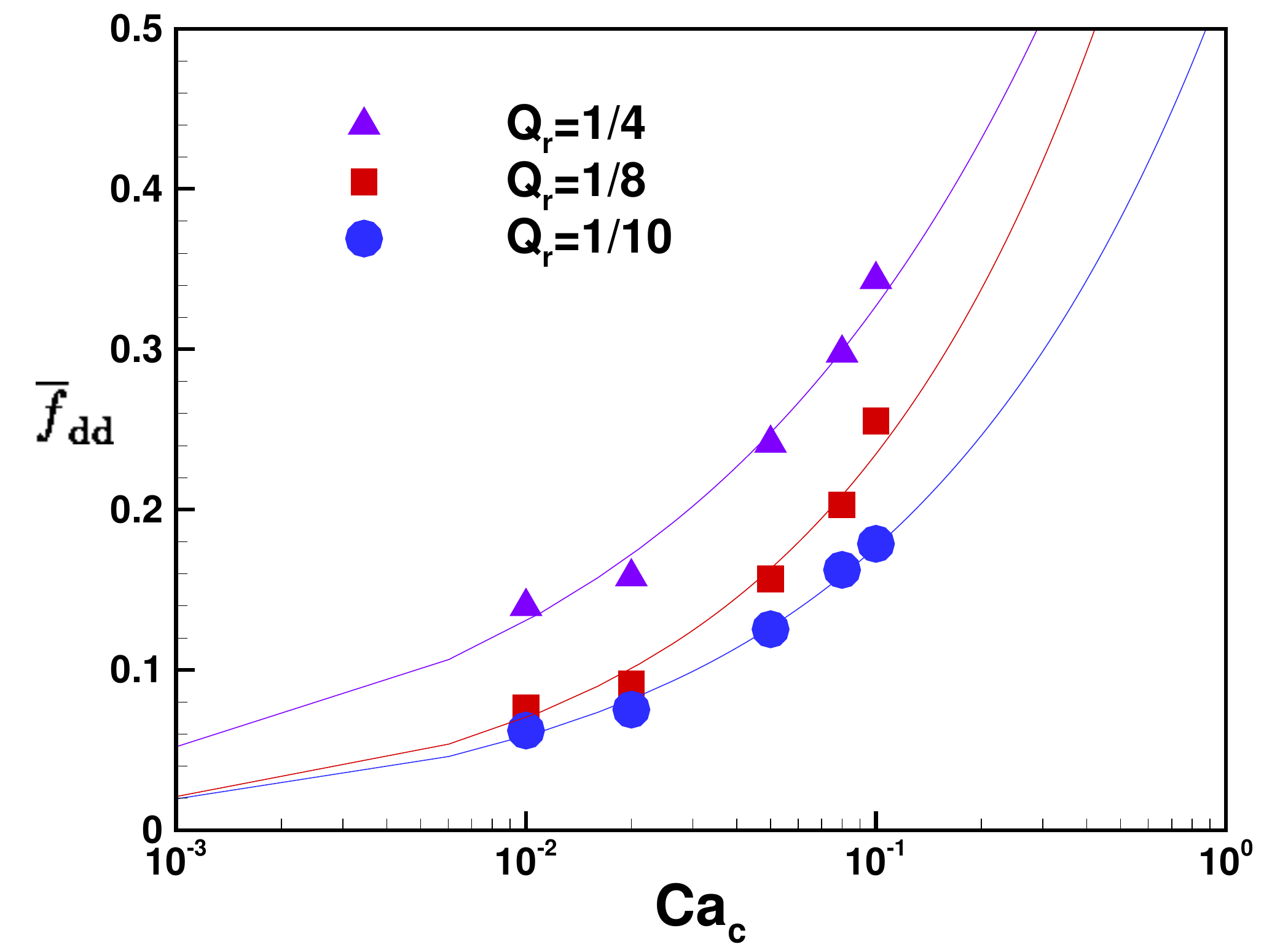}\label{fig:11a}}
	\\
	\subfloat[$\overline{f}_{\text{dd}}$ vs. $\qr$ ]{\includegraphics[width=0.48\linewidth]{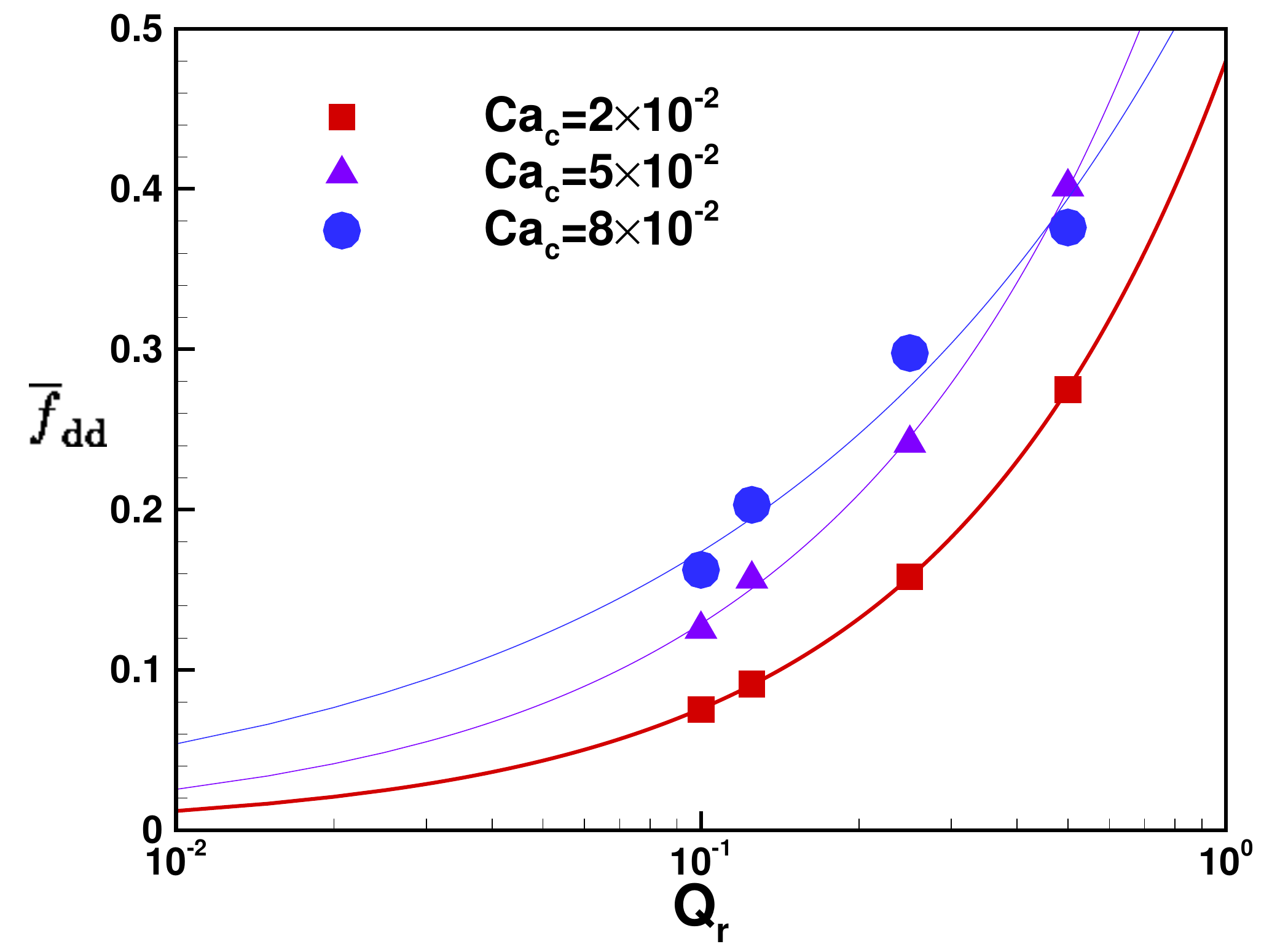}\label{fig:11b}}
	\subfloat[$\overline{f}_{\text{dd}}$ (\eqn \ref{eq:dfr3}) vs. $\overline{f}_{\text{dd}}$ (numerical)]{\includegraphics[width=0.48\linewidth]{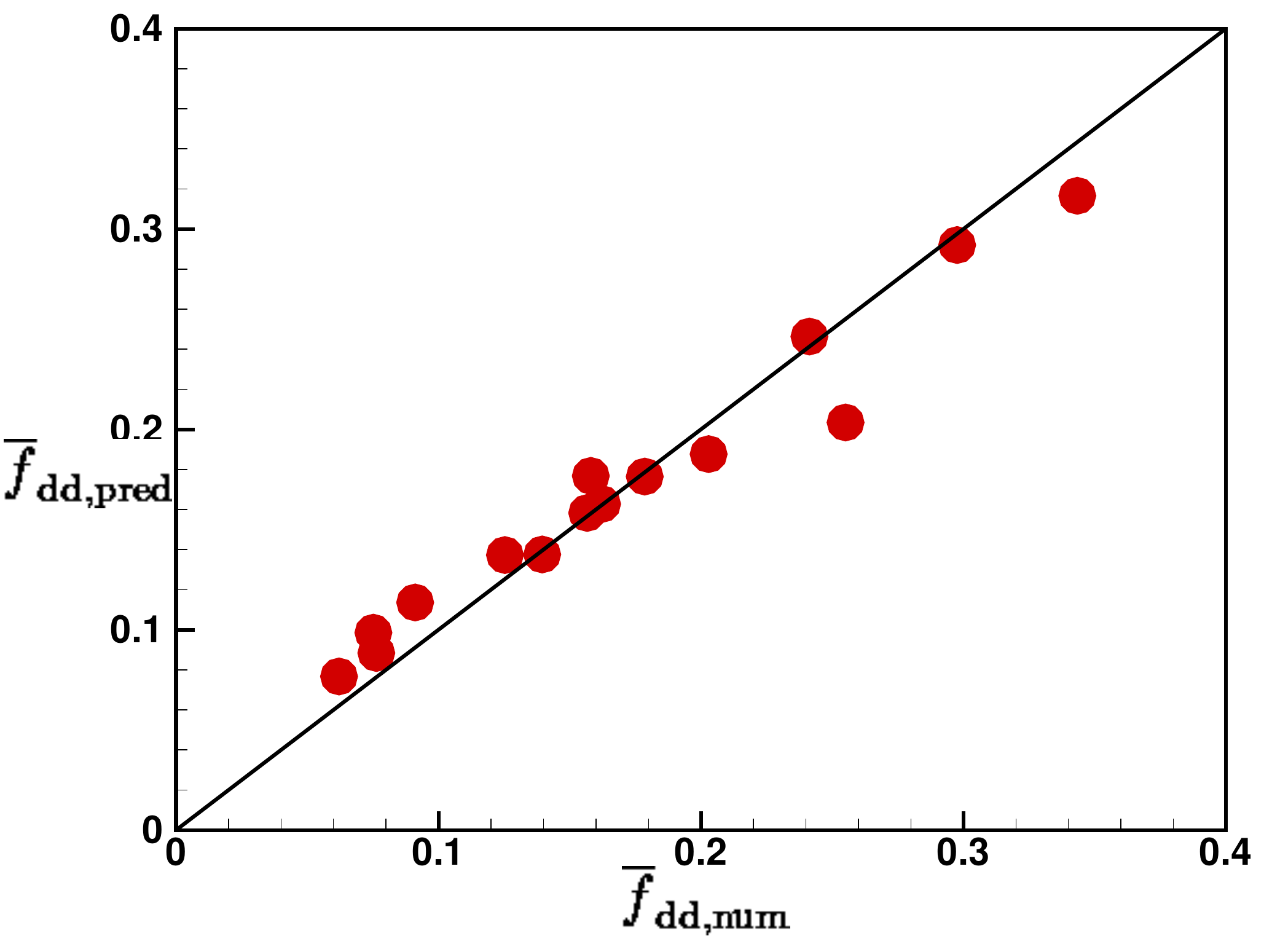}\label{fig:13}}
	\caption{Dependence of droplet frequency on $\cac$ and $\qr$. }
	\label{fig:11}
\end{figure}
%
\subsection{Droplet generation frequency}
\noindent 
The dimensionless droplet generation frequency ($\overline{f}_{\text{dd}}=1/\tau_{\text{dd}}$, \eqn\ref{eq:tfdd}) is calculated as the inverse of the detachment time of the droplet ($\tau_{\text{dd}}$, \figs \ref{fig:9} and \ref{fig:10}).  The dependence of $\overline{f}_{\text{dd}}$ on $\qr$ and $\cac$ under the droplet zone is shown in \fig \ref{fig:11}. 
When the flow rates of both the continuous and dispersed phases are equal (i.e., $\qr=1$), $\overline{f}_{\text{dd}}$ remains almost constant in the squeezing regime for the lower values of $\cac~(<10^{-2})$. 
There is, however, a sudden jump in $\overline{f}_{\text{dd}}$ at $\cac=2\times10^{-2}$ and it is decreasing beyond $\cac >10^{-2}$, as seen in \fig \ref{fig:12}. 
In the case of $\qr=1$ in the present study, the viscosities of both phases are equal. Therefore, the flow patterns transit into the parallel regime after forming few droplets with increasing $\cac$. The shear stress exerted by the continuous fluid on the dispersed phase interface is insufficient to overcome the interfacial forces to create a droplet.  Whereas for $\qr=1/2$, the frequency $\overline{f}_{\text{dd}}$ increases with the increase $\cac$ and attaining a maximum value at $\cac=5\times10^{-2}$, and it decreases sharply thereafter.  
Such a complex dependence of $\overline{f}_{\text{dd}}$ on $\qr$ and $\cac$ is attributed to the significantly larger elongated droplets ($L\ggg w_{\text{c}}$) observed under the squeezing and first transitional flow regimes.  
\\\noindent
In contrast to nature of  $\overline{f}_{\text{dd}}$  seen in  \fig \ref{fig:12}, the behaviour of $\overline{f}_{\text{dd}}$ is, however, quite different in \figs \ref{fig:11a} and \ref{fig:11b} under the dripping regime.  
In general, the frequency of droplet generation proportionally enhances, i.e., an increasing number of droplets per unit time, with both increasing  $\qr$ and $\cac$, as displayed by \figs \ref{fig:11a} and \ref{fig:11b}.
The present droplet frequency results are consistent with previous experimental and numerical studies \citep{Christopher2008,Gupta2010} for varied geometrical arrangements.  
\begin{table}[!b]
	\centering
	\caption{Statistical analysis for dependence of $\overline{f}_{\text{dd}}$ on $\cac$ and $\qr$.}
\label{tab:3}
	\subfloat[$\overline{f}_{\text{dd}}=\alpha \cac^{\beta}$\label{tab:3a}]{\scalebox{0.9}{\begin{tabular}[t]{cccc}
			\hline 
			$10\qr$   &$\alpha$  &$\beta$   & $R^{2}$  \\ \hline
			$2.50$     &$0.8202$  &$0.3994$  & $0.9758$  	\\
			$1.25$  	  &$0.7856$  &$0.5249$  & $0.9746$   	\\
			$1.00$ 	  &$0.5321$  &$0.4791$  & $0.9874$   \\ \hline          
		\end{tabular}
	}}\qquad
	\subfloat[$\overline{f}_{\text{dd}}=\alpha \qr^{\beta}$\label{tab:3b}]{\scalebox{0.9}{\begin{tabular}[t]{cccc}
			\hline 
			$10^2\cac$          &$\alpha$  &$\beta$   & $R^{2}$\\ \hline
			$2$  &$0.4799$  &$0.8022$  & $0.9990$  	\\
			$5$  &$0.6511$  &$0.7041$  & $0.9968$\\
			$8$  &$0.5607$  &$0.5091$  & $0.9677$  \\ \hline       
		\end{tabular}
	}}
	\vspace{-1em}
\end{table}
\\\noindent
The influence of interfacial tension ($\cac \ge 10^{-2}$) on the droplet generation frequency ($\overline{f}_{\text{dd}}$) is depicted in \fig \ref{fig:11a} for the fixed $\qr$. The numerical data presented in \fig \ref{fig:11a} has shown the power-law dependence of $\overline{f}_{\text{dd}}$  on $\cac$  for the fixed  $\qr$  as follows. 
\begin{equation}
\overline{f}_{\text{dd}}=\alpha \cac^{\beta}
\label{eq:dfr2}
\end{equation}
The values of the statistical constants ($\alpha$ and $\beta$) are shown in Table \protect\ref{tab:3a} for the ranges of conditions ($1/10\leq \qr\leq 1/4$, and $10^{-2}\leq \cac\leq 10^{-1}$) of dripping flow regime. 
\\\noindent
Further,  \fig \ref{fig:11b}  depicts the influence of flow rates ratio ($\qr$) on the droplet generation frequency ($\overline{f}_{\text{dd}}$) for the fixed $\cac$.
A statistical analysis of the numerical data presented in \fig \ref{fig:11b} has also shown the power-law dependence of $\overline{f}_{\text{dd}}$  on $\qr$ for the fixed  $\cac$ as follow. 
\begin{equation}
\overline{f}_{\text{dd}}=\alpha \qr^{\beta}
\label{eq:dfr1}
\end{equation}
The values of the statistical constants ($\alpha$ and $\beta$) are shown in Table \protect\ref{tab:3b} for the ranges of conditions ($1/10\leq \qr\leq 1/4$; $10^{-2}\leq \cac\leq 10^{-1}$) of dripping flow regime.
\\\noindent 
Furthermore, to determine the combined influences of interfacial, inertial and viscous forces, a new correlation is proposed to predict the functional dependence of $\overline{f}_{\text{dd}}$ on $\qr$ and $\cac$ under the dripping flow regime ($1/10\leq \qr\leq 1/2$ and $10^{-2}\leq \cac\leq 0.1$) as follows. 
\begin{equation}
\overline{f}_{\text{dd}}=\alpha \qr^{\beta} \cac^{\gamma}
\label{eq:dfr3} 
\end{equation}
where $\alpha=2.3$, $\beta=0.417$ and $\gamma=0.685$ with the coefficient of determination $R^2=0.9634$. 
A parity plot  in \fig \ref{fig:13}  displays an excellent agreement between the present numerical and predicted (using \eqn \ref{eq:dfr3}) values of droplet generation frequency ($\overline{f}_{\text{dd}}$). 
%
%
\section{Conclusions}
\noindent 
The microfluidic phenomena of droplet formation in a two-dimensional cross-flow T-junction geometry have been studied numerically for a wide range of conditions ($Re_{\text{c}}=0.1$, $10^{-4}\le \cac\le 1$, $ 0.1 \le \qr\le 10$, $\rho_{\text{r}}=1$ and $w_{\text{r}}=1$) by using the \rev{conservative} level set method in conjunction with the finite element modelling approach. In-depth insights into droplet generation and dynamics are presented in terms of instantaneous phase ($\phi$) profiles, dimensionless droplet size ($L/w_{\text{c}}$), detachment time ($\tau_{\text{dd}}$), and frequency  ($\overline{f}_{\text{dd}}$) as a function of flow governing parameters ($\cac$ and $\qr$).  Evidently, even at $\cac>10^{-2}$ where the interfacial force dominates the pressure build-up in the upstream region is responsible for the droplet generation. 
The two-phase microfluidic flow is characterized as squeezing, the first transition, dripping, second transition, parallel, and jet-type flow regimes based on the nature of flow as droplets of elongated or circular shape, slug, and parallel layers. A transitional capillary number ($Ca_{\text{trans}}$) is defined to express the changes in the droplet shape from a regular mono-dispersed to an irregular poly-dispersed.  
Interestingly, the ratio of transition capillary numbers of continuous and dispersed phases varied proportionally as a quadratic function of their flow rate ratio, $Ca_{\text{r,trans}} = \beta Q^{2}_{\text{r}}$. This splits $Ca_{\text{r,trans}}$ versus $\qr$ plane into two zones: (i) droplet zone ($Ca_{\text{r,trans}} > \beta Q^{2}_{\text{r}}$),  and (ii) non-droplet zone ($Ca_{\text{r,trans}} < \beta Q^{2}_{\text{r}}$).  The droplet zone comprises of the squeezing, first transition, and dripping flow regimes. On the other hand, the second transition, parallel and jet-type flow regimes fall under the non-droplet zone. This classification helps to predict the possibility of the droplet formation for a particular $\cac$ and $\qr$.
The droplet dynamics have also shown the complex dependence on the governing parameters ($\cac$ and $\qr$). The droplet length is linearly dependent on $\qr$ in the squeezing flow regime, whereas power-law dependence on  $\cac$ and $\qr$ in the dripping flow regime.
A new correlation has been proposed to predict the dimensionless frequency of the droplet generation in the dripping flow regime as a power-law function of  $\cac$ and $\qr$. 
The present results have been verified and shown excellent correspondence with the limited experimental and numerical studies available in the literature. Finally, the presented flow regimes and predictive correlations guide their use in the engineering and design of microfluidic droplet generators. 
%
\section*{Declaration of Competing Interest}
\noindent All authors declare that they have no conflict of interest. The  authors  certify  that  they  have  NO  affiliations  with  or  involvement  in  any  organization or entity with any financial interest (such as honoraria; educational grants; participation in speakers’ bureaus; membership, employment, consultancies, stock ownership, or other equity interest; and expert testimony or patent-licensing arrangements), or non-financial interest (such as personal or professional relationships, affiliations, knowledge or beliefs) in the subject matter or materials discussed in this manuscript.
\section*{Acknowledgements}
\noindent R.P. Bharti would like to acknowledge Science and Engineering Research Board (SERB), Department of Science and Technology (DST), Government of India (GoI) for providence of MATRICS grant (File No. MTR/2019/001598). 
%
%
{
\noindent 
\nomenclature[z0]{\textit{Abbreviations}}{}
\nomenclature[g0]{\textit{Greek letters}}{}
\nomenclature[d0]{\textit{Dimensionless groups}}{}
%
%
 \nomenclature[zbdf]{BDF}{backward differentiation formula}
 \nomenclature[zcfd]{CFD}{computational fluid dynamics}
 \nomenclature[zcp]{CP}{continuous phase}
 \nomenclature[zdp]{DP}{disperse phase}
 \nomenclature[zdae]{DAE}{differential algebraic equations}
 \nomenclature[zpfm]{PFM}{phase field method}
 \nomenclature[zlbm]{LBM}{lattice Boltzmann method}
 \nomenclature[zlsm]{LSM}{level set method}
 \nomenclature[zfdm]{FDM}{finite difference method}
 \nomenclature[zfem]{FEM}{finite element method}
 \nomenclature[zfvm]{FVM}{finite volume method}
 \nomenclature[zpardiso]{PARDISO}{parallel direct solver}
 \nomenclature[zvof]{VOF}{volume of fluid}
%
%
\nomenclature[adeff]{$d_{\text{eff}}$}{effective droplet diameter (\eqn\ref{eq:deff}), m}
\nomenclature[aDt]{$\mathbf{D}$}{rate of strain tensor (\eqn\ref{eq:tauD}), s$^{-1}$}
\nomenclature[aFsigma]{$\mathbf{F}_{\sigma}$}{interfacial force (\eqn\ref{eq:Fsigma}), N}
\nomenclature[aLu]{$L_{\text{u}}$}{upstream length of the main channel, m} 
\nomenclature[aLd]{$L_{\text{d}}$}{downstream length of the main channel, m}
\nomenclature[aLm]{$L_{\text{m}}$}{length of the main channel, m}
\nomenclature[aLs]{$L_{\text{s}}$}{length of the side channel, m}
\nomenclature[aP]{$p$}{pressure, Pa} 
\nomenclature[aQc]{$Q_{\text{c}}$}{flow rate of CP, m$^{3}$/s}
\nomenclature[aQd]{$Q_{\text{d}}$}{flow rate of DP, m$^{3}$/s}
\nomenclature[aQr]{$Q_{\text{r}}$}{flow rate ratio (\eqn\ref{eq:dimp2}), dimensionless}
\nomenclature[afdd]{$f_{\text{dd}}$}{droplet detachment frequency, s$^{-1}$}
\nomenclature[atdd]{$t_{\text{dd}}$}{droplet detachment time, s}
\nomenclature[aRec]{$Re_{\text{c}}$}{Reynolds number for CP (\eqn\ref{eq:dimp1}), dimensionless}
\nomenclature[aRed]{$Re_{\text{d}}$}{Reynolds number for DP (\eqn\ref{eq:dimp1}), dimensionless}
\nomenclature[aRer]{$Re_{\text{r}}$}{ratio of Reynolds numbers (\eqn\ref{eq:dimp2}), dimensionless}
\nomenclature[aCac]{$Ca_{\text{c}}$}{capillary number for CP (\eqn\ref{eq:dimp1}), dimensionless}
\nomenclature[aCad]{$Ca_{\text{d}}$}{capillary number for DP (\eqn\ref{eq:dimp1}), dimensionless}
\nomenclature[aCar]{$Ca_{\text{r}}$}{ratio of capillary numbers (\eqn\ref{eq:dimp2}), dimensionless}
\nomenclature[aCactrans]{$Ca_{\text{c,trans}}$}{transitional or threshold capillary number for CP (\eqn\ref{eq:10}), dimensionless}
\nomenclature[aCadtrans]{$Ca_{\text{d,trans}}$}{transitional capillary number for DP (\eqn\ref{eq:10}), dimensionless}
\nomenclature[aCartrans]{$Ca_{\text{r,trans}}$}{ratio of transitional capillary numbers (\eqn\ref{eq:10}), dimensionless}
\nomenclature[aU]{$\mathbf{u}$}{velocity vector, m/s}
\nomenclature[awc]{$w_{\text{c}}$}{width of the main channel, m}
\nomenclature[awd]{$w_{\text{d}}$}{width of the side channel, m}
\nomenclature[awr]{$w_{\text{r}}$}{channel width ratio (\eqn\ref{eq:dimp2}), dimensionless}
\nomenclature[ax]{$x$}{stream-wise coordinate}
\nomenclature[ay]{$y$}{transverse coordinate}
%
%
\nomenclature[ggamma]{$\gamma$}{re-initialization or stabilization parameter  (\eqn\ref{eqn:lsm}), m/s}
\nomenclature[gepsilon]{$\epsilon_{\text{ls}}$}{interface thickness controlling parameter  (\eqn\ref{eqn:lsm}), m}
\nomenclature[gmuc]{$\mu_{\text{c}}$}{viscosity of CP, Pa.s}
\nomenclature[gmud]{$\mu_{\text{d}}$}{viscosity of DP, Pa.s}
\nomenclature[gmur]{$\mu_{\text{r}}$}{viscosity ratio (\eqn\ref{eq:dimp2}), dimensionless}
\nomenclature[grhoc]{$\rho_{\text{c}}$}{density of CP, kg/m$^3$}
\nomenclature[grhod]{$\rho_{\text{d}}$}{density of DP, kg/m$^3$}
\nomenclature[grhor]{$\rho_{\text{r}}$}{density ratio (\eqn\ref{eq:dimp2}), dimensionless}
\nomenclature[gsigma]{$\sigma$}{interfacial tension, N/m}
\nomenclature[gtau]{$\tau$}{extra stress tensor (\eqn\ref{eq:tauD}),  N/m$^2$}
\nomenclature[gphi]{$\phi$}{level set function, dimensionless}
\nomenclature[gkappa]{$\kappa$}{curvature of the interface, m}
\nomenclature[gtheta]{$\theta$}{contact angle, degrees}
\nomenclature[gtaudd]{$\tau_{\text{dd}}$}{droplet detachment time, dimensionless}
%
%
\nomenclature[dRe]{$Re$}{Reynolds number (\eqn\ref{eq:dimp1}), dimensionless}
\nomenclature[dCa]{$Ca$}{Capillary number (\eqn\ref{eq:dimp1}), dimensionless}
%
%
\renewcommand{\nompreamble}{\vspace{1em}\fontsize{10}{8pt}\selectfont}
{\printnomenclature[5em]}}
%
%
%
\bibliographystyle{elsarticle/elsarticle-harv}\biboptions{authoryear}
\bibliography{references}
%
%
%
%
%
%
%
%
%
%
%
\end{document}